\documentclass[12pt]{article} 

\usepackage{PRIMEarxiv}
\usepackage{amsfonts,amsmath,amssymb,authblk,fullpage,graphicx,tabularx,threeparttable,tikz,pgfplots}
\usepackage[utf8]{inputenc} 
\usepackage[T1]{fontenc}    
\usepackage{hyperref}       
\usepackage{url}            
\usepackage{booktabs}       
\usepackage{amsfonts}       
\usepackage{nicefrac}       
\usepackage{microtype}      
\usepackage{lipsum}
\usepackage{graphicx}
\usepackage{bm}
\usepackage{amsmath}
\usepackage{amsmath}
\usepackage{enumerate}
\usepackage{subcaption}
\usepackage{epstopdf}
\usepackage{ragged2e}
\usepackage{grffile}
\usepackage[capitalise, noabbrev]{cleveref}
\crefname{equation}{}{} 
\crefname{definition}{Remark}{Remark}
\usepackage{caption}
\usepackage{subcaption}
\usepackage{amsthm}
\usepackage{tikz-dimline}
\usepackage{tikz,tikz-3dplot}
\usepackage{stmaryrd}
\usetikzlibrary{scopes}
\usetikzlibrary{tikzmark}
\usetikzlibrary{calc}
\usetikzlibrary{matrix}
\usetikzlibrary{backgrounds}
\usetikzlibrary{shapes.misc}
\usetikzlibrary{positioning}
\usetikzlibrary{arrows.meta,arrows}
\usetikzlibrary{decorations.shapes}
\usetikzlibrary{patterns}
\usetikzlibrary{hobby}
\usetikzlibrary{spy}
\usepgfmodule{nonlineartransformations}
\makeatletter
\usepackage{subfiles}
\usepackage{float}

\usepackage{algorithm}
\usepackage{algpseudocode}
\usepackage{caption}

\usepackage[normalem]{ulem} 
\usepackage{xcolor}

\newtheorem{definition}{Definition}[subsection]
\newtheorem{remark}[definition]{Remark}

\newcommand{\defgrad}{\bm{F}}

\newcommand{\cauchygreen}{\bm{C}}

\newcommand{\matderiv}{\nabla_{\bm{X}}}

\newcommand{\dispu}{\bm{u}}

\newcommand{\cordx}{\bm{x}}
\newcommand{\cordX}{\bm{X}}

\newcommand{\cauchystress}{\bm{\sigma}}

\newcommand{\secondPK}{\bm{S}}

\newcommand{\hff}{\Psi} 

\newcommand{\powi}{{(i)}}

\newcommand{\transpose}[1]{{#1}^{\text{T}}}
\newcommand{\transposeinv}[1]{{#1}^{-\text{T}}}
\newcommand{\inverse}[1]{{#1}^{-1}}

\newcommand{\magvector}[1]{\parallel {#1} \parallel}
\newcommand{\pd}[2]{\frac{\partial #1}{\partial #2}}

\newcommand{\dgamma}{\textrm{d}\gamma}

\newcommand{\normal}{\bm{n}}

\newcommand{\slavesurf}{\gamma_c^{(1)}}
\newcommand{\mastersurf}{\gamma_c^{(2)}}
\newcommand{\tractiontangnet}{\bm{t}_{\bm{\tau}} }

\newcommand{\tangent}{\bm{\tau}}
\newcommand{\tangentj}{\bm{\tau}_j}

\newcommand{\gap}{g_n}

\newcommand{\lm}{\bm{\lambda}_c}
\newcommand{\lmnormal}{{\lambda}_{c,n}}
\newcommand{\lmnormalj}{{\lambda}_{c,n,j}}

\newcommand{\intslave}{\int_{\gamma_c^{(1)}} }

\newcommand{\proj}[1]{\llbracket #1 \rrbracket}
\newcommand{\projoper}[1]{(#1)^{(1)}-[(#1)^{(2)} \circ \chi_t]}
\newcommand{\weightedgap}{\tilde{g}_{n,j}}

\newcommand{\lmms}{\bm{\lambda}_{m}}

\newcommand{\ga}[2]{\frac{ #1 \cdot #2}{ \parallel #1 \parallel \parallel #2 \parallel}}

\newcommand{\direcij}{\bm{v}^i_j}
\newcommand{\direcAvg}{\Bar{\bm{v}}^i}

\newcommand{\consEdge}{G_E}
\newcommand{\consEdgeAvg}{\Bar{G}_E}
\newcommand{\consEdgeInd}{\widehat{G}_E}
\newcommand{\consAngle}{G_A}

\newcommand{\rlength}{l^i_{\text{r}}}
\newcommand{\rlengthi}[1]{l^{#1}_{\text{r}}}

\newcommand\myeq{\mathrel{\stackrel{\makebox[0pt]{\mbox{\normalfont\tiny !}}}{=}}}

\title{A novel mesh regularization approach based on finite element distortion potentials: Application to material expansion processes with extreme volume change}

\author{ Abhiroop Satheesh, Christoph P. Schmidt, Wolfgang A. Wall, Christoph Meier  \\
Institute for Computational Mechanics  \\
Technical University of Munich  \\
Garching, Germany, 85748\\
\texttt{abhiroop.satheesh@tum.de, christoph.schmidt@tum.de, wolfgang.a.wall@tum.de,christoph.anton.meier@tum.de}
} 

\begin{document}
\maketitle
\begin{abstract}
The accuracy of finite element solutions is closely tied to the mesh quality. In particular, geometrically nonlinear problems involving large and strongly localized deformations often result in prohibitively large element distortions. In this work, we propose a novel mesh regularization approach allowing to restore a non-distorted high-quality mesh in an adaptive manner without the need for expensive re-meshing procedures. The core idea of this approach lies in the definition of a finite element distortion potential considering contributions from different distortion modes such as skewness and aspect ratio of the elements. The regularized mesh is found by minimization of this potential. Moreover, based on the concept of spatial localization functions, the method allows to specify tailored requirements on mesh resolution and quality for regions with strongly localized mechanical deformation and mesh distortion. In addition, while existing mesh regularization schemes often keep the boundary nodes of the discretization fixed, we propose a mesh-sliding algorithm based on variationally consistent mortar methods allowing for an unrestricted tangential motion of nodes along the problem boundary. Especially for problems involving significant surface deformation (e.g., frictional contact), this approach allows for an improved mesh relaxation as compared to schemes with fixed boundary nodes. To transfer data such as tensor-valued history variables of the material model from the old (distorted) to the new (regularized) mesh, a structure-preserving invariant interpolation scheme for second-order tensors is employed, which has been proposed in our previous work and is designed to preserve important properties of tensor-valued data such as objectivity and positive definiteness. As a practically relevant application scenario, we consider the thermo-mechanical expansion of materials such as foams involving extreme volume changes by up to two orders of magnitude along with large and strongly localized strains as well as thermo-mechanical contact interaction. For this scenario, it is demonstrated that the proposed regularization approach preserves a high mesh quality at small computational costs. In contrast, simulations without mesh adaption are shown to lead to significant mesh distortion, deteriorating result quality, and, eventually, to non-convergence of the numerical solution scheme.
\end{abstract}

\section{Introduction}
The accuracy of finite element solutions is closely tied to the quality of the underlying finite element mesh. Excessive element distortions can result in unreliable solutions or even lead to divergence. Therefore, it is important to keep mesh distortions small to achieve accurate and reliable results. To address this issue, dynamic mesh treatment techniques have been developed and studied. In general, replacing the old mesh by a newly generated mesh (i.e., remeshing) is typically a computationally inefficient approach, especially in 3D, as it results in high computational costs and challenges regarding parallel communication. In addition, the inherent challenges of producing high-quality meshes for complex geometries are re-occurring in every remeshing step, which also means that it requires additional means for maintaining mesh quality. In contrast, mesh adaptation, i.e. moving element nodes while keeping the number of nodes and their connectivity fixed in the sense of r-refinement, allows for updating the mesh without consuming excessive computational resources. For many applications, this seems to be a more efficient approach. Mesh adaptation typically consists of two major steps viz.~finding the new nodal positions (new mesh) and transferring data from the old to the new mesh. In this work, we term the first step as \textit{mesh refitting} and the second step as \textit{data transfer}, while the overall procedure including both steps is termed \textit{mesh adaptation}.  

Mesh refitting techniques have been studied in the past, e.g., in the context of fluid-structure interaction based on arbitrary Lagrangian-Eulerian (ALE) discretizations. These methods can be classified as interpolation and physical analogy-based methods. A detailed review of mesh refitting methods can be found in~\cite{selim2016mesh}. In interpolation-based schemes, an interpolation function is used to obtain the new mesh and in general they do not require nodal connectivity information, which enables their application to polyhedral elements or unstructured grids. The most common methods in this category are transfinite interpolation~\cite{thompson1985tfi}, the algebraic damping method~\cite{zhao2003algebraic}, and radial basis function interpolation~\cite{de2007mesh}. Transfinite interpolation has the disadvantage of (potentially) producing inverted elements, while the algebraic damping method may yield non-smooth aspect ratio distributions in the domain. Additionally, the use of radial basis function methods can be computationally expensive. In contrast, physical analogy-based methods use the element connectivity information and find the new nodal positions by solving a system of equations formulated according to a physical process. The first approach in this class is the linear spring analogy scheme proposed by Batina~\cite{batina1990unsteady}. In this method, a fictitious spring is added to the discretization with a stiffness inversely proportional to the element edge length. This method frequently results in inverted elements and is less suited for large deformation problems. To prevent element inversion, modified spring analogies such as torsional spring~\cite{farhat1998torsional}, semi-torsional spring~\cite{blom2000considerations}, ball-vertex~\cite{bottasso2005ball}, and ortho-semi-torsional spring approaches were proposed. However, these methods are either limited to triangular elements (respectively, tetrahedral elements in 3D) or have been shown to exhibit poor performance for large deformation problems. Moreover, these methods can also result in inverted elements or boundary nodes that are not relaxed. Another set of methods in this group is given by Laplacian methods~\cite{crumpton1997implicit,burg2006analytic}, in which a Laplace equation is solved in the interior of the domain. These schemes allow for a certain degree of regularization for distorted meshes and they ensure that the interior nodes remain confined by the domain boundaries. However, Laplacian methods often result in a limited mesh movement, i.e., a limited mesh regularization, and may also lead to inverted elements (see~\cite{burg2006analytic}). Finally, in an elastostatic approach~\cite{johnson1994elastic,wall1999fluid}, the discretized domain is considered as an elastic body and the new nodal locations are obtained through the solution of an elasticity problem. Here, the new nodal positions are dependent on the values of Young's modulus and Poisson's ratio underlying the pseudo-elastic problem. Yet, a linear elastostatic equation can lead to inverted elements and non-linear constitutive equations can lead to poor element quality for large deformation problems. Improved elastostatic approaches can be found in~\cite{freitag1999tetrahedral,knupp1999matrix,yang2007mesh}. Still, these methods typically suffer from insufficient element quality control and are not well-suited for large deformation problems with localized mesh distortion.

In summary, there is still a need for mesh refitting techniques that can produce a high quality mesh for problems involving very large and strongly localized deformation. Existing methods tend to give either inverted elements or elements with low quality in such scenarios. Moreover, controlling the element size and quality at specific regions of the domain is not easily feasible with existing approaches. In addition, the mesh movement for boundary nodes is either limited or for some approaches even impossible, leading to distorted elements at the boundary.

Once the new mesh is obtained, the primary and history variables must be mapped from the old to the new mesh. The accuracy of the data mapping between the meshes is critical for all adaptive FEM procedures. The most common data types that arise as history variables are scalars and tensors. On the one hand, the mapping of scalar variables is well investigated, and methods such as moving least squares can be adopted~\cite{zienkiewicz1987simple,zienkiewicz1992superconvergent1,zienkiewicz1992superconvergent2,gu2004modified,brancherie2008consistent, kindo2014toward}. On the other hand, the mapping of tensor-valued data was not well-studied before~\cite{satheesh}. To bridge this gap, structure preserving tensor interpolation schemes have been proposed and evaluated in our previous work~\cite{satheesh}. These interpolation schemes fulfill essential properties of the underlying continuum mechanics problem such as objectivity, positive definiteness, and monotonicity of invariants of the interpolated tensors along with higher-order spatial convergence.

To overcome the aforementioned limitations of existing approaches, we propose a novel mesh adaptation scheme. The underlying mesh refitting approach is based on the definition of an element distortion potential considering contributions from different distortion modes such as skewness and aspect ratio of the elements. The regularized mesh is obtained by minimizing this potential. Moreover, based on the concept of spatial localization functions, the method allows to specify tailored requirements on mesh resolution and quality for regions with strongly localized mechanical deformation and mesh distortion. To address also problems involving significant surface deformation, we propose the usage of a mortar mesh-sliding scheme to allow for a tangential motion of boundary nodes without changing the boundary topology. It is demonstrated that this approach enables a significantly improved mesh relaxation as compared to schemes with fixed boundary nodes. To transfer tensor-valued history data from the old to the new mesh, we utilize the structure-preserving interpolation methods proposed in our previous work~\cite{satheesh}. The effectiveness of the proposed scheme is validated by means of several selected numerical examples. In particular, as a practically relevant application scenario, we consider the thermo-mechanical expansion of materials such as foams involving extreme volume changes by up to two orders of magnitude along with large and strongly localized strains as well as thermo-mechanical contact interaction. For this scenario it is demonstrated that the proposed regularization approach preserves a high mesh quality at small computational costs. In the investigated numerical examples, the computation time for mesh adaptation was typically in the order of only a few percent of the total simulation time. In contrast, simulations without mesh adaption are shown to lead to significant mesh distortion and, eventually, to non-convergence of the numerical solution scheme.

The remainder of this work is organized as follows: While the main novelty of the work, the overall mesh adaptation approach, can be applied to all kinds of problems, we demonstrate it using thermomechanical problems driven by the demands from specific applications. To ease presentation, we first introduce this problem class and related methods in~\cref{sec:probelm_defintion} before presenting the mesh adaptation in~\cref{sec:mesh_adapt}. In particular, the finite-strain inelastic material model is introduced in~\cref{sec:problem_material}, followed by the considered mortar methods for thermo-mechanical contact in~\cref{sec:thermomechanical_contact}, and finally discretization in space and time in~\cref{sec:space_time_discretization}. Next, the mesh adaptation method is detailed in~\cref{sec:mesh_adapt}, in which the mesh refitting problem is formulated in~\cref{sec:new_nodal_location}, and the data transfer methods are described in~\cref{sec:mesh_adapt_data_transfer}. Finally, selected numerical examples are presented in~\cref{sec:numerical_results} and the main novelties and findings of this work are summarized in~\cref{sec:conclusion}.

\section{Thermo-mechanical problem}\label{sec:probelm_defintion}
In this work, we propose a novel mesh adaption method, which is able to regularize strongly distorted meshes without the need for a complete remeshing. While the proposed method is very general, in this work as a demonstration example it is applied to a finite deformation thermo-mechanical problem involving thermally activated materials undergoing very large (inelastic) volume expansion as well as thermo-mechanical contact interaction, which typically results in prohibitively large mesh distortions if no mesh adaption is applied. We start with a description of the underlying thermo-mechanical problem. Next, the thermo-mechanical constitutive model for the inelastic expansion is presented, followed by the thermo-mechanical contact formulation. 
\subsection{Thermo-mechanical initial boundary value problem}
\label{sec:thermomech}
    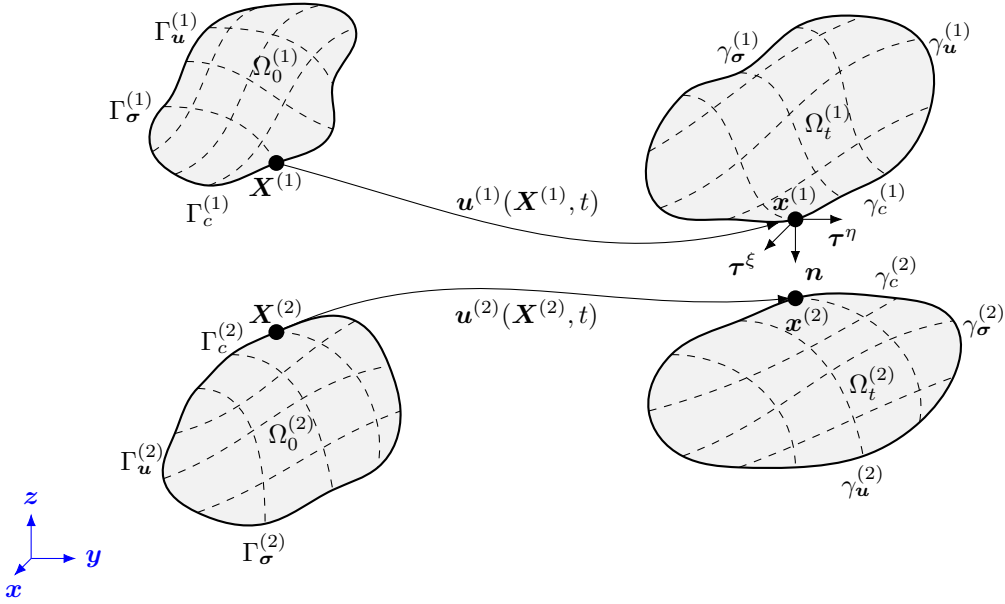
\begin{figure}
      \centering
      \scalebox{1}{\begin{tikzpicture}[scale=0.6]
    \def\ysby{0.75};
    
		\coordinate (0) at (1, 11+\ysby);
		\coordinate (1) at (1.5, 12+\ysby);
		\coordinate (2) at (2, 12.75+\ysby);
		\coordinate (3) at (3.5, 13.25+\ysby);
		\coordinate (4) at (4.5, 13.25+\ysby);
		\coordinate (5) at (5.25, 12.75+\ysby);
		\coordinate (6) at (4.75, 11.5+\ysby);
		\coordinate (7) at (4.75, 10.5+\ysby);
		\coordinate (8) at (3.5, 9.75+\ysby);
		\coordinate (9) at (2, 9.25+\ysby);
		\coordinate (10) at (1.25, 9.5+\ysby);
		\coordinate (11) at (0.75, 10+\ysby);
		
		\coordinate (12) at (1.75, 6.25-\ysby);
		\coordinate (13) at (2.5, 7-\ysby);
		\coordinate (14) at (3.5, 7.5-\ysby);
		\coordinate (15) at (5.25, 7.75-\ysby);
		\coordinate (16) at (6, 6.75-\ysby);
		\coordinate (17) at (6.25, 5.75-\ysby);
		\coordinate (18) at (5.75, 4.5-\ysby);
		\coordinate (19) at (4.75, 4-\ysby);
		\coordinate (20) at (3.25, 3.25-\ysby);
		\coordinate (21) at (1.75, 3.5-\ysby);
		\coordinate (22) at (1, 4.25-\ysby);
		\coordinate (23) at (1.25, 5.25-\ysby);
		
		\coordinate (24) at (12.5, 12);
		\coordinate (25) at (13.75, 12.5);
		\coordinate (26) at (15, 13.5);
		\coordinate (27) at (16.5, 13.75);
		\coordinate (28) at (17.75, 13.25);
		\coordinate (29) at (18, 11.75);
		\coordinate (30) at (17, 10.25);
		\coordinate (31) at (16, 9.75);
		\coordinate (32) at (15, 9.25);
		\coordinate (33) at (13.5, 9.25);
		\coordinate (34) at (12, 9.5);
		\coordinate (35) at (11.75, 10.75);
		\coordinate (36) at (12.25, 6.25);
		\coordinate (37) at (13.75, 7);
		\coordinate (38) at (15, 7.5);
		\coordinate (39) at (17.25, 7.5);
		\coordinate (40) at (18.5, 7);
		\coordinate (41) at (18.5, 5.75);
		\coordinate (42) at (17.75, 4.75);
		\coordinate (43) at (16.5, 4);
		\coordinate (44) at (15, 3.75);
		\coordinate (45) at (13.75, 3.75);
		\coordinate (46) at (12.5, 4);
		\coordinate (47) at (11.75, 5);
		
	\begin{scope}
	\usetikzlibrary{arrows,decorations.pathmorphing,backgrounds,positioning,fit,petri}
    \usetikzlibrary{decorations.pathreplacing}
    \usetikzlibrary{decorations.markings}
    \usetikzlibrary{decorations.shapes}
    \usetikzlibrary{arrows.meta}
    \usetikzlibrary{quotes,angles}
    \usetikzlibrary{positioning}
    \usetikzlibrary{patterns} 
    \usetikzlibrary{chains}
    \usetikzlibrary{hobby}
    
    \draw[fill=black!5,thick] (0) to [pattern=north east lines, closed, curve through = {(0) (1) (2) (3) (4) (5) (6) (7) (8) (9) (10) (11)}] (0);
    \draw[dashed] (0) to[out=0,in=120] (8);
    \draw[dashed] (1) to[out=0,in=160] (7);
    \draw[dashed] (2) to[out=0,in=115] (6);
    \draw[dashed] (11) to[out=30,in=180] (3);
    \draw[dashed] (10) to[out=20,in=190] (4);
    \draw[dashed] (9) to[out=20,in=200]  (5);

    \draw[fill=black!5,thick] (12) to [pattern=north east lines, closed, curve through = {(12) (13) (14) (15) (16) (17) (18) (19) (20) (21) (22) (23)}] (12);
    \draw[dashed] (12) to[out=0,in=90] (20);
    \draw[dashed] (13) to[out=0,in=100] (19);
    \draw[dashed] (14) to[out=0,in=80] (18);
    \draw[dashed] (23) to[out=15,in=210] (15);
    \draw[dashed] (22) to[out=20,in=190] (16);
    \draw[dashed] (21) to[out=20,in=200]  (17);

    \draw[fill=black!5,thick] (24) to [pattern=north east lines, closed, curve through = {(24) (25) (26) (27) (28) (29) (30) (31) (32) (33) (34) (35)}] (24);
    \draw[dashed] (24) to[out=0,in=180] (32);
    \draw[dashed] (25) to[out=0,in=150] (31);
    \draw[dashed] (26) to[out=0,in=150] (30);
    \draw[dashed] (35) to[out=30,in=200] (27);
    \draw[dashed] (34) to[out=20,in=190] (28);
    \draw[dashed] (33) to[out=20,in=200] (29);

   \draw[fill=black!5,thick] (36) to [pattern=north east lines, closed, curve through = {(36) (37) (38) (39) (40) (41) (42) (43) (44) (45) (46) (47)}] (36);    

    \draw[dashed] (36) to[out=0,in=90] (44);
    \draw[dashed] (37) to[out=0,in=100](43);
    \draw[dashed] (38) to[out=0,in=80] (42);
    \draw[dashed] (47) to[out=15,in=210] (39);
    \draw[dashed] (46) to[out=20,in=190] (40);
    \draw[dashed] (45) to[out=20,in=200] (41);
	\end{scope}	
	
	\begin{scope}	 
	    \tikzset{d/.style={minimum width=6pt,inner sep=0pt,circle,fill=black}}
        
    	\node[d] (x1) at (8) {};
		\node[d] (x2) at (14) {};
		\node[d] (x11) at (32) {};
		\node[d] (x22) at (38) {};

    	\node[below] at (x1) { $\bm{X}^{(1)}$ };
    	\node[above] at (x2) { $\bm{X}^{(2)}$ };
    	\node[above] at (x11) { $\bm{x}^{(1)}$ };
    	\node[below] at ([xshift=8pt]x22) { $\bm{x}^{(2)}$ };
    	
    	\draw [in=-165, out=-15,-Latex] (8.center) to node[midway,yshift=2pt,above, color=black,align=center]  {$\bm{u}^{(1)}(\bm{X}^{(1)},t)$} ([xshift=-10pt,yshift=-2pt]32.center) ;
    	\draw [in=-175, out=25,-Latex] (14.center) to node[midway,below, color=black,align=center]  {$\bm{u}^{(2)}(\bm{X}^{(2)},t)$} ([xshift=-4pt,xshift=1pt]38.center);

    	\draw[-Latex] (32) -- ([xshift=0pt,yshift=-28pt]32) node[below right,yshift=2pt, color=black,align=center ]    {$\bm{n}$};
		\draw[-Latex] (32) -- ([xshift=30pt,yshift=0pt]32) node[below, color=black,align=center ]     {$\bm{\tau}^{\eta}$};
		\draw[-Latex] (32) -- ([xshift=-20pt,yshift=-20pt]32) node[below left,yshift=3pt,xshift=1pt, color=black,align=center ]  {$\bm{\tau}^{\xi}$};
    	
       \node[] (100) at (0) {};
       \node[left] at (100) { $\Gamma_{\bm{\sigma}}^{(1)}$ };
       \node[] (100) at (2) {};
       \node[left] at (100) { $\Gamma_{\bm{u}}^{(1)}$ };
       \node[] (100) at (20) {};
       \node[below] at (100) { $\Gamma_{\bm{\sigma}}^{(2)}$ };
       \node[] (100) at (23) {};
       \node[below left] at (100) { $\Gamma_{\bm{u}}^{(2)}$ };
       \node[] (100) at (25) {};
       \node[above] at (100) { $\gamma_{\bm{\sigma}}^{(1)}$ };
       \node[] (100) at (28) {};
       \node[right] at (100) { $\gamma_{\bm{u}}^{(1)}$ };
       \node[] (100) at (40) {};
       \node[right] at (100) { $\gamma_{\bm{\sigma}}^{(2)}$ };
       \node[] (100) at (43) {};
       \node[below] at (100) { $\gamma_{\bm{u}}^{(2)}$ };

      \node[] (100) at ([yshift=45pt]8) {};
      \node[above] at (100) { $\Omega_{0}^{(1)}$ };
      \node[] (100) at ([yshift=45pt]32) {};
      \node[above right] at (100) { $\Omega_{t}^{(1)}$ };
      
      \node[] (100) at ([yshift=-80pt,xshift=10pt]14) {};
      \node[above ] at (100) { $\Omega_{0}^{(2)}$ };
      \node[] (100) at ([yshift=-70pt]38) {};
      \node[above right] at ([xshift=28pt]100) { $\Omega_{t}^{(2)}$ };
        
       \node[] (100) at (9) {};
       \node[below] at (100) { $\Gamma_{c}^{(1)}$ };
       \node[] (100) at (13) {};
       \node[above] at ([xshift=-5pt,yshift=-5pt]100) { $\Gamma_{c}^{(2)}$ };
       \node[] (100) at (30) {};
       \node[below] at (100) { $\gamma_{c}^{(1)}$ };
       \node[] (100) at (39) {};
       \node[above] at (100) { $\gamma_{c}^{(2)}$ };

	\end{scope}
	
\begin{scope}[yshift=35pt,xshift=-55pt]
    \draw[-Latex,blue] (0,0.5,0) -- (0,1.5,0) node[above, color=blue,align=center ]{$\bm{z}$};
	\draw[-Latex,blue] (0,0.5,0) -- (1,0.5,0) node[right, color=blue,align=center ]{$\bm{y}$};
	\draw[-Latex,blue] (0,0.5,0) -- (0,0.5,1) node[below, color=blue,align=center ]{$\bm{x}$};
\end{scope}

\end{tikzpicture}}  \caption{ Notation and kinematics to depict the interaction between two deformable bodies.}  \label{fig:contact_Kinematics}
    \end{figure}
     Consider the deformation of bodies $i=\{1,2\}$ with reference configuration $\cordX^\powi \in \Omega_0^\powi$ and current configuration $\cordx^\powi \in\Omega_t^\powi$ at time $t$ described by the mapping $\varphi_t^\powi: \cordX^\powi \mapsto \cordx^\powi$ as illustrated in~\cref{fig:contact_Kinematics}. The displacement $\dispu^\powi$ at material point $\cordX^\powi$ is given by $\dispu^\powi(\cordX^\powi,t)=\cordx^\powi(\cordX^\powi,t)-\cordX^\powi$ and the temperature is denoted by $T^\powi(\cordX^\powi,t)$. The thermo-mechanical initial boundary value problem (IBVP) summarizing the set of equations required to determine the displacement and temperature field, i.e. the primary variables $\dispu^\powi$ and $T^\powi$ in the time interval $t \in [0,t^E]$, reads:
        \begin{equation}\label{strong_form}
            \begin{aligned}
                \matderiv \cdot(\bm{F}^\powi \bm{S}^\powi)+\hat{\bm{b}}^\powi_{0} & =\rho_{0}^\powi \ddot{\bm{u}}^\powi &  & \textrm{ in } \Omega_0^\powi \times\left[0, t^E\right],          \\
                -\matderiv \cdot \bm{Q}^\powi+\hat{R}^\powi_{0}             & =c_{v}^\powi \dot{T}^\powi     &  & \textrm{ in } \Omega_0^\powi \times\left[0, t^E\right],          \\
                (\bm{F}^\powi \bm{S}^\powi) \bm{N}^\powi                                & =\hat{\bm{t}}_{0}^\powi       &  & \textrm{ on } \Gamma_{\sigma}^\powi \times\left[0, t^E\right], \\
                \bm{Q}^\powi \cdot \bm{N}^\powi                                   & =\hat{Q}^\powi_{0}            &  & \textrm{ on } \Gamma_{q}^\powi \times\left[0, t^E\right],      \\
                \bm{u}^\powi                                                & =\hat{\bm{u}}^\powi           &  & \textrm{ on } \Gamma_{u}^\powi \times\left[0, t^E\right],      \\
                T^\powi                                                     & =\hat{T}^\powi                &  & \textrm{ on } \Gamma_{T}^\powi \times\left[0, t^E\right],      \\
                \cauchystress^\powi \bm{n}^\powi                                & =\bm{t}_c^\powi       &  & \textrm{ on } \gamma_{c}^\powi \times\left[0, t^E\right], \\
                 \bm{q}^\powi \cdot \bm{n}^\powi                                   & ={q}^\powi_{c}            &  & \textrm{ on } \gamma_{c}^\powi \times\left[0, t^E\right],  \\
                                 {\bm{u}}^\powi                                              & ={\bm{u}}^\powi_{0}           &  & \textrm{ in } \Omega_0^\powi \textrm{ for } t=0,                             \\
                \dot{\bm{u}}^\powi                                          & =\dot{\bm{u}}^\powi_{0}       &  & \textrm{ in } \Omega_0^\powi \textrm{ for } t=0,                             \\
                T^\powi                                                     & =T_{0}^\powi                  &  & \textrm{ in } \Omega_0^\powi \textrm{ for } t=0,
            \end{aligned}
        \end{equation}
        where $\dot{(\cdot)}$ is the total time derivative, $\matderiv$ the gradient with respect to the material position vector $\cordX$, $\hat{\bm{b}}_{0}^\powi$ the body forces per unit reference volume, $\rho_{0}^\powi $ the mass density per unit reference volume, $\hat{R}^\powi_{0}$ the heat source per unit reference volume, and $c^\powi_{v}$ the specific heat capacity. $\bm{N}$ and $\normal$ represent outward unit-normal vectors onto the surfaces of the considered bodies in material and spatial description, respectively. Furthermore, $\defgrad$, $\secondPK$, $\cauchystress$, $\bm{Q}$, and $\bm{q}$ are the deformation gradient, second Piola--Kirchhoff stress tensor, Cauchy stress tensor, material heat flux, and spatial heat flux, respectively, and are detailed in the following sections. The first two equations in~\eqref{strong_form} are the momentum balance and the heat conduction equation, respectively. The Neumann boundary conditions for the mechanical problem on $\Gamma_\sigma^\powi$ and for the thermal problem on $\Gamma_q^\powi$ involve the prescribed fluxes $\hat{\bm{t}}^\powi_{0}$ and $\hat{Q}^\powi_{0}$. The Dirichlet boundary conditions for the mechanical problem on $\Gamma_u^\powi$ and for the thermal problem on $\Gamma_T^\powi$ are represented by the prescribed values $\hat{\bm{u}}^\powi$ and $\hat{{T}}^\powi$, respectively. The next two equations correspond to the Cauchy traction $\bm{t}_c^\powi$ and Cauchy heat flux $q_c^\powi$ at the contact surfaces $\gamma^\powi_c$. Finally, ${\bm{u}}_{0}^\powi$, $\dot{\bm{u}}_{0}^\powi$, and $T_{0}^\powi$ depict the initial conditions of the displacement, velocity, and temperature, respectively.
        
        For the subsequent finite element formulation, first the weak form of the coupled thermo-mechanical IBVP is formulated. The weak form is obtained by multiplying the linear momentum and heat conduction equations in~\cref{strong_form} with test functions $\delta \dispu^\powi$ and $\delta T^\powi$, respectively, then integrating over the domains, and applying Green’s theorem. Eventually, the weak forms of the mechanical and thermal problem without contact contribution read
        \begin{align}
           \underbrace{\int_{\Omega^\powi} \rho_{0}^\powi\ \delta\dispu^\powi \cdot \ddot{\dispu}^\powi \mathrm{~d} \Omega }_{\delta \mathcal{W}_u^{\text{iner}}}+\underbrace{\int_{\Omega^\powi} \nabla_{\cordX} \delta\dispu^\powi:(\bm{F}^\powi \bm{S}^\powi) \ \mathrm{d} \Omega}_{\delta \mathcal{W}_u^{\text{int}}}
            -\underbrace{\left(\int_{\Omega^\powi} \delta\dispu^\powi \cdot \hat{\bm{b}}^\powi_{0} \mathrm{~d} \Omega+\int_{\Gamma_{\sigma}^\powi} \delta\dispu^\powi \cdot \hat{\bm{t}}^\powi_{0} \mathrm{~d} \Gamma\right)}_{\delta \mathcal{W}_u^{\text{ext}}} &=0 \quad \label{mechanical_weakform}\\
         \text{and}\quad {\underbrace{\int_{\Omega^\powi}\delta T^\powi c_{v}^\powi \dot{T}^\powi \mathrm{~d} \Omega}_{\delta \mathcal{W}_T^{\text{iner}}}+\underbrace{\int_{\Omega^\powi} \nabla_{\cordX} \delta T^\powi \cdot \bm{Q}^\powi \mathrm{~d} \Omega \ }_{\delta \mathcal{W}_T^{\text{int}}}}  -\underbrace{\left(\int_{\Omega^\powi} \delta T^\powi \hat{R}_0^\powi \mathrm{~d} \Omega+\int_{\Gamma_{q}^\powi} \delta T^\powi \hat{Q}_{0}^\powi \mathrm{~d} \Gamma\right)}_{\delta \mathcal{W}_T^{\text{ext }}}&=0,\label{thermal_weakform}
        \end{align}
         respectively, wherein $\delta \mathcal{W}_u^{\text{iner}}$, $\delta \mathcal{W}_u^{\text{int}}$, and $\delta \mathcal{W}_u^{\text{ext}}$ are the mechanical inertia, internal, and external virtual work contributions, respectively, and the corresponding virtual work contributions of the thermal problem are denoted by $\delta \mathcal{W}_T^{\text{iner}}$, $\delta \mathcal{W}_T^{\text{int}}$, and $\delta \mathcal{W}_T^{\text{ext}}$.
        
\subsection{Kinematics and constitutive model for large deformation thermo-mechanical problem}\label{sec:problem_material}
    The mesh adaptation method proposed in Section~\ref{sec:mesh_adapt} is motivated by one of our current research questions involving material behavior with extreme volume expansion. Since the proposed mesh adaption scheme is independent of the specific form of the material law, and also for reasons of confidentiality by our industrial partner, the constitutive law governing the inelastic volume expansion, i.e., the function $f(T,\secondPK,\alpha)$ in equation~\eqref{evolution}, will be stated in a generic form below.
    
    Following the framework of nonlinear continuum mechanics, the local deformation at a material point $\cordX$ is described by the deformation gradient $\defgrad=\pd{\cordx}{\cordX}$. To account for inelastic deformations, we adopt the multiplicative split of the deformation gradient into an elastic part $\defgrad_e$ and an inelastic part $\defgrad_{in}$ as proposed by Lee~\cite{lee1967finite} in the context of plasticity:
    \begin{align}
    \label{multiplicative_split}
        \defgrad=\defgrad_e \defgrad_{in}.
    \end{align}
    We allow for an anisotropic inelastic volume expansion with respect to the principal stretch directions $\bm{N}_{C}^i$, which are given by the eigenvectors of the right Cauchy--Green stretch tensor $\cauchygreen=\transpose{\bm{F}}\bm{F}$, according to
    \begin{align}
        \bm{F}_{in} = \sum^{3}_{i=1}  \lambda^i_{in} \bm{N}_{C}^i \otimes \bm{N}_{C}^i,
    \end{align}
    where $\lambda^i_{in}$ is the magnitude of the inelastic expansion in the direction $\bm{N}_{C}^i$, governed by an evolution equation
        \begin{align}
        \label{evolution}
        \dot{\lambda}^i_{in}=f(T,\secondPK,\alpha),
    \end{align}
    accounting for dependencies of the inelastic volume expansion on the current temperature, stress state and on the scalar-valued internal variable $\alpha$ governing the material history. Exemplarily, the scalar $\alpha$ can be a material degradation factor which reduces from $1$ to $0$ during the expansion process. In the numerical examples studied in this work, we prescribe $\alpha$ as an explicit function of time. 
    
    In analogy to the (total) right Cauchy--Green stretch tensor $\cauchygreen$, the \textit{elastic} right Cauchy--Green stretch tensor is defined as $\cauchygreen_e=\transpose{\defgrad_e}\defgrad_e$. Based on $\cauchygreen_e$, we define a hyperelastic strain-energy function $\hff_e(\cauchygreen_e)$ under the assumption that the elastic response does not explicitly depend on the temperature. Based on this assumption and the multiplicative split~\eqref{multiplicative_split}, the second Piola--Kirchhoff stress tensor $\secondPK$ can be computed as (see~\cite{holzapfel2002nonlinear} for more details)
    \begin{align}
        \secondPK= 2 \inverse{\defgrad_{in}} \frac{\partial \hff_e}{\partial \cauchygreen_e} \transposeinv{\defgrad_{in}}. 
    \end{align}
    Furthermore, the spatial Cauchy stress tensor $\cauchystress$ follows as $\cauchystress=\inverse{(\text{det}\defgrad)} \defgrad \secondPK \transpose{\defgrad}$. In the examples presented in this paper, the hyperelastic strain-energy function is based on a Neo--Hookean model as presented in~\cite{holzapfel2002nonlinear}.

    Finally, the thermal constitutive equation relating heat flux and temperature gradient via the isotropic heat conductivity $k_0$ is formulated on the basis of Fourier's law. The latter can be stated using either the (material) second Piola--Kirchhoff heat flux $\bm{Q}$ or the spatial heat flux $\bm{q}=\inverse{(\text{det}\defgrad)} \defgrad \bm{Q}$ according to
    \begin{align}
    \label{heat_fluxes}
        \bm{Q} = -k_0 \bm{C}^{-1} \matderiv T \quad \text{or} \quad
        \bm{q} = - \frac{k_0}{\text{det} \bm{F} }\nabla_{\cordx}T,
    \end{align}
    where we distinguish between the material gradient $\matderiv=\frac{\partial}{\partial \cordX}$ and the spatial gradient $\nabla_{\cordx}=\frac{\partial}{\partial \cordx}$.
    
\subsection{Thermo-mechanical contact}
\label{sec:thermomechanical_contact}
    In this section, we discuss the thermo-mechanical contact formulation used in this work. The underlying contact constraints along with the basics of mortar methods for constraint enforcement and regularization are briefly summarized below. For a detailed description of these mortar methods in the context of thermomechanical contact interaction, the interested reader is referred to exemplary literature~\cite{dittmann2014isogeometric,pantuso2000finite}. While the general formulation accounts for frictional contact interaction, for simplicity, only the frictionless case will be recapitulated in the following.
    \subsubsection{Mechanical contact problem: Kinematics and contact forces}
         In the following, we distinguish the contacting surfaces as master and slave side denoted by the sets $\mathcal{M}$ and $\mathcal{S}$, respectively. Here, a superscript (1) refers to the slave side and $\gamma^{(1)}_c$ represents the contact surface on the slave side. The relative motion between these interfaces at any time instant $t$ is quantified by the smooth mapping (see~\cref{fig:contact_Kinematics}) 
        \begin{align} \label{slave_to_master_map}
            \chi_t (\cordx^{(1)}): \slavesurf \to \mastersurf, \cordx^{(1)} \mapsto \cordx^{(2)}.
        \end{align}
        This mapping projects any point $\cordx^{(1)}$ from the slave surface $\slavesurf$ onto the master surface $\mastersurf$ along the outward normal $\bm{n}(\cordx^{(1)})$ or in short $\bm{n}$ (see~\cref{fig:contact_Kinematics}). The unit vectors spanning the tangential plane at the contact point are denoted as $\tangent^{\eta}$ and $\tangent^{\xi}$. The mapping is assumed to exist in the zone of closed contact and its close vicinity. For (potentially) interacting points on the slave surface $\slavesurf$, the normal gap is defined as
        \begin{align}
        \label{normal_gap}
            \gap (\cordx^{(1)}) = -\bm{n} \cdot [\cordx^{(1)}-\cordx^{(2)}].
        \end{align}
        Moreover, the traction vectors acting on the contact surfaces are denoted as $\bm{t}^{(i)}_c$. Based on a balance of linear momentum, the traction vectors on the slave and master side of the contact surface are related according to $\bm{t}^{(1)}_c =-\bm{t}^{(2)}_c=:\bm{t}_c$. Furthermore, the contact traction can be decomposed into a normal component ${p}_n^{(i)}$ and a tangential component $\bm{t}_{\bm{\tau}}^{(i)}$:
        \begin{align}
            {p}_n^{(i)}              & = \bm{n} \cdot \bm{t}^{(i)}_c,                     \\
            \bm{t}_{\bm{\tau}}^{(i)} & = (\bm{I}-  \bm{n} \otimes \bm{n} )\bm{t}^{(i)}_c.
        \end{align}

\subsubsection{Mechanical contact problem: Constraints and virtual work}\label{sec:problem_contact}
    The (frictionless) mechanical contact constraints are given by the Hertz--Signorini--Moreau conditions:
    \begin{align}\label{KKT}
        \gap \geq 0, \quad p_n \leq 0, \quad p_n \gap =0 \quad \text{on } \slavesurf. 
    \end{align}
    As basis for a variational statement of the contact problem, the slave side traction vector $\bm{t}_c$ is introduced as an additional primary field, which is identified as Lagrange multiplier $\lm=-\bm{t}_c$ associated with the contact constraint. If the normal component of the Lagrange multiplier is denoted as $\lmnormal=\normal \cdot \lm$, the contact virtual work can be shown to yield:
    \begin{align}\label{contact_weak_form}
        \delta\mathcal{W}^{c}_{u}=- \intslave  \proj{\delta \bm{u}} \cdot \lm \ \dgamma = \intslave  \delta \gap \lmnormal \ \dgamma,
    \end{align}
    where $\proj{\cdot} =\projoper{\cdot}$ is the jump operator (cf.~\cref{slave_to_master_map}) and $\delta \gap$ the variation of the normal gap in~\eqref{normal_gap}. 
    
    \subsubsection{Thermal contact problem}\label{sec:problem_contact_thermal}
    
    Next, the thermal effects at the contacting surfaces have to be addressed. The heat balance at the interface reads
    \begin{align}
        q^{(1)}_c +   q^{(2)}_c =0,
    \end{align}
    with $ q^{(1)}_c$ and $q^{(2)}_c $ being the slave and master side heat fluxes across the contact interface defined according to (cf.~\cref{heat_fluxes})
    \begin{align}
        q^{(i)}_c = \bm{q}^{(i)} \cdot \bm{n}^{(i)}.
    \end{align}
    Within this work, an interface heat flux model with a linear dependence on the contact pressure is used according to
    \begin{equation}\label{heat_flux_equation}
        \begin{aligned}
            q^{(1)}_c & = \beta_c  |{p_n}| \llbracket T \rrbracket,\quad q^{(2)}_c & \!\! = -\beta_c |{p_n}| \llbracket T \rrbracket,
        \end{aligned}
    \end{equation}
    where $\beta_c \geq 0$ is the contact heat conductivity. Finally, the virtual work of the interface heat conduction problem reads: 
    \begin{align} \label{thermal_contact_weakform}
        \delta \mathcal{W}^c_{T}=- \intslave  q^{(1)}_c \proj{\delta T} \dgamma,
    \end{align}
    Therein, all contact integrals are transformed into pure slave side integrals using~\cref{slave_to_master_map}. As studied in~\cite{dittmann2014isogeometric,pantuso2000finite,gitterlethesis}, it is not necessary to introduce thermal Lagrange multipliers $\lambda_T = - q_c^{(1)}$ to enforce the thermal interface constraints. Instead, a direct substitution of the heat flux model~\cref{heat_flux_equation} into~\cref{thermal_contact_weakform} allows to express the interface heat fluxes as function of the temperature, which is the primary variable of the thermal problem.     
    \begin{remark}
        The drawback of this direct heat flux substitution method is that the problem becomes ill-conditioned for very large values of $\beta_c$, i.e., in the limit $\beta_c \to \infty$. In this case, alternative approaches such as the Lagrange multiplier method~\cite{seitz2018computational} or Nitsche's method~\cite{seitz2019nitsche} are well suited.
    \end{remark}

\subsection{Discretization in space and time}\label{sec:space_time_discretization}

The displacement and temperature field are approximated in space through trial functions defined on basis of discrete nodal values $\mathbf{d}_j$ and $\mathbf{T}_j$ and ansatz functions $N_j$, whereas the Lagrange multiplier field is approximated in space through trial functions defined on basis of discrete nodal values $\mathbf{\Lambda}_{c,j}$ and ansatz functions $\phi_j$, viz 
\begin{align}
    \dispu^h=\sum^n_{j=1} N_j \mathbf{d}_j, \quad T^h=\sum^n_{j=1} N_j \mathbf{T}_j, \quad \bm{\lambda}_c^h=\sum_{j\in \mathcal{S}} \phi_j \mathbf{\Lambda}_{c,j}.\label{shape_funtion}
\end{align}
In~\cref{shape_funtion} the global vectors $\mathbf{{d}}$ and $\mathbf{{T}}$ contain all displacement and temperature degrees of freedom, respectively and the vector $\mathbf{\Lambda}_{c}$ all nodal Lagrange multipliers. The corresponding test functions can be written as 
\begin{align}
    \delta \dispu^h=\sum^n_{j=1} N_j \delta\mathbf{d}_j, \quad \delta T^h=\sum^n_{j=1} N_j \delta\mathbf{T}_j, \quad \delta \bm{\lambda}_c^h=\sum_{j\in \mathcal{S}} \phi_j \delta \mathbf{\Lambda}_{c,j}.\label{var_shape_funtion}
\end{align}
For more information on the choice of the ansatz functions $\phi_j$ for the Lagrange multiplier field, the interested reader is referred to~\cite{popp2010,seitz2018computational}. Eventually, the semi-discrete solid mechanics problem can be obtained by substituting~\cref{shape_funtion} and~\cref{var_shape_funtion} into the weak form~\cref{mechanical_weakform}, resulting in:
\begin{align}
    \mathbf{M}_{u} \ddot{\mathbf{d}}+\mathbf{f}_{u}^{\mathrm{int}}\left(\mathbf{d}, \mathbf{T}\right)-\mathbf{f}_{u}^{\mathrm{ext}}+\mathbf{f}^c_u=\mathbf{0}, \label{semi_discrete_structural}
\end{align}
where $\mathbf{M}_u$ represents the constant mass matrix and $\mathbf{M}_{u} \ddot{\mathbf{d}}$ corresponds to the nodal force vector resulting from the inertia virtual work contribution $\delta\mathcal{W}_u^{\text{iner}}$, $\mathbf{f}_{u}^{\mathrm{int}}$ is the nodal internal force vector resulting from $\delta\mathcal{W}_u^{\text{int}}$, and the nodal external force vector is denoted as $\mathbf{f}_{u}^{\mathrm{ext}}$ and associated with $\delta\mathcal{W}_u^{\text{ext}}$. Lastly, $\mathbf{f}^c_u$ in~\cref{semi_discrete_structural} is the nodal contact force vector which is associated with the contact virtual work~\cref{contact_weak_form} and given as
\begin{align}\label{contact_force_discretized}
    \mathbf{f}^c_u = [\bm{0}, -\bm{M}(\mathbf{d}),\bm{D}(\mathbf{d})]^{\text{T}} \boldsymbol{\Lambda}_c,
\end{align}
where $\bm{D}$ and $\bm{M}$ are the well-known mortar matrices~\cite{popp2010}. Equation~\cref{contact_force_discretized} is obtained after rearranging the global displacement vector in a set of inactive $\mathcal{I}$, master $\mathcal{M}$, and slave $\mathcal{S}$ degrees of freedom. 

The spatial discretization of the mechanical problem is completed by discretizing also the contact constraints~\cref{KKT}. The constraints are discretized using the Lagrange multiplier ansatz functions $\phi_j$~\cref{shape_funtion}, resulting in
\begin{align}\label{discretized_kkt}
    \weightedgap  := \int_{\gamma_c^{(1),h}} \phi_j g^h_{n} \ \text{d}\gamma \geq 0, \quad  \lambda_{c,n,j} \leq 0, \quad  \lambda_{c,n,j} \tilde{g}_{n,j} =0 \quad \forall j \in \mathcal{S} \quad \text{on} \quad \slavesurf,
\end{align}
where $\weightedgap$ is referred to as weighted nodal gap and $\lambda_{c,n,j}$ is the normal component of the nodal Lagrange multiplier $\mathbf{\Lambda}_{c,j}$ (see~\cite{popp2010}). To enforce the normal contact constraint~\cref{discretized_kkt}, we employ a penalty regularization as detailed in~\cite{popp2010}. Accordingly, the contact pressure $\lambda_{c,n}$ and weighted gap $\tilde{g}_n$ at every slave node $j$ are related by introducing a penalty parameter $0<\epsilon_c< \infty$ according to 
\begin{align}\label{contact_regularization}
    \lmnormalj = \epsilon_c \langle -\weightedgap \rangle,
\end{align}
where $\langle \cdot \rangle$ denotes the Macauley bracket. As a consequence of the regularization, the nodal Lagrange multipliers are no longer primary variables. For a comprehensive treatment of penalty-regularized mortar finite element methods, the interested reader is referred to~\cite{dittmann2014isogeometric}.

\begin{remark}
    The choice of the penalty parameter affects the accuracy of the contact problem. To circumvent prohibitively large penetration for too low values of the penalty parameter as well as ill-conditioning for too high values of the penalty parameter, an adaptive penalty parameter scaling can be employed as shown in~\cite{seitz2019nitsche}.
\end{remark}

The semi-discrete thermal problem is achieved by substituting~\cref{shape_funtion} and~\cref{var_shape_funtion} in~\cref{thermal_weakform} and reads:
\begin{align}
    \mathbf{M}_{T} \dot{\mathbf{T}}+\mathbf{f}_{T}^{\text {int }}(\mathbf{d}, \mathbf{T})-\mathbf{f}_{T}^{\text {ext}}+\mathbf{f}^c_T=\mathbf{0}, \label{semi_discrete_thermal}
\end{align}
where $\mathbf{M}_{T}$ is the constant heat capacity matrix, $\mathbf{f}_{T}^{\mathrm{int}}$,
$\mathbf{f}_{T}^{\mathrm{ext}}$, and $\mathbf{f}^c_T$ are the nodal thermal internal, external, and contact forces,
respectively. In~\cref{semi_discrete_thermal}, $\mathbf{M}_{T}\dot{\mathbf{T}}$, $\mathbf{f}_{T}^{\mathrm{int}}$, $\mathbf{f}_{T}^{\mathrm{ext}}$, and $\mathbf{f}^c_T$ result from the virtual work contributions $\delta\mathcal{W}^{\text{iner}}_T$, $\delta\mathcal{W}^{\text{int}}_T$, $\delta\mathcal{W}^{\text{ext}}_T$, and $\delta\mathcal{W}^{\text{c}}_T$, respectively. For simplicity, the external forces ( $\mathbf{f}_{u}^{\mathrm{ext}}$ and $\mathbf{f}_{T}^{\mathrm{ext}}$) are assumed to be independent of the displacement and temperature field.

Next, the temporal discretization of the solid dynamics problem is achieved using a generalized-$\alpha$ time integration scheme. The discrete solid mechanics problem for the time interval $[t^n, t^{n+1}]$ with step $\Delta t$ reads
\begin{align}
    \mathbf{r}_{u}\left(\mathbf{d}^{n+1}, \mathbf{T}^{n+1}\right)=\mathbf{M}_{u} \ \ddot{\mathbf{d}}^{n+1-\alpha_{u,M}}+\mathbf{f}_{u}^{\mathrm{int},{n+1-\alpha_{u,f}}}-\mathbf{f}_{u}^{\mathrm{ext},{n+1-\alpha_{u,f}}}+\mathbf{f}^{c,{n+1}}_u=\mathbf{0}, \label{discrete_structural}
\end{align}
where the superscript ${n+1-\alpha_{u,(\cdot)}}$ denotes quantities evaluated at generalized mid-points within the time interval $[t^n, t^{n+1}]$ based on generalized-$\alpha$ parameters $\alpha_{u,(\cdot)}$ (see, e.g.,~\cite{seitz2018computational,danowski2013monolithic} for more details). In~\cref{discrete_structural}, all terms except $\mathbf{f}_{u}^{\mathrm{int},{n+1-\alpha_{u,f}}}$ and $\mathbf{f}^{c,{n+1}}_u$ are evaluated in a standard manner, i.e., the value at the generalized mid-point is obtained by linearly combining the values at $t^n$ and $t^{n+1}$ (see, e.g.,~\cite{seitz2018computational,popp2010}). In contrast, the contact force $\mathbf{f}^{c}_u$ is computed at $t^{n+1}$ to avoid an undesirable violation of energy conservation in the discrete system in the event of a changing active contact surface as stated in~\cite{seitz2018computational}. The evaluation of the deformation gradient, elastic right Cauchy--Green tensor and stress tensor, as required to compute the nodal internal force vector $\mathbf{f}_{u}^{\mathrm{int},{n+1}}$ from $\delta\mathcal{W}_u^{\text{int}}$, is conducted as follows: The total deformation gradient at $t_{n+1}$ is given as
\begin{align}
    \defgrad^{n+1}=\defgrad_e^{n+1}\defgrad_{in}^{n+1}.
\end{align}
To avoid numerically involved linearizations of the principal stretch directions $\bm{N}_{C}$, the inelastic deformation gradient $\defgrad^{n+1}_{in}$ at $t^{n+1}$ is approximated according to
\begin{align}
    \defgrad^{n+1}_{in}= \sum^{3}_{i=1}  \lambda^{i,n+1}_{in} \bm{N}_{C}^{i,n} \otimes \bm{N}_{C}^{i,n},
\end{align}
where the magnitude of the inelastic expansion is found by explicit time integration of the corresponding rate equation~\eqref{evolution} using an explicit Euler scheme, i.e., $\lambda^{i,n+1}_{in}=\lambda^{i,n}_{in}+\dot{\lambda}^{i,n}_{in} \Delta t$ with $\dot{\lambda}^{i,n}_{in}=f(T^n, \secondPK^{n}, \alpha^n)$. Here, $\bm{N}_{C}^{i,n}$ is computed from the spectral decomposition of $\cauchygreen^{n}$. The elastic deformation gradient results from $\defgrad^{n+1}_e=(\defgrad_{in}^{n+1})^{-1}\defgrad^{n+1}$, and the right Cauchy--Green tensor is determined according to 
\begin{align}
    \cauchygreen^{n+1}_e = (\defgrad^{n+1}_e)^{\text{T}}\defgrad^{n+1}_e.
\end{align}
Finally, the second Piola--Kirchhoff stress tensor is computed as 
\begin{align}
    \secondPK^{n+1}= 2 \inverse{(\defgrad^{n+1}_{in})} \frac{\partial \hff_e (\cauchygreen_e^{n+1})}{\partial \cauchygreen^{n+1}_e} \transposeinv{(\defgrad^{n+1}_{in})}. 
\end{align}

In a similar fashion, the fully discrete thermal problem is achieved based on a generalized-$\alpha$ time integration scheme and reads
\begin{align}
    \mathbf{r}_{T}\left(\mathbf{d}^{n+1}, \mathbf{T}^{n+1}\right)= \mathbf{M}_{T} \ \dot{\mathbf{T}}^{n+\alpha_{T,M}}+\mathbf{f}_{T}^{\text{int},{n+\alpha_{T,f}}}-\mathbf{f}_{T}^{\text{ext},{n+\alpha_{T,f}}}+\mathbf{f}^{c,{n+\alpha_{T,f}}}_T=\mathbf{0}. \label{discrete_thermal}
\end{align}
Finally, the solution of the coupled non-linear problem \cref{discrete_structural,discrete_thermal} is found in a
monolithic manner using the classical Newton--Raphson method with consistent linearization. The  linearized system is solved using iterative solvers based on preconditioners such as AMG(BGS) as studied in~\cite{danowski2013monolithic,verdugo2016unified}. 
\section{Mesh adaptation}\label{sec:mesh_adapt}
 The proposed mesh adaptation consists of two steps: mesh refitting and subsequent data transfer. In the mesh refitting step we construct a "new mesh" with improved quality compared to the "old mesh". The "old mesh" is the starting point of this procedure and usually exhibits heavily distorted elements, i.e., a low mesh quality. In the second step, the associated data, e.g., nodal primary variables and history variables of the material model defined at quadrature points, is transferred from the old mesh to the new one. These two steps are described in the following. All methods presented throughout this article are implemented in our in-house parallel multi-physics research code BACI~\cite{Baci2022}. 

\subsection{Mesh refitting} \label{sec:new_nodal_location} 
     The objective of the mesh refitting (MR) step is to achieve high-quality elements which are less distorted while preserving the topology of the boundary, i.e., the total volume of the domain. Thereto, we define an element distortion 
     potential and solve for the minimum of this potential during the mesh refitting step, thus minimizing the distortion of the finite element mesh. As mentioned before, an additional requirement for the MR step is the preservation of the boundary topology, which requires to prohibit motion of boundary nodes in the direction normal to the boundary of the problem. However, motion of the boundary nodes should be permitted in tangential direction to allow for an optimal mesh relaxation also in the domains close to the boundary. This aspect is crucial for problems involving interface phenomena such as contact mechanics, which often result in a strong mesh distortion at the boundaries of the interacting bodies. These two requirements, i.e., prohibiting the normal displacement component while allowing free tangential movement for boundary nodes will be fulfilled by employing a novel mortar mesh sliding approach.   
     
     In the following, the mortar mesh sliding approach (\cref{sec:meshsliding}) is presented, followed by the definition of the element distortion potential in~\cref{sec:constraints_potential}. Finally, in~\cref{sec:meshadaptation_subproblem}, the complete description of the MR problem is given.
    
    \subsubsection{Mesh sliding approach}\label{sec:meshsliding}
        Mesh sliding denotes a relative motion at the interface of two meshes that allows for free tangential sliding without detachment. To formulate such a constraint, the two bodies in~\cref{fig:contact_Kinematics} shall be considered, which are initially in contact. To enforce non-detachment, the normal gap has to remain zero during the motion, i.e.,
        \begin{align}\label{mshl_gap}
            g_n  = 0. 
        \end{align}
        Since no resistance with respect to tangential relative motion shall be applied, the tangential component of the interface traction vector has to vanish, similar to the friction-less contact scenario discussed before: 
        \begin{align}\label{mshl_tangent}
            \tractiontangnet \cdot \tangent = 0.
        \end{align}
        Identical to the mechanical contact case in~\cref{sec:thermomechanical_contact}, the contribution of the mesh sliding constraint to the virtual work of the mesh refitting problem in terms of the Lagrange multiplier $\lmms$ can be written as:
        \begin{align}\label{meshsliding_contact_weak_form}
            \delta \mathcal{W}_{u} = - \intslave  \proj{\delta \bm{u}} \cdot \lmms \ \dgamma = \intslave  \delta \gap \lambda_{m,n} \ \dgamma = 0,
        \end{align}
        where the normal component of the Lagrange multiplier is $\lambda_{m,n} = \lmms \cdot  \normal $. Following the mortar finite element formulation presented above, the spatially discretized mesh sliding force vector $\mathbf{f}^m_u$ is identical to the vector $\mathbf{f}^c_u$ in~\cref{contact_force_discretized}. The only difference between the mesh sliding approach and the friction-less contact case presented above is given through the constraints~\eqref{KKT} and~\eqref{mshl_gap}, where the former represents an inequality constraint and the latter an equality constraint. Keeping this difference in mind, the mesh sliding constraints are enforced through a regularization based on the discretized weighted nodal gap $\weightedgap$ (see~\cref{discretized_kkt}) and a penalty parameter $0< \epsilon_m< \infty$, which yields for node $j$:
        \begin{align}\label{meshsliding_regularization}
            \lambda_{m,n,j}=\bm{\lambda}_{m,j}\cdot \bm{n}_j = -\epsilon_m \weightedgap , \quad \bm{\lambda}_{m,j} \cdot \tangentj = 0 \quad \forall j \in \mathcal{S}.
        \end{align}
        The significant difference between the penalty regularization of the mesh sliding~\cref{meshsliding_regularization} and the mechanical contact~\cref{contact_regularization} is the Macaulay brackets. They must be used in contact mechanics because contact forces only exist if the gap is negative. In contrast, in mesh sliding, the forces exist regardless of the sign of the gap to prevent penetration and detachment.
       
        \begin{remark}
        We employ mesh sliding to allow free tangential sliding on relatively smooth surfaces to minimize mesh distortions close to the interface. However, due to the construction of the mesh sliding constraints, it is inherent to the method that sharp edges or corners in the mesh sliding interface can lead to a penetration of the bodies at this interface. To explain this issue, we consider a square body~$\Omega_S$  with a corner node $j$ and an L-shaped body $\Omega_M$ as shown in~\cref{meshsliding_remark}. As it is not straightforward to define the normal at a corner node since the normals of the adjacent edges are not parallel ($\bm{n}_j^1, \bm{n}_j^2$), we apply a common strategy from computational contact mechanics, i.e.~we construct an averaged normal ($\bm{n}_j$) at the corner. The averaged normal ($\bm{n}_j$) at the corner $j$ points along the diagonal of the square (green vector), and the tangential vector (red vector) are constructed as shown in~\cref{meshsliding_remark}. According to the definition of the mesh sliding constraint, the nodes are free to move in the tangent plane. As a consequence, the node $j$ can penetrate into $\Omega_M$, keeping the gap zero in a weighted sense, i.e., $\weightedgap=0$.
        \begin{figure}[htb]
            \centering
            \scalebox{1}{         \begin{tikzpicture}[scale=0.6]
          \tikzset{d/.style={minimum width=4pt,inner sep=0pt,circle,fill=black}}
    	\node [] (0) at (0, 0) {};
		\node [] (1) at (0, -2) {};
		\node [] (2) at (0, -4) {};
		\node [] (3) at (-2, -2) {};
		\node [] (4) at (-2, 0) {};
		\node [] (5) at (-4, 0) {};
		\node [] (6) at (-4, -2) {};
		\node [] (7) at (-4, -4) {};
		\node [] (8) at (-2, -4) {};
		\node [] (9) at (-4, 1) {};
		\node [] (10) at (1, 1) {};
		\node [] (11) at (-0.75, 1) {};
		\node [] (12) at (-2.5, 1) {};
		\node [] (13) at (1, -1) {};
		\node [] (14) at (1, -4) {};
		\node [] (15) at (1, -2.5) {};
		\node [] (16) at (-4, 2.5) {};
		\node [] (17) at (2.5, 1) {};
		\node [] (18) at (2.5, -1) {};
		\node [] (19) at (2.5, -2.5) {};
		\node [] (20) at (2.5, -4) {};
		\node [] (21) at (2.5, 2.5) {};
		\node [] (22) at (1, 2.5) {};
		\node [] (23) at (-0.75, 2.5) {};
		\node [] (24) at (-2.5, 2.5) {};
		\node [] (25) at (1.75, 0) {};
		\node [] (26) at (0, 1.75) {};
		\node [] (27) at (-0.5, 0.5) {};
		\node [] (28) at (0.5, -0.5) {};
		\node [] (29) at (1.25, 1.25) {};

        \draw[thick] (5.center) to (0.center);
		\draw[thick] (0.center) to (2.center);
		\draw[thick] (2.center) to (7.center);
		\draw[thick] (7.center) to (5.center);
		\draw[thin,gray] (4.center) to (8.center);
		\draw[thin,gray] (6.center) to (1.center);
		\draw[thick] (9.center) to (10.center);
		\draw[thick] (10.center) to (14.center);
		\draw[thick] (16.center) to (21.center);
		\draw [thick](21.center) to (20.center);
		\draw [thick](20.center) to (14.center);
		\draw [thick](16.center) to (9.center);
		\draw [thin,gray](24.center) to (12.center);
		\draw [thin,gray] (23.center) to (11.center);
		\draw [thin,gray] (22.center) to (10.center);
		\draw [thin,gray](10.center) to (17.center);
		\draw [thin,gray](13.center) to (18.center);
		\draw [thin,gray](15.center) to (19.center);

		\draw [thin,blue,-Latex](0.center) to (25.center)node[below, color=black,align=center ]  {$\bm{n}^1_j$};
		
		\draw [thin,blue,-Latex](0.center) to (26.center)node[right, color=black,align=center ]  {$\bm{n}^2_j$};
		\draw [thin,green,-Latex](0.center) to (29.center)node[right, color=black,align=center ]  {$\bm{n}_j$};;
		
		\node [] (27) at (-0.75, 0.75) {};
		\node [] (28) at (0.75, -0.75) {};
		\draw [thin,red,-Latex](27.center) to (28.center);
		\draw [thin,red,-Latex](28.center) to (27.center)node[left, color=black,align=center,yshift=-3pt]  {$\bm{\tau}_j$};
		
		\node [d] (0) at (0, 0) {};
        
        \node [] (322) at (-2.4, -2.4) {$\Omega_S$};
        \node [] (322) at (2.0, -2) {$\Omega_M$};
        
        \draw[shorten <=2pt,very thin] (-4,0) to[out=90,in=-90] ++ (150:0.8)  node[left]{$\Gamma^{(1)}_m$};
        
        \draw[shorten <=2pt,very thin] (-4,0.9) to[out=90,in=-90] ++ (140:0.8)  node[left]{$\Gamma^{(2)}_m$};
        
        \node [] (322) at (-0.15,0)[below,xshift=-1pt] {$j$};
\end{tikzpicture}} \caption{Geometry with sharp edge}\label{meshsliding_remark}
        \end{figure}
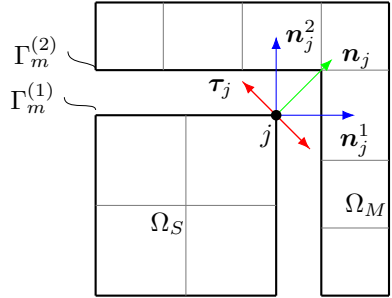
        
        In this work for the numerical examples showcased in~\cref{sec:numerical_results}, the discussed phenomenon is avoided by applying a displacement Dirichlet boundary condition $\bm{u}=\bm{0}$ (equivalent to mesh tying) to the slave and master nodes of sharp corners or edges. As a result, there is no relative motion between the slave and master side.
        \end{remark}    

    \subsubsection{Element distortion potential} \label{sec:constraints_potential}
        Generally, mesh quality can be measured in geometric quantities such as skewness and aspect ratio. The skewness quantifies the angles enclosed by the element faces or edges, whereas the aspect ratio measures the ratio of the dimensions of elements in different spatial directions. In this work, we control the element size and aspect ratio by enforcing constraints on the element edge lengths, denoted as $\consEdge$, and the skewness by enforcing constraints on the angles enclosed by the element edges, denoted as $\consAngle$.

        \begin{remark}
            It is known that certain finite element formulations are sensitive to specific element distortions (e.g. trapezoidal vs parallelogram). In such cases, it is straightforward to include this knowledge in the definition of the element distortion potential.
        \end{remark}
            
\begin{figure}[htb]
            \centering
            \begin{tikzpicture}
    [cube/.style={very thick,black},
        grid/.style={very thin,gray},
        axis/.style={->,blue,thick}]

    \def\dx{3}
    \def\ax{4.5}

    \coordinate [label={above right :{$0$}}](0) at (0,0,0);
    \coordinate [label={above right :{$1$}}](1) at (\dx,0,0);
    \coordinate [label={below right :{$2$}}](2) at (\dx,0,\dx);
    \coordinate [label={[label distance=1mm] left :{$3$}}](3) at (0,0,\dx);
    \coordinate [label={[label distance=1mm] left :{$4$}}](4) at (0,\dx,0);
    \coordinate [label={[label distance=0mm] right:{$5$}}](5) at (\dx,\dx,0);
    \coordinate [label={[label distance=2mm] right:{$6$}}](6) at (\dx,\dx,\dx);
    \coordinate [label={[label distance=2mm] left :{$7$}}](7) at (0,\dx,\dx);

    \draw[axis, -Latex] (0,0,0) -- (\ax,0,0) node[anchor=west]{$e^1$};
    \draw[axis, -Latex] (0,0,0) -- (0,\ax,0) node[anchor=west]{$e^2$};
    \draw[axis, -Latex] (0,0,0) -- (0,0,\ax) node[anchor=west]{$e^3$};
    
    \draw[] (0) -- (1);
    \draw[] (1) -- (2);
    \draw[] (3) -- (2);
    \draw[] (0) -- (1);
    
    \draw[] (4) -- (5);
    \draw[] (7) -- (6);
    \draw[] (6) -- (5);
    \draw[] (4) -- (7);

    \draw[] (1) -- (5);
    \draw[] (2) -- (6);
    \draw[] (7) -- (3);
    \draw[] (4) -- (0);
    
    \draw[-Latex] ([xshift=10pt,yshift=30pt]1) -- ([xshift=10pt,yshift=-30pt]5)   node[midway,right, color=black,align=center ]  {$\bm{v}^2_{4}$};
    \draw[-Latex] ([xshift=10pt,yshift=40pt]2) -- ([xshift=10pt,yshift=-20pt]6)   node[midway,right, color=black,align=center ]  {$\bm{v}^2_{3}$};
    \draw[-Latex] ([xshift=-10pt,yshift=30pt]3) -- ([xshift=-10pt,yshift=-30pt]7)  node[midway,left, color=black,align=center ]{$\bm{v}^2_{2}$};
    \draw[-Latex] ([xshift=-10pt,yshift=20pt]0) -- ([xshift=-10pt,yshift=-40pt]4)  node[midway,left, color=black,align=center ] {$\bm{v}^2_{1}$};
    
    \draw[-Latex,dashed] ([xshift=20pt,yshift=5pt]0) -- ([xshift=-40pt,yshift=5pt]1) node[midway,above, color=black,align=center ]  {$\bm{v}^1_{1}$};
    \draw[-Latex,dashed] ([xshift=30pt,yshift=5pt]3) -- ([xshift=-30pt,yshift=5pt]2) node[midway,above, color=black,align=center ]  {$\bm{v}^1_{4}$};
    \draw[-Latex,dashed] ([xshift=30pt,yshift=5pt]4) -- ([xshift=-30pt,yshift=5pt]5) node[midway,above, color=black,align=center ]  {$\bm{v}^1_{2}$};
    \draw[-Latex,dashed] ([xshift=40pt,yshift=5pt]7) -- ([xshift=-20pt,yshift=5pt]6) node[midway,above, color=black,align=center ]  {$\bm{v}^1_{3}$};

    \tikzset{d/.style={minimum width=6pt,inner sep=0pt,circle,fill=black}}
    \foreach \i in {0,...,7}
    \node [d](\i) at (\i) {};
    
    \def\dzo{0.80}
    \def\dxo{0.25}
    \coordinate [](0) at (-\dxo,0,\dzo);
    \coordinate [](1) at (\dx+\dxo,0,\dzo);
    \coordinate [](2) at (\dx+\dxo,0,\dx-\dzo);
    \coordinate [](3) at (-\dxo,0,\dx-\dzo);
    
    \coordinate [](4) at (-\dxo,\dx,\dzo);
    \coordinate [](5) at (\dx+\dxo,\dx,\dzo+0.1);
    \coordinate [](6) at (\dx+\dxo,\dx,\dx-\dzo);
    \coordinate [](7) at (-\dxo,\dx,\dx-\dzo);
    
    \draw[-Latex,dotted] (4) -- (7)node[midway,above left, color=black,align=center ]   {$\bm{v}^3_{2}$};
    \draw[-Latex,dotted] (5) -- (6)node[midway,xshift=4pt,below, color=black,align=center ]  {$\bm{v}^3_{3}$};
    \draw[-Latex,dotted] (0) -- (3)node[midway,above, color=black,align=center ]  {$\bm{v}^3_{1}$};
    \draw[-Latex,dotted] (1) -- (2)node[midway,right, color=black,align=center ]  {$\bm{v}^3_{4}$};
    
    \begin{scope}[xshift=225pt,yshift=50pt]
       \matrix (m) [matrix of nodes,
               nodes in empty cells,
               column sep=-\pgflinewidth, 
               row sep=-\pgflinewidth, 
               nodes={minimum width=3em, 
                      minimum height=1.5em, 
                      outer sep=0, 
                      anchor=center}, 
               ]
              {
                $\bm{v}^1_{\{1,2,3,4\}} = \bm{x}_{\{1,5,6,2\}}-\bm{x}_{\{0,4,7,3\}}$ \\
                $\bm{v}^2_{\{1,2,3,4\}} = \bm{x}_{\{4,7,6,5\}}-\bm{x}_{\{0,3,2,1\}}$\\
                $\bm{v}^3_{\{1,2,3,4\}} = \bm{x}_{\{3,7,6,2\}}-\bm{x}_{\{0,4,5,1\}}$\\
              };
    \end{scope}

\end{tikzpicture}\caption{ Illustration of edge vectors defined to formulate included angle and edge constraints for a hexahedral element.}
            \label{fig:direction_vectors}
        \end{figure}
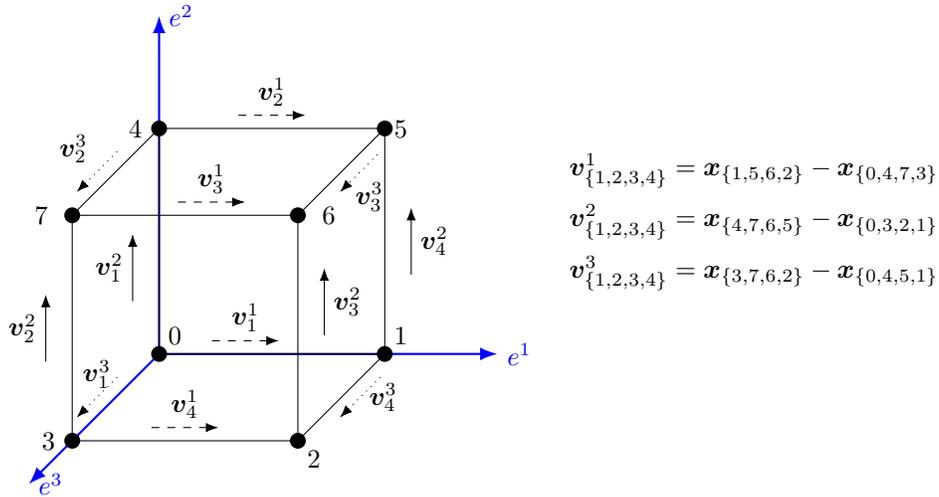

        We demonstrate the formulation of these constraints using an 8-noded hexahedral element, which is the finite element type employed in the numerical examples of this work. As shown in~\cref{fig:direction_vectors}, we first define $12$ edge vectors $\direcij$. Here, the superscript $i\in\{1,2,3\}$ denotes the direction $\bm{e}^i$ associated with the orientation of the edge in parameter space. The subscript $j\in\{1,2,3,4\}$ represents a counting index for the four individual edge vectors pointing to a given direction $i$. As illustrated in Figure~\ref{fig:direction_vectors}, each edge vector is defined as the difference vector between the (current) spatial position vectors of the two nodes associated with the edge vector. From these individual edge vectors, an averaged edge vector $\direcAvg$ is defined as 
        \begin{align}\label{average_direction_vector}
            \direcAvg = \frac{1}{4} \sum^4_{j=1} \direcij.
        \end{align}
        Based on these definitions, the edge and angle constraints will be formulated in the following.

    \subsubsection*{\textit{Edge constraints $\consEdge^i$}} 
        To impose specific edge lengths on elements, we apply constraints on both, the average edge vectors $\direcAvg$ and the individual edge vectors $\direcij$, which are denoted as $\consEdgeAvg^i$ and $\consEdgeInd^i$, respectively. The constraint $\consEdgeAvg^i$ on the average edge vector in direction $\bm{e}^i$ is defined as
        \begin{align}\label{constraint_avg}
            \consEdgeAvg^i = \frac{\sqrt{\direcAvg \cdot \direcAvg}}{\rlength} - 1 \myeq 0 \quad 
            \text{for} \quad i=1,2,3,
        \end{align}
        with $\rlength$ denoting a target element edge length in direction $\bm{e}^i$ to be prescribed. This constraint enforces the length $\rlength$ on the average edge vector $\direcAvg$. In a next step, the constraint $\consEdgeInd^i$ is defined according to
        \begin{align}\label{constraint_equal_edges}
            \big(\consEdgeInd^i\big)_j = \frac{ \direcij \cdot  \direcij}{\direcAvg \cdot \direcAvg}- 1 \myeq 0 \quad 
            \text{for} \quad i=1,2,3, \quad j=1,2,3,4,
        \end{align}
        which enforces that each individual edge vector equals the associated average edge vector. Prescribing a spatial distribution function for the element size will be called \textit{mesh localization} throughout this work. It can be achieved by prescribing spatial functions for the target lengths, i.e., $l^i_{\text{r}}=l^i_{\text{r}}(\bm{X})$. To sum up, so far we have defined $15$ constraints associated with element size and aspect ratio, $3$ on the average edge vectors and $12$ on the individual edge vectors. 
        
    \subsubsection*{\textit{Angle constraints $G^{mn}_{A}$}} 
        Next, we construct angle constraints considering the angle enclosed by a pair of edge vectors $\bm{v}^i_j$ with shared node. For example, the angle constraints $G^{mn}_{A}$ for node "$0$" are formulated as (c.f.~\cref{fig:direction_vectors})
        \begin{align}\label{constratint_angle}
            G^{12}_{A} &= \ga{\bm{v}^1_1}{\bm{v}^2_1} - \cos{\theta^{12}_{\text{r}}}\myeq 0 , \quad \notag \\
            G^{13}_{A} &= \ga{\bm{v}^1_1}{\bm{v}^3_1} -\cos{\theta^{13}_{\text{r}}}\myeq 0, \quad \\
            G^{23}_{A} &= \ga{\bm{v}^2_1}{\bm{v}^3_1} -\cos{\theta^{23}_{\text{r}}}\myeq 0, \notag
        \end{align}
        where $\theta^{12}_{\text{r}}$, $\theta^{13}_{\text{r}}$, and $\theta^{23}_{\text{r}}$ are the enclosed target angles to be achieved. To achieve perpendicular edges, we set
        $\theta^{12}_r=\theta^{13}_r=\theta^{23}_r=\pi/2$. Likewise, angle constraints can be formulated for all remaining nodes. To conclude, we have formulated $24$ angular constraints in (i.e., three constraints for each of the eight nodes).
        
        \subsubsection*{\textit{Constraint enforcement}} 
        Finally, we enforce these constraints on basis of a distortion potential with penalty parameters $\bar{\varepsilon}_E$, $\widehat{\varepsilon}_E$ and
        $\varepsilon_A$ given as
        \begin{align}\label{distortion_potential}
            \pi_{d} = \frac{1}{2} \bar{\varepsilon}_{E} \sum^{ndir}_{i=1} \bar{G}^i_{E} \bar{G}^i_{E} +  \frac{1}{2} \widehat{\varepsilon}_E \sum^{ndir}_{i=1} \sum^{4}_{j=1}  \big(\widehat{G}^i_{E}\big)_j  \big(\widehat{G}^i_{E}\big)_j +  \frac{1}{2}  \varepsilon_{A} \sum^{nnode}_{n=1} \big(G^{12}_{A}G^{12}_{A}+G^{13}_{A}G^{13}_{A}+G^{23}_{A}G^{23}_{A}\big)_n,
        \end{align}
        where $ndir=3$ is the number of spatial directions and $nnode=8$ is the number of nodes. In the following, the meaning of the different constraint contributions shall be briefly discussed. Clearly, the constraints $\widehat{G}^i_{E}$ penalize deviations of the individual edge lengths from the average edge length in a given direction, i.e., they enforce the element shape to equal a parallelepiped in the limit of $\widehat{\varepsilon}_E \rightarrow \infty$. If additionally also the angular constraints $G_{A}$ with target angles $\theta^{12}_r=\theta^{13}_r=\theta^{23}_r=\pi/2$ are enforced, the element will tend towards a cuboid shape. Eventually, if equal target lengths are chosen for the constraints $\bar{G}^i_{E}$, i.e., $l_r^1=l_r^2=l_r^3=:l_r$, the element will approach a cubic shape. Finally, the absolute value of $l_r$ determines how a shape-preserving scaling of the element size will be penalized. For example, by choosing $l_r$ as the initial edge length of a regular mesh with cubic elements, every (even shape-preserving) deviation from the initial element size will be penalized. Clearly, the chosen set of $15+24$ constraints per element is redundant, since an hexahedral element with $8$ nodes only exhibits $24$ degrees of freedom (including $6$ rigid body modes changing neither the shape nor the size of the element). However, this over-constraining is no problem when employing a penalty potential for constraint enforcement. Moreover, this specific choice of (redundant) constraints allows to independently control different modes of element distortion (with different effect on the mesh quality), as elaborated above. In addition, the chosen set of constraints leads to a distortion potential that is symmetric with respect to the node numbering, i.e., the result will not change if the node numbering is changed for a given mesh. In conclusion, it is emphasized that the definition of an element distortion potential is not unique. The specific choice presented above has proven effective in the numerical test cases we have investigated so far. In particular, the specification $\bar{\varepsilon}_{E}=\widehat{\varepsilon}_{E}=:{\varepsilon}_{E}$ turned out as a robust choice and will be used in the remainder of this work.
        
        \begin{figure}[htb]
            \centering
            \input{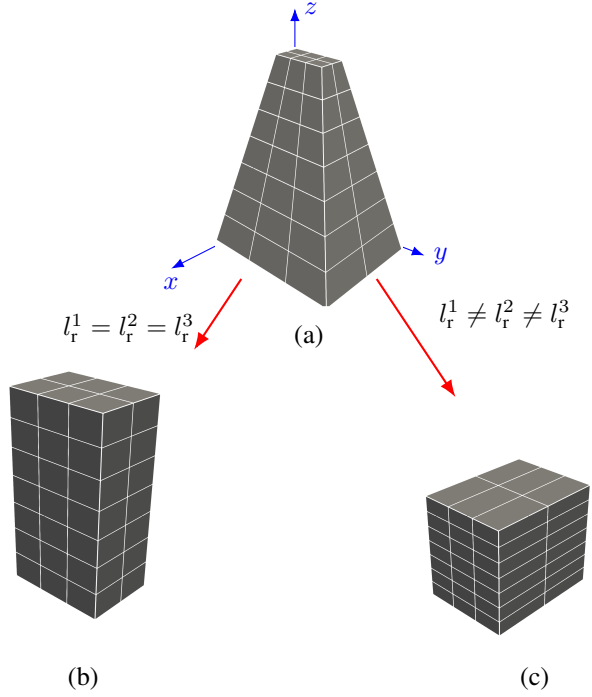}
            \caption{ Illustration of constraints: (a) initial geometry (b) resulting geometry when elemental constraints with $\rlengthi{1}=\rlengthi{2}=\rlengthi{3}$ and $\theta^{mn}_r=\pi/2$ are applied on (a), and (b) when $\rlengthi{1}\neq\rlengthi{2}\neq\rlengthi{3}$ with $\theta^{mn}_r=\pi/2$ are applied on (a).}\label{fig:distrotion_potential_example}
        \end{figure}
        
        \subsubsection*{\textit{Demonstration example: Effect of target element edge lengths}} 
        To conclude this section, the effect of the element constraints shall be demonstrated using the numerical example shown in~\cref{fig:distrotion_potential_example}. Consider the truncated rectangular pyramid-shaped body in~\cref{fig:distrotion_potential_example}a meshed with 42 hexahedral elements. The body is freely supported in the first quadrant, so that rigid body modes are suppressed. We solve the constraint equations~\cref{distortion_potential} (see next section, for details of the solution procedure) with equal target lengths (see~\cref{constraint_avg}) in all directions, i.e. $\rlengthi{1}=\rlengthi{2}=\rlengthi{3}$ and $\theta^{mn}_r=\pi/2$. The resulting geometry is shown in~\cref{fig:distrotion_potential_example}b. As expected the obtained geometry is rectangular with cubic elements. When $\rlengthi{1}\neq\rlengthi{2}\neq\rlengthi{3}$ and $\theta^{mn}_r=\pi/2$ the resulting geometry remains rectangular and contains rectangular elements as illustrated in~\cref{fig:distrotion_potential_example}c. This shows that the suggested approach is suitable to obtain elements of a specific desired shape. It is emphasized that this example has been designed to visualize the isolated effect of the element distortion potential without imposing mesh sliding constraints at the boundaries. Thus, in contrast to actual mesh refitting problems, the mesh in this demonstration example is not required to preserve the boundary contour of the discretized body.
        
        \subsubsection{Problem description of mesh refitting method}\label{sec:meshadaptation_subproblem}
        \begin{figure}[htb]
            \centering
            \scalebox{1.0}{\begin{tikzpicture}[scale=0.75]
    \pgfdeclarelayer{nodelayer}
    \pgfdeclarelayer{edgelayer}
    \pgfsetlayers{edgelayer,nodelayer,main}
    \begin{pgfonlayer}{nodelayer}
        \node [] (1) at (-1.25, 6.5) {};
        \node [] (2) at (-1, 5.25) {};
        \node [] (3) at (-1, 3.75) {};
        \node [] (4) at (-1.5, 2.75) {};
        \node [] (27) at (-1.75, 2) {};
        \node [] (22) at (-1.25, 7) {};

        \node [] (5) at (-2.25, 3) {};
        \node [] (6) at (-3, 3.25) {};
        \node [] (10) at (-2.75, 6.25) {};
        \node [] (11) at (-2.75, 5.25) {};
        \node [] (12) at (-2.75, 4) {};
        \node [] (13) at (-2, 5.25) {};
        \node [] (21) at (-2, 4) {};

        \node [] (24) at (-2, 6.25) {};
        \node [] (25) at (-2.25, 7) {};
        \node [] (26) at (-3, 2.5) {};
        \node [] (28) at (0, 6.5) {};
        \node [] (29) at (0, 7) {};
        \node [] (30) at (0, 6.5) {};
        \node [] (32) at (0.25, 5.25) {};
        \node [] (34) at (0.25, 3.75) {};
        \node [] (36) at (-0.25, 2.75) {};
        \node [] (37) at (-0.5, 2) {};
        \node [] (38) at (1, 7.0) {};
        \node [] (40) at (1, 6.5) {};
        \node [] (43) at (1.25, 5.25) {};
        \node [] (44) at (1.25, 3.5) {};
        \node [] (45) at (0.75, 2.5) {};
        \node [] (46) at (0.5, 1.75) {};

        \node [] (35) at (-0.25, 2.75) {};
        \node [] (33) at (0.25, 3.75) {};
        \node [] (39) at (1, 6.5) {};
        \node [] (41) at (0, 6.5) {};
    \end{pgfonlayer}
    \begin{pgfonlayer}{edgelayer}
        \draw [in=90, out=-90, looseness=0.75, thick, line width=1pt] (1.center) to (2.center);
        \draw [bend left=15, looseness=0.75, thick, line width=1pt] (2.center) to (3.center);
        \draw [in=75, out=-90, looseness=0.50, thick, line width=1pt] (3.center) to (4.center);
        \draw [thick, line width=1pt](22.center) to (1.center);
        \draw [thick, line width=1pt](4.center) to (27.center);
        \node[inner sep=2pt,label=above:$\Omega_m$] at (-2.5, 4.25)  {};  
        \draw[shorten <=2pt,very thin] (-1.25, 6.5) to[out=90,in=-90] ++ (-100:-0.8) node[above]{$\Gamma^{(1)}_m$};

        \draw [opacity=0.25](4.center) to (5.center);
        \draw [opacity=0.25](5.center) to (6.center);
        \draw [opacity=0.25](2.center) to (13.center);
        \draw [opacity=0.25](13.center) to (11.center);
        \draw [in=90, out=-90,opacity=0.25] (13.center) to (21.center);
        \draw [in=345, out=165,opacity=0.25] (3.center) to (21.center);
        \draw [opacity=0.25](12.center) to (21.center);
        \draw [opacity=0.25](5.center) to (26.center);
        \draw [opacity=0.25](21.center) to (5.center);
        \draw [opacity=0.25](10.center) to (24.center);
        \draw [opacity=0.25](24.center) to (1.center);
        \draw [in=90, out=-90, looseness=0.75,opacity=0.25] (24.center) to (13.center);
        \draw [opacity=0.25](25.center) to (24.center);

        \draw [thick, line width=1pt](29.center) to (28.center);
        \draw [bend left=15, looseness=0.75,thick, line width=1pt] (32.center) to (33.center);
        \draw [in=75, out=-90, looseness=0.50,thick, line width=1pt] (34.center) to (35.center);
        \draw [thick, line width=1pt] (36.center) to (37.center);
        \draw [in=90, out=-90, looseness=0.75,thick, line width=1pt] (30.center) to (32.center);



        \draw[shorten <=2pt,very thin] (0, 6.5) to[out=90,in=-90] ++ (100:0.8) node[above]{$\Gamma^{(2)}_m$};
        

    \end{pgfonlayer}
\end{tikzpicture}} \caption{Illustration of mesh sliding interface surfaces with auxiliary boundary $\Gamma^{(2)}_m$.}\label{fig:auxilliary_boundary}
        \end{figure}
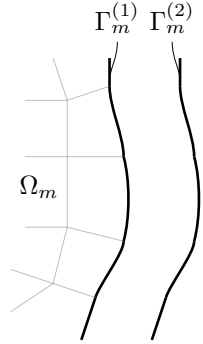
        Let the old mesh be defined on the domain $\Omega_m \subset \Omega_t$ with boundary $\Gamma^{(1)}_m := \partial\Omega_m$. We apply the mesh sliding approach from above (see \cref{sec:meshsliding}) to preserve the boundary contour of a given body while allowing for tangential sliding. This will be done only for the boundary nodes without prescribed Dirichlet boundary condition, i.e., only for nodes on the boundary $\Gamma^{(1)}_{m} \setminus \Gamma_{u}$. From a technical point of view, to enable the mesh sliding approach in the same sense as typical for mortar interface problems, we replicate the discretized boundary $\Gamma^{(1)}_m$ and denote this auxiliary boundary as $\Gamma^{(2)}_m$ (see~\cref{fig:auxilliary_boundary}). Since $\Gamma^{(2)}_m$ and $\Gamma^{(1)}_m$ coincide, a one-to-one mapping between the nodes on these boundaries exists, i.e., nodal positions can be transferred in a straight-forward manner from $\Gamma^{(1)}_m$ to $\Gamma^{(2)}_m$. It is emphasized that the auxiliary boundary $\Gamma^{(2)}_m$ is only required to represent a fictitious interaction partner to apply the standard mesh-sliding method to the boundary nodes of the original mesh on $\Gamma^{(1)}_m$. This means, that the nodal position and displacement values on the auxiliary boundary $\Gamma^{(2)}_m$ remain fixed, i.e., are prescribed per Dirichlet boundary condition, and the mesh-refitting problem does not need to be solved for these nodes. 
        \begin{remark}
             To prevent changes in the topology of the body, the movement of the corner nodes must be avoided during mesh refitting. In this work, it is achieved by applying a displacement Dirichlet boundary condition $\bm{u}=\bm{0}$.
        \end{remark}
        
        The complete MR problem is performed after solving the fully discretized thermo-mechanical problem for time step $t^{n+1}$, as described in Section~\ref{sec:thermomech}. The solution of the mesh refitting problem is defined as the stationary value of the following total potential
         \begin{align}\label{mesh_adaptation_potential}
            \pi_d +  \pi_m \rightarrow \text{stat.},
        \end{align}
        where the element distortion potential $\pi_d$ is given in~\eqref{distortion_potential} and $\pi_m$ represents an abstract potential for mesh sliding constraint enforcement (e.g., a Lagrange multiplier or a penalty potential; the latter approach is used in the examples presented in the following) whose variation is given by the discretized form of~\eqref{meshsliding_contact_weak_form}. Similar to the weak form of our physical (thermo-mechanical) problem, as necessary condition for a stationary value the variation of the discrete potential~\eqref{mesh_adaptation_potential} has to vanish, leading to the following system of nonlinear (residual) equations:
        \begin{align}\label{mesh_adaptation_subproblem}
             \mathbf{f}^d_u +  \mathbf{f}^m_u = \mathbf{0},
        \end{align}
        where $\mathbf{f}^d_u=\pd{\pi_m}{\mathbf{d}}$ is the gradient of the discrete element distortion potential~\eqref{distortion_potential} and $\mathbf{f}^m_u$ is the nodal mesh sliding force vector according to~\cref{sec:meshsliding} (see~\cref{appendix_rhs_sysmat} for more details). The solution of the non-linear system~\cref{mesh_adaptation_subproblem} is found using a Newton-Raphson scheme based on a consistent linearization. 

        \begin{remark}
            It should be noted that sometimes it is useful to execute an "artificial time step" after mesh adaptation (i.e. after mesh refitting and data transfer) to rebuild the dynamic equilibrium.
        \end{remark}
        
        \begin{remark}
            In this work, the reference configuration of the MR problem is updated with the (converged) current configuration of the old mesh. 
            This means that the current state of the old mesh becomes the reference state of the MR problem. By updating the reference configuration of the MR problem with the current configuration of the old mesh the MR procedure gets more robust because the MR problem is thereby independent of the original reference configuration. 
        \end{remark}
        
        \subsubsection*{\textit{Demonstration example: Influence of mesh sliding approach}}
        \begin{figure}[htb]
            \centering
            \input{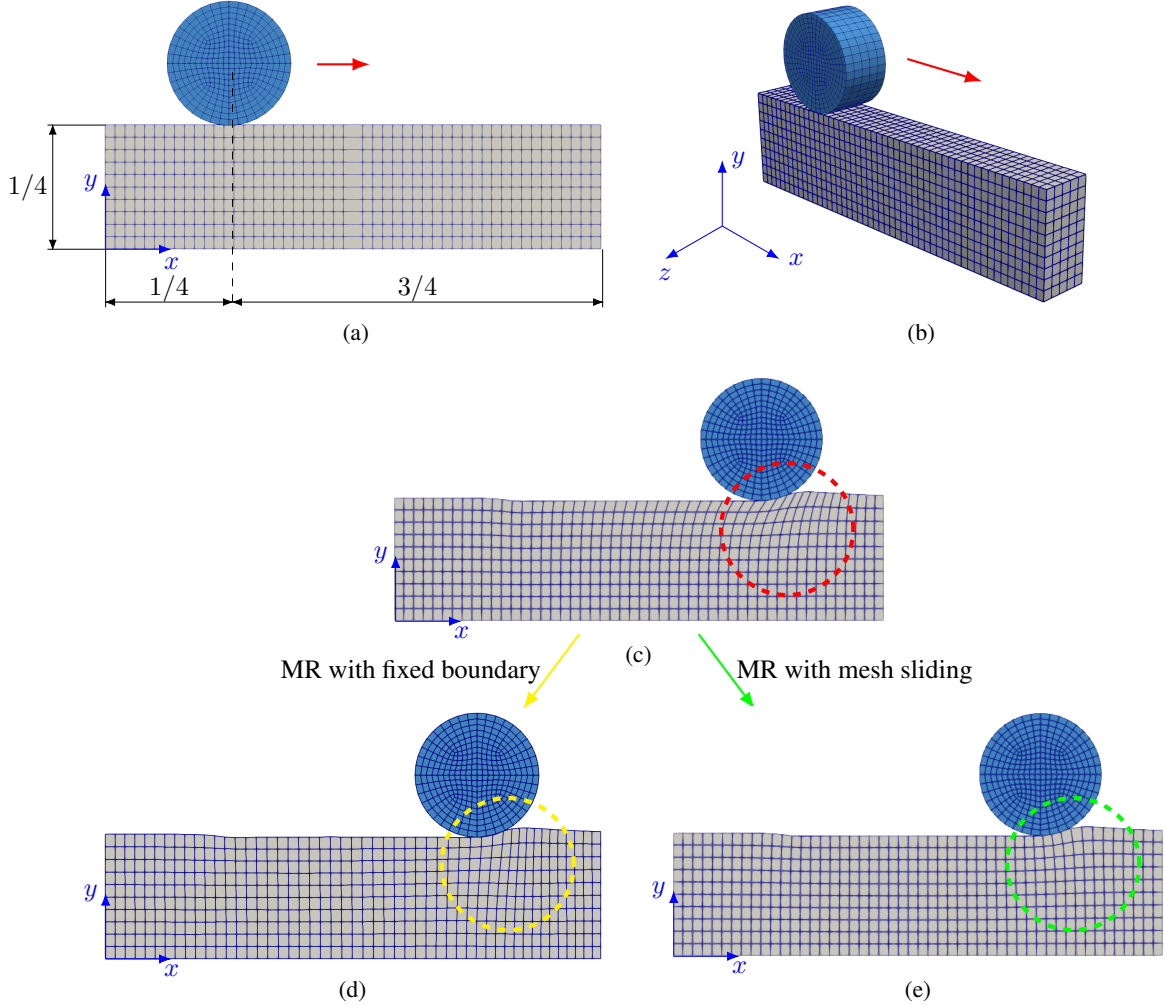}
            \caption{ Frictional sliding of a rigid cylinder with a diameter of $0.25$ over a rectangular block of dimensions $1\times0.25\times0.125$: (a) and (b) show the initial configuration. (c) illustrates the final deformed state without mesh refitting. (d) shows MR with fixed boundaries for the deformed state in (c). (e) presents MR with mesh sliding for the deformed state in (c).} \label{fig:mesh_adaptation_example_0}
        \end{figure}
        
\begin{figure}[htb]
            \centering
            \input{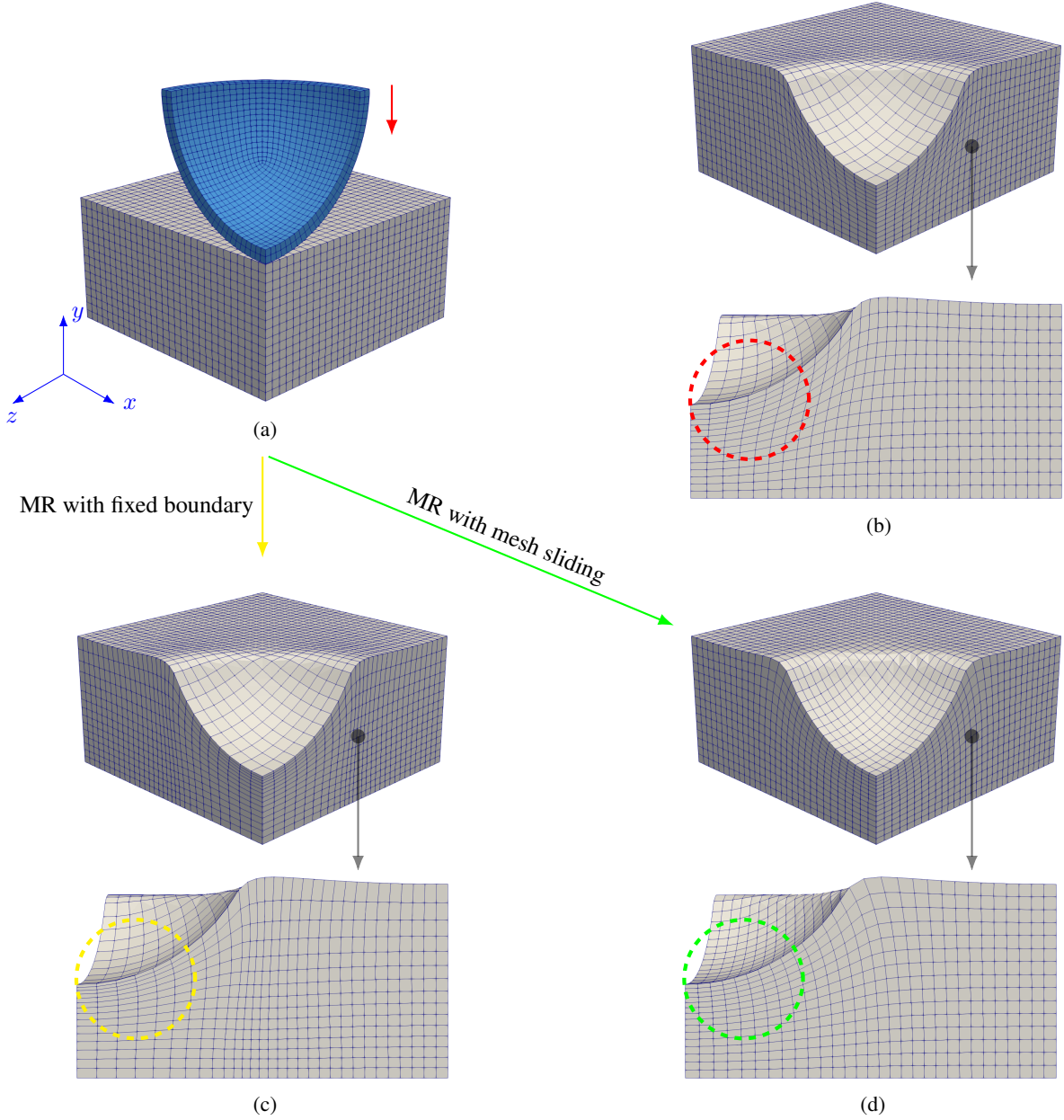}
            \caption{ Elasto-plastic punching of a rectangular block of dimensions $0.5\times0.5\times0.25$ by a rigid sphere of outer radius~$0.25$: (a) reference configuration. (b) final deformed state without mesh refitting. (c) The solution of the same physical problem with accompanying mesh refitting after every time step with mesh sliding and (d) with fixed boundary, respectively. In the subfigures (b)-(d), the rigid sphere is hidden to provide a clear view of the deformed zone.} \label{fig:mesh_adaptation_example_0_II}
        \end{figure}

       Next, we demonstrate the benefits arising from the use of the mesh sliding algorithm compared to keeping boundary nodes fixed using two numerical examples in~\cref{fig:mesh_adaptation_example_0,fig:mesh_adaptation_example_0_II}. These examples are provided for illustration and a better understanding of these specific effects only and are not designed to demonstrate the extreme cases that our approach can handle. First, consider the isothermal frictional sliding of a rigid cylinder over a rectangular block as shown in~\cref{fig:mesh_adaptation_example_0}. The rigid cylinder has a diameter of $0.25$ with height 0.125, and the rectangular block has the dimensions $1\times0.25\times0.125$ (see~\cref{fig:mesh_adaptation_example_0_a,fig:mesh_adaptation_example_0_b}). The rectangular block is meshed with $2500$ cubic 8-noded hexahedral elements and is modeled with a finite strain hyperelastoplastic material model as studied in Section 3.4.2.5 of~\cite{danowski2014thesis}. The isothermal frictional contact is modeled using the mortar penalty method presented in~\cite{puso2004mortar}. The motion of the cylinder is completely displacement-controlled. First, the cylinder is pressed onto the rectangular block by displacing the cylinder through 0.005 in the \(-y\) direction. Next, the cylinder is moved from \(x=0.25\) to \(x=0.75\) in a straight line. This procedure induces elastic and plastic deformation on the top surface of the block as shown in~\cref{fig:mesh_adaptation_example_0_a_woma}. It can be seen that the boundary elements underneath the cylinder (shown in a red dashed circle) have experienced shear distortion and this boundary is under consideration for mesh sliding. At this deformed state, the presented mesh refitting is performed with fixed boundary and mesh sliding and the resulting new mesh configurations are illustrated in~\cref{fig:mesh_adaptation_example_0_a_wma_mt,fig:mesh_adaptation_example_0_a_wma}. When MR is performed with a fixed boundary, the original cubic shape of the elements cannot be restored, as is evident from the elements within the yellow dashed circle in \cref{fig:mesh_adaptation_example_0_a_wma_mt}. However, in MR with mesh sliding (see~\cref{fig:mesh_adaptation_example_0_a_wma}), the refitted mesh in the whole domain, and in particular in the region below the cylinder (shown in a green dashed circle), resembles the initial uniform mesh, i.e., cubic elements. This is achieved by the free tangential sliding of the boundary nodes in the mesh refitting step. In short, in case of pure shear at the boundary, the quality of the refitted mesh is close to the initial mesh after applying the mesh sliding approach.
        
       The second example investigates the punching of a rectangular block by a rigid sphere. Only a quarter part is modeled exploiting the symmetry of the problem configuration (see ~\cref{fig:mesh_adaptation_example_0_ref}). The quarter rectangular block of size $0.5\times0.5\times0.25$ is modeled with a hyperelastoplastic model, the same as in the first example, and meshed with $13500$ cubic 8-noded hexahedral elements. The rigid quarter sphere has an outer radius of $0.25$ and is represented by the blue colored body in~\cref{fig:mesh_adaptation_example_0_ref}. Isothermal frictionless contact is applied as presented in~\cref{sec:thermomechanical_contact}. The sphere segment is moved vertically downwards by a distance equal to half of the block thickness, i.e., $0.125$, in $125$ load steps. This induces elastic and plastic deformations in the rectangular block as depicted in~\cref{fig:mesh_adaptation_example_0_b_woma}. The elements near the upper surface of the block are largely distorted, especially in the transition region between the contact and non-contact areas (highlighted by the red dashed circle in~\cref{fig:mesh_adaptation_example_0_b_woma}). This boundary is considered for mesh sliding. Mesh refitting was carried out at every time step using two different variants, one with fixed boundary and one with mesh sliding. In MR with mesh sliding, as shown in the~\cref{fig:mesh_adaptation_example_0_b_wma_ms}, mesh contains less distorted elements at the boundary and inside the volume. Furthermore, the contact area has more elements than the original problem. Moreover, the gradual transition of element shape from the boundary to the volume, as highlighted in the green dashed circle in~\cref{fig:mesh_adaptation_example_0_b_wma_ms}, shows a significant improvement. MR with fixed boundary nodes can lead to a strong mesh distortion in the boundary region, which can in turn deteriorate the convergence of nonlinear solvers. In this example, convergence using the Newton--Raphson scheme could not be achieved with the same number of load steps when boundary nodes were fixed. Instead, the number of load steps had to be increased to 250, i.e., doubled, to achieve convergence. In addition, even the penalty parameters for MR had to be reduced. The resulting mesh, illustrated in~\cref{fig:mesh_adaptation_example_0_b_wma_mt}, reveals that the quality of the mesh at the boundary does not show significant improvement when compared to MR with mesh sliding, as evident in the yellow dashed circle.
       
       For both shown examples, no significant improvement of the mesh quality could be achieved if the boundary nodes where kept fixed. Thus, mesh sliding is required to obtain a proper mesh relaxation also in boundary regions, especially for problems with large boundary distortion.
        
        \subsubsection{Algorithmic aspects of the mesh refitting method}\label{sec:meshadaptation_algorithm}
        In the following, more detailed algorithmic aspects of the overall mesh refitting approach will be presented.
        
        \subsubsection*{\textit{Target shape incrementation scheme}}
        The non-linear problem in~\cref{mesh_adaptation_subproblem} may not be solvable in one step if the old mesh is heavily distorted. To improve convergence of the Newton-Raphson scheme we employ an incrementation approach to the target lengths in~\cref{constraint_avg} and target angles in~\cref{constratint_angle}. Thereto, we define $N$ incrementation steps for the mesh refitting algorithm during which the incrementation factor $\alpha_{n_{inc}} \in [0,1]$, with $n_{inc} \in {1,...,N}$, is increased from $0$ to $1$. Let $l^{i}_{e0}$ and $\theta^{ij}_{e0}$ be the element average lengths and angles of the original distorted mesh at the beginning of the MR algorithm (updated reference configuration). Moreover, $l^{i}_{\text{r}}$ and $\theta^{ij}_{\text{r}}$ are the target element edge lengths and angles to achieve. Then, for the current incrementation step ${n_{inc}}$, the elemental target lengths and angles are defined as 
        \begin{align}
            \big(l^{i}_{\text{r}}\big)_{n_{inc}}     & = l^{i}_{e0} + \alpha_{n_{inc}} \big( l^{i}_{\text{r}} -l^{i}_{e0} \big),                    \\
            \big(\theta^{ij}_{\text{r}})_{n_{inc}}   & =\theta^{ij}_{e0} + \alpha_{n_{inc}} \big( \theta^{ij}_{\text{r}} - \theta^{ij}_{e0} \big).
        \end{align}
        For the numerical examples considered in this work, the target angles are set to $\theta^{ij}_{\text{r}}=\pi/2$. 
        \begin{remark}
            The target incrementation scheme is similar to a classical substepping procedure. When a Newton step is not converged the step size is subdivided. However, when a prescribed number of consecutive subs-steps converges in a few iterations, the sub-step size can be increased again.
        \end{remark}
     
        \subsubsection*{\textit{Uniform mesh regularization}}
            The aim of a purely uniform mesh regularization is to achieve uniform element sizes and shapes inside the entire problem domain. To achieve this goal, we prescribe the target element edge lengths as the average element edge length within the total problem domain determined for the (original) distorted mesh. The average edge length $\bar{l}^i_{\text{r}}$ is defined as 
            \begin{align}
                \bar{l}^i_{\text{r}} =\frac{1}{N_e} \sum^{N_e}_j \magvector{(\bar{\bm{v}}^i)_j},\label{average_lengths}
            \end{align}
            where $N_e$ is the number of elements in $\Omega_m$ and the index $i$ refers to the direction $\bm{e}^i$ in parameter space.
            
        \subsubsection*{\textit{Mesh localization}}
            To achieve different target element edge lengths at different locations of the problem domain, we define a continuous spatial distribution function for ${l}^i_{\text{r}}$:
            \begin{align}\label{mesh_localization}
                {l}^i_{\text{r}}(\bm{X}) = {l}^i_{\text{r0}} f(\cordX),
           \end{align} 
            where $f(\cordX)$ is a spatial function and ${l}^i_{\text{r0}}$ is a reference value of the target element edge length; for instance, it can be an average length as defined in~\cref{average_lengths}. In this work, we employ the following exponential function:
            \begin{align}\label{spatial_exp_average_length}
                f(\cordX) = 1+\exp{(-c \magvector{\bm{X}-\bm{X}_0}^2)},
            \end{align}
            where the parameter $c$ controls the rate of decay of the function when departing from the reference point $\bm{X}_0$. According to~\eqref{spatial_exp_average_length}, the distribution function ${l}^i_{\text{r}}(\bm{X})$ reproduces the reference value ${l}^i_{\text{r0}}$ when evaluated at the reference point $\bm{X}_0$. In practice, the reference point $\bm{X}_0$ typically represents a location of the physical problem, which is characterized by strong gradients of the primary variables accompanied by a significant mesh distortion. Thus, the mesh localization approach allows to have smaller elements, i.e., a higher mesh resolution, at this location. Moreover, to enforce a high mesh quality, i.e., small element distortions, at this critical location, also the penalty parameters may be prescribed as spatial functions based on~\eqref{spatial_exp_average_length}, i.e., $\varepsilon_E(\cordX)=\varepsilon_{E0} \ f(\cordX)$ and $\varepsilon_A(\cordX)=\varepsilon_{A0} \ f(\cordX)$.
            
            \subsubsection*{\textit{Demonstration example: Uniform mesh regularization and localization}}

        \begin{figure}[htb]
            \centering
            \input{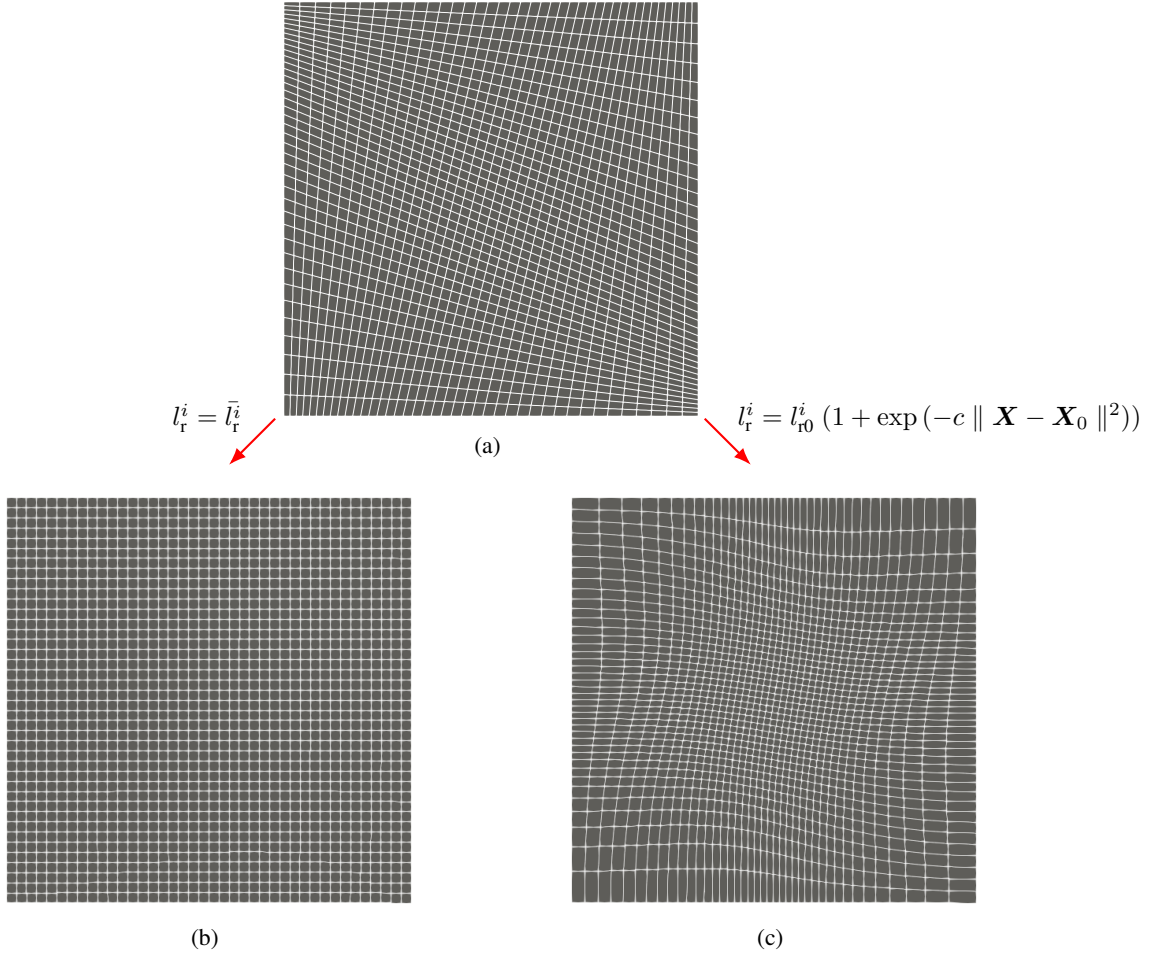}
            \caption{Illustration of mesh regularization and localization: (a) initial distorted mesh (b) uniformly regularized mesh (c) localized mesh with smaller elements at center of the square.} \label{fig:mesh_adaptation_example}
        \end{figure}
        
        Uniform mesh regularization and localization shall be illustrated by a 2-dimensional numerical example defined on a $2\times2$ domain. The initial distorted mesh is created by 40 unequal divisions of the edges resulting in 1600 elements as illustrated in~\cref{fig:mesh_adaptation_example_aux_initial_mesh}. To achieve a uniform regularized mesh, we set the target lengths to an average length according to~\cref{average_lengths}. The resulting mesh is shown in~\cref{fig:mesh_adaptation_example_reg_mesh}, which clearly confirms the underling idea of the uniform mesh regularization approach. On the other hand, a localization is achieved using an exponential function according to~\cref{spatial_exp_average_length} with ${l}^i_{\text{r0}}=0.025$,  $c=0.1$, and $\cordX_0$ representing the center of the domain. The resulting mesh is portrayed in~\cref{fig:mesh_adaptation_example_loc_mesh}. As desired, this approach allows to achieve a higher mesh resolution with very regular elements at the location of interest, i.e., the center of the domain. Of course this ansatz compromises the mesh quality in other regions of the problem setup. Nevertheless, it can be very helpful in scenarios with very strict requirements on the mesh quality in certain regions of the problem.
    
\subsection{Transfer of mesh data}\label{sec:mesh_adapt_data_transfer}
    \begin{figure}[htbp]
        \centering
        \scalebox{1.0}{\def\latticetilt{%
    \pgf@xa=\pgf@x%
    \pgf@ya=\pgf@y%
    \pgfmathsetmacro{\myx}{\pgf@xa+\pgfkeysvalueof{/tikz/lattice/amplitude}*sin((\pgf@ya/\pgfkeysvalueof{/tikz/lattice/spacing})*360/\pgfkeysvalueof{/tikz/lattice/superlattice period})}%
    \pgf@x=\myx pt%
    \pgfmathsetmacro{\myy}{\pgf@ya+\pgfkeysvalueof{/tikz/lattice/amplitude}*sin((\pgf@xa/\pgfkeysvalueof{/tikz/lattice/spacing})*360/\pgfkeysvalueof{/tikz/lattice/superlattice period})}%
    \pgf@y=\myy pt}

\begin{tikzpicture}[lattice/.cd,spacing/.initial=5,superlattice
        period/.initial=30,amplitude/.initial=3]
    \begin{scope}[xshift=-5.0cm]
        \tikzset{d/.style={minimum width=4pt,inner sep=0pt,circle,fill=black}}
        \tikzset{cross/.style={cross out, draw=black, fill=none, minimum size=2*(#1-\pgflinewidth), inner sep=0pt, outer sep=0pt}, cross/.default={2pt}}

        \draw[black] (-0.5,-0.5) grid (4.5,4.5);

        \def\nx{4} \def\ny{4}
        \foreach \i in {0,...,\nx}
        \foreach \j in {0,...,\ny}
        \node[d] at (\i,\j) {};

        \draw[dashed,red] (2,2) circle(1.85);                                           
        \draw[-latex] (2,2) -- ++(22.5:1.85)  node [midway,sloped,fill=white] {$\color{red} r_p$};  
        \node[inner sep=2pt,label=above:$\color{red} \Omega_p$] at (3.25,-0.125)  {};   
        \node[circle,red,fill,inner sep=2pt,label=above left:$\color{red} \bm{x}_p$] at (2,2) {};        

        \draw[-latex] (5.25,2) to[bend left] ++(1,0);
    \end{scope}
    
    \begin{scope}[xshift=-5.30cm,yshift=-0.20cm]
        \tikzset{d/.style={minimum width=4pt,inner sep=0pt,circle,fill=black,opacity=0.5}}
        \pgftransformnonlinear{\latticetilt}
        \draw[black,step=1.125cm, opacity=0.25] (-0.125,-0.125) grid (5.125,5.125);;

        \def\nx{4} \def\ny{4}
        \foreach \i in {0,...,\nx}
        \foreach \j in {0,...,\ny}
        \node[d] at (1.125*\i,1.125*\j) {};
    \end{scope}
    
    \begin{scope}[xshift=2.0cm,yshift=-0.20cm]
        \tikzset{d/.style={minimum width=4pt,inner sep=0pt,circle,fill=black,opacity=0.5}}
        \pgftransformnonlinear{\latticetilt}
        \draw[black,step=1.125cm, opacity=0.25] (-0.125,-0.125) grid (5.125,5.125);;

        \draw[black,step=1.125cm] (0.575,0.575) grid (4.0,4.0);;
        \def\nx{3} \def\ny{3}
        \foreach \i in {1,...,\nx}
        \foreach \j in {1,...,\ny}
        \node[d] at (1.125*\i,1.125*\j) {};
        
        \draw[shorten <=2pt,very thin] (1.125*3,1.125*0.75) to[out=-90,in=90] ++ (110:-0.8) node[right]{$\text{element patch}$};
    \end{scope}

    \begin{scope}[xshift=-9.125cm,yshift=-0.25cm]
        \coordinate [](A) at (4, -1.0);
        \draw[ opacity=0.25, color=black] (3.5, -1.0)--(A);
        \tikzset{d/.style={minimum width=4pt,inner sep=0pt,circle,fill=black,opacity=0.75}}
        \node[d] at (3.75,-1.0) {};

        \tikzset{d/.style={minimum width=4pt,inner sep=0pt,circle,fill=black}}
        \draw[](A) node[anchor=west,align=center]  {old mesh ($\Omega_m$)} ;
        \coordinate [](A) at (4, -1.5);
        \draw[color=black] (3.5, -1.5)--(A);
        \node[d] at (3.75,-1.5) {};
        \draw[](A) node[anchor=west,align=center]  {new mesh ($\Omega^\prime_m$)} ;
    \end{scope}

\end{tikzpicture}} \caption{ Illustration of element patch for transferring data from old mesh to new mesh.}\label{fig:data_mapping_illustrative}
    \end{figure}
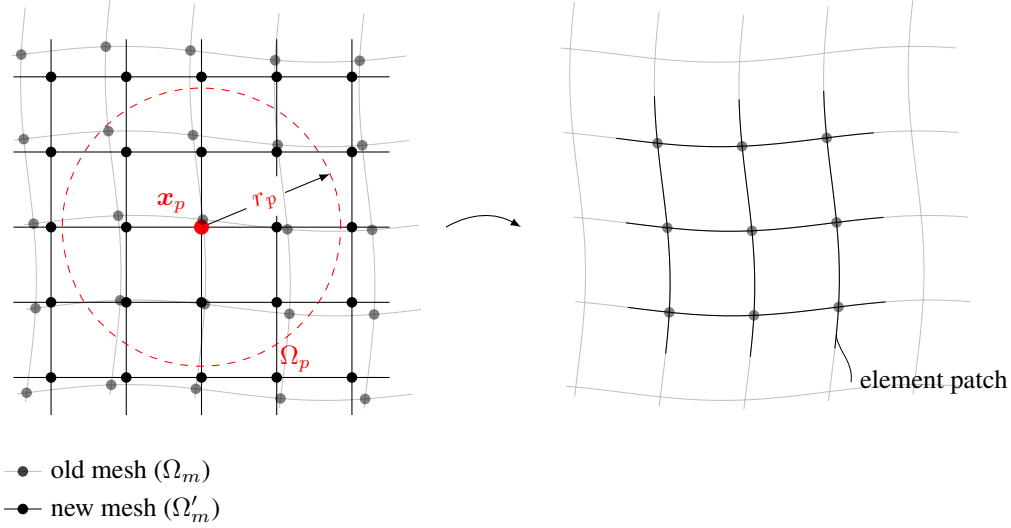
    Within our overall mesh refitting approach, the transfer of data from the old (distorted) mesh on $\Omega_m$ to the new (regularized) mesh on $\Omega^\prime_m$ is a critical aspect. The variables to be transferred include nodal primary variables (i.e., displacement and temperature field) but also internal material variables (e.g., the inelastic deformation gradient) defined at quadrature points. These variables can be broadly classified as scalars, vectors, and tensors, whereas the latter represents the most challenging case from a data transfer point of view. This section presents the main strategy for data transfer as employed in this work including a tensor interpolation scheme proposed in our recent contribution~\cite{satheesh}. It is emphasized that the proposed data transfer schemes are independent of the mesh refitting scheme proposed in the previous sections, and can be combined with arbitrary mesh regularization, mesh refinement and remeshing schemes.
    
    Consider an element node or quadrature point located at $\bm{x}_p$ in the new mesh $\Omega^\prime_m$. To determine the new data at $\bm{x}_p$, we interpolate data from an element patch $\Omega_p \subset \Omega_m$ within a radius of $r_p$ around this point (see~\cref{fig:data_mapping_illustrative}). Let $\bm{x}_j \in \{ \bm{x}_1,\dots,\bm{x}_N \}$ be a set of position vectors in $\Omega_p$, while $ \alpha_j \in \{ \alpha_1,\dots,\alpha_N \}$ and $\bm{T}_j  \in \{\bm{T}_1,\dots,\bm{T}_N\}$ represent scalar- and tensor-valued data associated with these points. The methods presented in the following rely on a relative weighting of data considering the distance of the data points from the interpolation point. For this purpose, we employ the normalized weighting function $\tilde{w}(\bm{x}_j)$ according to:
    \begin{align}
      \tilde{w}(\bm{x}_j) := \frac{{w}(\bm{x}_j)}{\sum_{j=1}^N {w}(\bm{x}_j)} \quad \textrm{such that} \quad \sum_j \tilde{w}(\bm{x}_j) =1. \label{weighting_function}
    \end{align}
    Here the weighting function ${w}(\bm{x}_j)$ can be any monotonic continuous function that decreases as it moves away from the interpolation point $\bm{x}_p$. For example, an exponential weighting function with control parameter $c$ reads:
    \begin{align}
      {w}(\bm{x}_j) = \exp\left(-c || \bm{x}_j - \bm{x}_p||^2\right). \label{exponential-weighting-function}
    \end{align}
    In the following subsection, we demonstrate methods to compute scalars and tensors at $\bm{x}_p$, denoted as $\alpha_{p}$ and $\bm{T}_{p}$. 
    
        \subsubsection{Scalar interpolation}\label{sec:scalar_interpolation}
        We employ two different schemes for scalar interpolation, namely the moving least squares (MLS) and the logarithmic moving least squares (LOGMLS) method. Importantly, the LOGMLS scheme preserves strict positivity of strictly positive data ($\alpha_j > 0$), but, in turn, is limited to data exhibiting this property. Furthermore, both methods preserve important additional properties such as monotonicity of the data. The two schemes are briefly outlined below.
        
        \begin{enumerate}
            \item \textit{Moving least squares (MLS)}: This variant employs a spatial polynomial approximation which reads
            \begin{align}
                \alpha(\bm{x}):= \bm{p}(\cordx) \bm{a},
            \end{align}
             where $\bm{p}(\bm{x}) \in \mathbb{R}^m$ is the vector of polynomial basis functions of order $m$, and $\bm{a} \in \mathbb{R}^m$ is the corresponding vector of coefficients. The unknown coefficient vector $\bm{a}$ is obtained by minimizing the residual
            \begin{align}
                 r = \sum_{j=1}^N \tilde{w}(\bm{x}_j) \big(\bm{p} (\bm{x}_j) \bm{a} - \alpha_j \big)^2. \label{mls_definition}
            \end{align}
            As a prerequisite, the order of the polynomial function $m$ must be chosen such that $m \leq N$.
            \item \textit{Logarithmic moving least squares (LOGMLS))}: This method ensures non-negative interpolation of positive quantities. It employs a moving least squares approximation of a logarithmically transformed scalar field:
            \begin{align}
                \alpha(\bm{x}):= \exp(\bm{p}(\cordx) \bm{a}).
            \end{align}
            In this approach, the unknown vector of coefficients $\bm{a}$ is found by minimizing the residual 
            \begin{align}
                  r = \sum_{j=1}^N \tilde{w}(\bm{x}_j) \big(\bm{p} (\bm{x}_j) \bm{a} - \ln (\alpha_j) \big)^2. \label{logmls_definition}
            \end{align}
        \end{enumerate}
        For a detailed analysis of these two methods the reader may refer to our previous work~\cite{satheesh}. 

        \begin{figure}[htbp]
            \centering
            \input{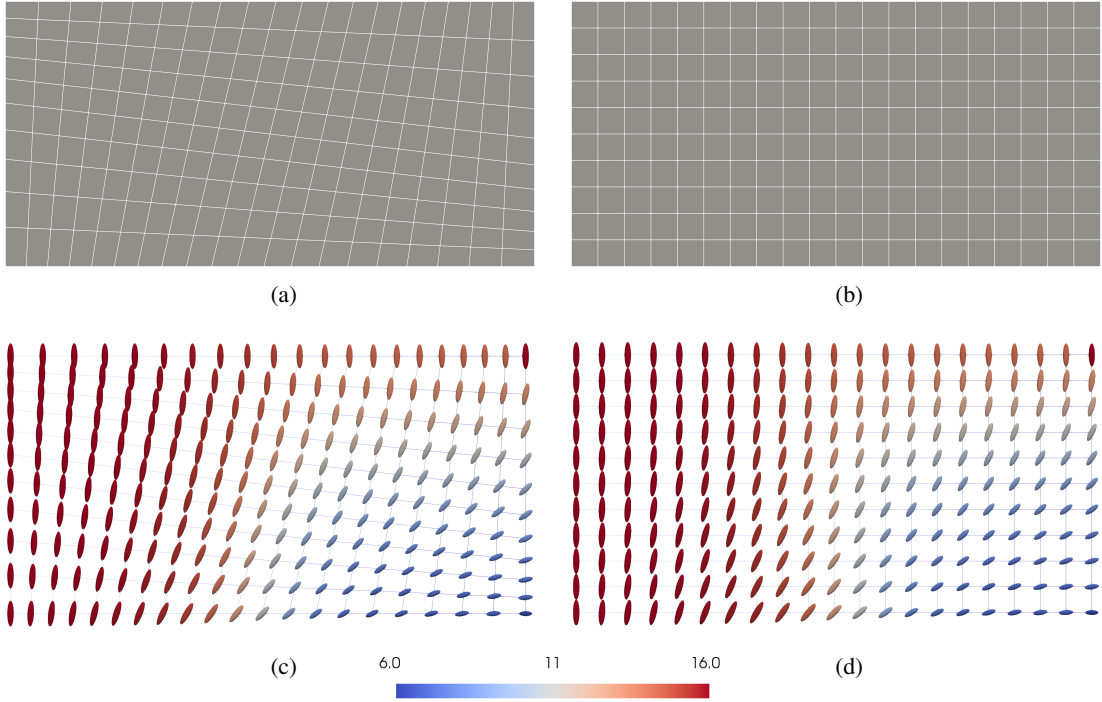}
            \caption{ Tensor interpolation example: (a) input mesh (b) regularized mesh (c) ellipsoidal representation of input tensor field (d) ellipsoidal representation of mapped tensor field. The color represents the determinant of the tensor.} \label{fig:tensor_interpolation_example}
        \end{figure}
        \subsubsection{Tensor interpolation}\label{sec:tensor_interpolation}
        We employ rotation vector-based methods for tensor interpolation proposed in our previous work~\cite{satheesh}. These methods exploit the polar and spectral decomposition of the tensor data according to $\bm{T}_j=\bm{R}_j\bm{Q}_j^T\bm{\Lambda}_j\bm{Q}_j$, where $\bm{R}_j, \ \bm{Q}_j \in \mathbb{SO}(3)$ are rotation tensors and $\bm{\Lambda}_j$ is the positive definite diagonal tensor containing the eigenvalues of $\bm{T}_j$. The general strategy for tensor interpolation relies on an individual interpolation of the rotation tensors and the eigenvalues contained in $\bm{\Lambda}_j$. First, the scalar eigenvalues are individually interpolated using the schemes from~\cref{sec:scalar_interpolation} to finally reconstruct $\bm{\Lambda}_p$. For interpolation of the rotation tensors $\bm{R}_j, \ \bm{Q}_j$ specific schemes are employed that preserve, among others, the objectivity of the underlying mechanical problem and are well-established, e.g., in the field of geometrically exact beam theories~\cite{meier2019geometrically}. This step results in the interpolated rotation tensors $\bm{R}_p$, $\bm{Q}_p$. Finally, the interpolated tensor $\bm{T}_p$ at $\bm{x}_p$ is reconstructed according to $\bm{T}_p=\bm{R}_p\bm{Q}_p^T\bm{\Lambda}_p\bm{Q}_p$.
        
        These interpolation methods are well suited for any invertible second-order tensor. In the context of finite element discretizations for problems of nonlinear continuum mechanics, tensor-valued history often arises for material models involving, e.g., inelastic constitutive behavior~\cite{prakash2015multiscale,frydrych2019solution,dittmann2020phase} or phase change~\cite{proell12b,proell12c,proell2023highly}. Generally, these methods have been proven to preserve important properties of the tensor during interpolation (e.g., positive definiteness, objectivity, etc.) and allow for higher-order spatial convergence~\cite{satheesh}.
        
        \subsubsection*{\textit{Demonstration example: Transfer of tensor data}}
        
        The tensor interpolation as part of the mesh adaptation approach is portrayed in~\cref{fig:tensor_interpolation_example}. Consider an initial mesh as visualized in~\cref{fig:tensor_interpolation_example_initialmesh} with a tensor field as illustrated in~\cref{fig:tensor_interpolation_example_iptensor} (ellipsoidal representation, see~\cite{satheesh}). Now, the mesh regularization is performed as described in~\cref{sec:meshadaptation_subproblem} and the resulting mesh is shown in~\cref{fig:tensor_interpolation_example_regmesh}. For tensor interpolation, we employed the "R-MLS" variant as defined in our previous work~\cite{satheesh} to transfer tensor data from the old to the new mesh. From~\cref{fig:tensor_interpolation_example_inttensor,fig:tensor_interpolation_example_iptensor} it is evident that the method delivers a smooth interpolation while preserving the magnitude and orientation of the tensor data when mapped.
\section{Numerical results}\label{sec:numerical_results}
In this section we show the capabilities of the proposed mesh adaptation scheme using different numerical examples.
\subsection{Expansion past a rigid obstruction}\label{sec:example_1}
\begin{figure}[htb]  
    \centering
  \begin{subfigure}{0.45\textwidth}
    \centering
    \begin{tikzpicture}[scale=0.75]
	\def\ax{1};
	\def\lx{3};
    \def\nx{3};
    
    \def\xshift{0}
    \def\yshift{3}
    \def\thic{1.25}
    \def\rad{0.35}
    \coordinate [](origin) at (\xshift,\yshift);
\begin{scope}
    \fill [black!30] (origin) rectangle (\lx+\xshift,\lx+\yshift);

    \draw[-latex,color=blue] (0,0) -- (\ax,0) node[anchor=west]{$x$};
    \draw[-latex,color=blue] (0,0) -- (0,\ax) node[anchor=west]{$y$};

	\tikzset{d/.style={minimum width=4pt,inner sep=0pt,circle,fill=black}}
	
	\coordinate [] (1) at (origin) {};
    \coordinate [] (2) at (\lx+\xshift,\yshift) {};
	\coordinate [] (3) at (\lx+\xshift,\lx+\yshift) {};
	\coordinate [] (4) at (\xshift, \lx+\yshift) {};

	\foreach \i in {0,...,\nx}
	\draw [fill=black] ([xshift=-4pt]\xshift,0.75*\i+\yshift+0.35) circle [radius=3pt];
	\foreach \i in {0,...,\nx}
	\draw [fill=black] (0.75*\i+\xshift+0.35,0.125+\lx+\yshift) circle [radius=3pt];

	\draw[]([yshift=7pt,xshift=-7pt]4) -- ([yshift=7pt]3);
	\path[pattern=north west lines, shift={(3cm,3cm)}] ([yshift=7pt,xshift=-7pt]4) rectangle ([yshift=12pt]3);

    \draw[]([yshift=7pt,xshift=-7pt]4) -- ([xshift=-7pt,yshift=-100pt]1);
    \path[pattern=north west lines, shift={(3cm,3cm)}] ([yshift=7pt,xshift=-7pt]4) rectangle ([xshift=-12pt,yshift=-100pt]1);
	
	\draw[yellow, line width=2pt] ([yshift=1pt]1) -- ([yshift=1pt]2);
	\draw[green, line width=2pt] ([xshift=-1pt]2) -- ([xshift=-1pt]3);
	\draw[red, line width=2pt] ([yshift=-1pt]3) -- ([yshift=-1pt]4);
	\draw[] (4) -- (1);	
	
      
      \draw[shorten <=2pt,very thin] ([xshift=-60pt]3) to[out=-90,in=90] ++ (-110:0.8) node[below]{$\Gamma_T$};
      
  \dimline[color=black]{([yshift=25pt]4)}{([yshift=25pt]3)}{$w_e$};
     \dimline[color=black]{([shift={(-10mm,0mm)}]1)}{([shift={(-10mm,0mm)}]4)}{\rotatebox{-90}{$h_e$}};

	\coordinate [] (7) at (\lx+\xshift,0+\rad) {};
	\coordinate [] (8) at (\lx+\xshift+\thic,0+\rad) {};
	\coordinate [] (9) at (\lx+\xshift,\lx+\yshift) {};
	\coordinate [] (10) at (\lx+\xshift+\thic,\lx+\yshift) {};
	
	\coordinate [] (20) at (\lx+\xshift+\rad,0+\rad) {};
	\coordinate [] (21) at (\lx+\xshift+\thic-\rad,0+\rad) {};

	\coordinate [] (11) at (\lx+\xshift+\rad,0) {};
	\coordinate [] (12) at (\lx+\xshift+\thic-\rad,0) {};

	\coordinate [] (96) at (\lx+\xshift,0) {};
	\coordinate [] (97) at  (\lx+\xshift+\thic,0) {};
	
    \draw[rounded corners=8pt,fill=blue!30] ([xshift=3pt]96) -- ([xshift=3pt]97) -- ([xshift=3pt]10) -- ([xshift=3pt]9) -- cycle;
    \draw[fill=blue!30,blue!30] ([xshift=3pt]9) rectangle ([xshift=3pt,yshift=-30pt]10);
    
	\draw[line width=2pt,orange] ([xshift=3pt]7) -- ([xshift=3pt]9);
	 \draw[shorten <=2pt,very thin] ([xshift=4pt,yshift=120pt]7) to[out=90,in=-90] ++ (-110:-0.8) node[right]{$\Gamma^{(1)}_c$};
 	 \draw[shorten <=2pt,very thin] ([xshift=-5pt,yshift=120pt]7) to[out=90,in=-90] ++ (110:0.8) node[left]{$\Gamma^{(2)}_c$};
   	 \draw[shorten <=2pt,very thin] ([xshift=-4pt,yshift=110pt]7) to ([xshift=-20pt,yshift=110pt]7) node[left]{$\Gamma^{(1)}_m$};
   	 \draw[shorten <=2pt,very thin] ([xshift=-50pt,yshift=75pt]7) to[out=90,in=-90] ++ (-110:-0.8) node[above]{};
	\draw[line width=2pt,orange] ([xshift=3pt]7) arc  (0:90:-\rad);
	\draw[line width=2pt,orange] ([xshift=3pt]12) arc (90:180:-\rad);
	\draw[line width=2pt,orange] ([xshift=3pt]11) -- ([xshift=3pt]12);
	\draw[thick] ([xshift=3pt]9) -- ([xshift=3pt]10);
	\draw[thick] ([xshift=3pt]10) -- ([xshift=3pt]8);

    \dimline[color=black]{([xshift=0pt,yshift=25pt]9)}{([xshift=3pt,yshift=25pt]10)}{$w_o$};
    
	\coordinate [] (22) at (\lx+\xshift-\rad+\thic,\lx+\yshift) {};
	\coordinate [] (23) at (\lx+\xshift-\rad+\thic,0) {};
    \dimline[color=black]{([shift={(12mm,0mm)}]22)}{([shift={(12mm,0mm)}]23)}{\rotatebox{90}{$h_o$}};
    
	\coordinate [] (24) at (\lx+\xshift,0) {};
	\coordinate [] (25) at (\lx+\xshift+\thic,0) {};
    \draw[color=black,-latex,label={1}] (24) -- ++(45:0.35*\rad) ;
    \draw[color=black,label={1}] (24) -- ++(45:-0.5*\rad) node[label={[xshift=-1mm, yshift=-8mm]$R=\frac{w_0}{4}$}] {};
    
    \draw[shorten <=2pt,very thin] ([xshift=5pt]24) to[out=90,in=-90] ++ (-110:-0.8) node[above right]{$\bm{X}_c$};

\end{scope}

\end{tikzpicture}
    \caption{}\label{fig:example_1_geometry}
  \end{subfigure}
\begin{subfigure}{0.45\textwidth}
    \centering
    \begin{tikzpicture}
      \node[inner sep=0pt] (mytable) {
                    \begin{tabular}{cc}
                      \hline
                      Quantity & Value\\
                      \hline
                   height $h_e$ & 1\\
                    width $w_e$ & 1\\
                    thickness $t_e$ &0.1\\
                   height $h_o$ & 2\\
                    width  $w_o$ & $1/4$\\
                      thickness $t_o$ &0.1\\
                      \hline
                    \end{tabular}
             };
        \node[] (dummy) at ([yshift=-90pt]mytable){};
    \end{tikzpicture}
     \caption{}\label{fig:example_1_dimension}
  \end{subfigure}
    \caption{ Example 1: Problem setup with a grey body denoting the expanding material and a violet body representing the rigid obstruction. (a) geometry and boundary conditions (b) dimensions. A temperature surface Dirichlet boundary condition is applied at $\Gamma_T$ (red line). Moreover, all other boundaries of expanding material are modeled adiabatic. The thermo-mechanical contact slave ($\Gamma^{(1)}_c$) and master ($\Gamma^{(2)}_c$) boundaries are represented using in orange and green lines. For the mesh adaptation the mesh sliding surfaces ($\Gamma^{(1)}_m$) are denoted by green and yellow lines. }\label{fig:numerical_exmaple_1_problem_dimension}
\end{figure}
\begin{minipage}{0.45\textwidth}
  \centering
    \begin{table}[H]
        \centering
        \caption{Material parameters of the expanding material}\label{material_parameters}
            \begin{tabular}[t]{cc}
            \hline
            Parameter & Value\\ 
            \hline
            Young's modulus ($E$) &$1\times10^4$\\
            Poisson's ratio ($\nu$) & $0$ \\
            thermal conductivity ($k_0$) &$7.55\times 10^6$\\
            Heat capacity ($c_v$)  & $1.4\times10^{-2}$\\ 
            \hline
        \end{tabular}
    \end{table}%
\end{minipage}
\hfill
\begin{minipage}{0.45\textwidth}
  \centering
     \begin{table}[H]
        \centering
        \caption{Mesh refitting parameters}\label{meshapatation_parameters}
            \begin{tabular}[t]{lc}
            \hline
            Parameter&Value\\
            \hline
             $\Bar{\varepsilon}_E,\Hat{\varepsilon}_E,{\varepsilon}_A$ & $1\times10^{-2}$\\
             $\epsilon_m$ &$2\times10^8$\\
             Maximum $n_{inc}$ & 20\\
             Displacement tolerance & $1\times10^{-5}$\\ 
            \hline
        \end{tabular}
     \end{table}%
\end{minipage}
\begin{figure}[htbp]
    \input{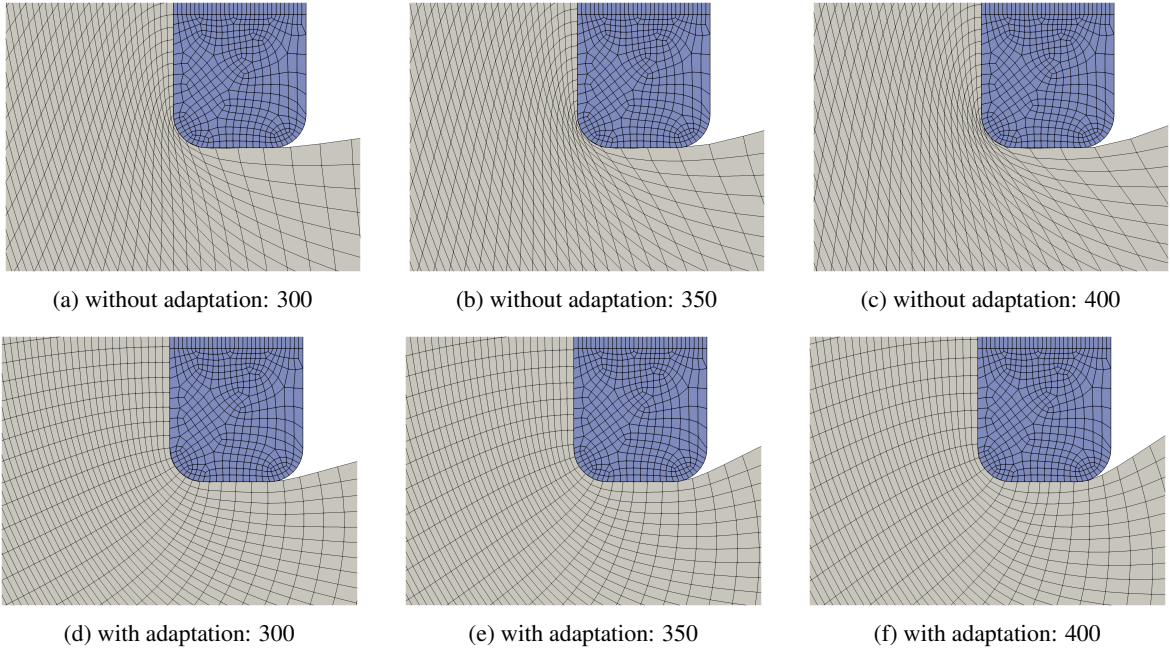}
    \caption{Example 1: Mesh around the corner (a)-(c) without mesh adaptation and (d)-(f) with mesh adaptation for time steps 300, 350, and 400, respectively}\label{fig:Example_1_mesh_at_corner}
\end{figure}

As the first numerical example, we explore a pseudo 2-dimensional expansion past a rigid body as illustrated in~\cref{fig:numerical_exmaple_1_problem_dimension}. The geometry and the boundary conditions are depicted in~\cref{fig:example_1_geometry} and the corresponding dimensions are given in~\cref{fig:example_1_dimension}. The expanding material is modeled as presented in~\cref{sec:problem_material}. It's elastic behavior described by $\Psi_e$ is modeled using a Neo-Hookean material model with parameters listed in~\cref{material_parameters}. The expansion is restricted to $+x$ and $-y$ direction by arresting the normal displacements as shown in~\cref{fig:example_1_geometry}. Furthermore, the initial temperature $T_0$ is set to $198$ and a temperature surface Dirichlet boundary condition $T= 148+ 345 \log_{10}(1+[(8\times(t+3))/60])$ is applied to $\Gamma_T$. Moreover, all other boundaries are modeled adiabatic. Finally, to avoid 3-dimensional effects, the displacements in $z-$direction are arrested. Both, the expanding and the rigid body are discretized with 8-noded hexahedral elements with $22500$ elements (45602 nodes) and $6414$ elements (13092 nodes) with one element in the thickness direction, respectively. For the mortar thermo-mechanical contact, the outer surface of the rigid body is chosen as the slave side and the surface of the expanding material as the master side. The contact interface is discretized with 4-noded quadrilateral elements, where the contact penalty parameter is set to $\epsilon_c=1\times10^8$ and the interface conductivity to $\beta_c=0$, i.e., modeling adiabatic contact. The thermo-mechanical problem is analyzed for $1000$ steps using a time step size $\Delta t=0.05$. The nonlinear system of equations resulting in each time step is solved using a Newton-Raphson scheme with a tolerance on the combined residual and increment of $10^{-8}$. The linear monolithic thermomechanical system to be solved in each Newton iteration is approached by means of an iterative GMRES method with AMG(BGS) preconditioner. The convergence tolerance for the linear solver is set to $10^{-10}$. 

The accompanying mesh adaptation problem is formulated as follows: to achieve a good quality mesh around the corner $\bm{X}_c$ (see~\cref{fig:example_1_geometry}), we employ a mesh localization as in~\cref{mesh_localization,spatial_exp_average_length} with $c=100$. The target element edge length ($l^i_{\text{r0}}$) in each time step are estimated as in~\cref{average_lengths} and mesh refitting parameters as listed in~\cref{meshapatation_parameters}. The mesh sliding surfaces $\Gamma^{(1)}_m$ in reference configuration are portrayed in the~\cref{fig:example_1_geometry}. For the Newton-Raphson scheme the convergence tolerance for the residual and the displacement increment is set to $10^{-5}$. Moreover, the linearized system is solved with the "SuperLU"~\cite{li2005overview} direct solver. After the mesh refitting step, data needs to be transferred from the old to the new mesh. For the material model as presented in~\cref{sec:problem_material}, the computation of the inelastic deformation gradient $\defgrad^{n+1}_{in}$ requires $\defgrad^{n}_{in}$, $\defgrad^{n}$, $\secondPK^{n}$, and $T^{n}$. These quantities have to be transferred to the new mesh. The tensor data $\defgrad^{n}_{in}$ is transferred by the "R-MLS" method with quadratic basis (see~\cref{{sec:tensor_interpolation}}), the deformation gradient $\defgrad^{n}$ is reconstructed from nodal displacements, and the stress $\secondPK^{n}$ and temperature $T^{n}$ are interpolated as scalar using moving least square with quadratic basis (see~\cref{sec:scalar_interpolation}). The mesh adaptation is carried out every $15^{\text{th}}$ step starting from step 210. Finally, the computation is carried out on 2 nodes (48 CPUs) of a computing cluster with Intel Xeon E5-2680v3 2.5GHz processors.

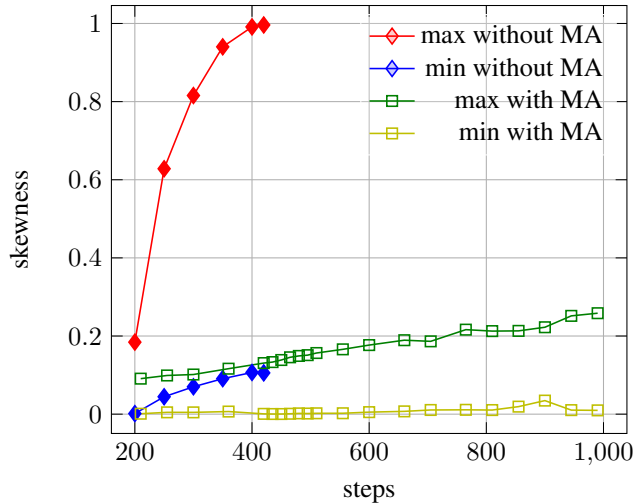
\begin{figure}[htbp]
        \centering
\begin{tikzpicture}

\definecolor{darkgray176}{RGB}{176,176,176}
\definecolor{goldenrod1911910}{RGB}{191,191,0}
\definecolor{green01270}{RGB}{0,127,0}

\begin{axis}[
legend cell align={right},
legend style={
  fill opacity=0.2,
  draw opacity=1,
  text opacity=1,
  draw=white
},
xmajorgrids,
ymajorgrids,
tick pos=both,
x grid style={darkgray176},
xmin=160, xmax=1040,
xtick style={color=black},
ylabel={skewness},
xlabel={steps},
y grid style={darkgray176},
ymin=-0.0494224863633597, ymax=1.04597453029848,
ytick style={color=black}
]
\addplot [semithick, red, mark=diamond*, mark size=3, mark options={solid}]
table {%
200 0.184276777025988
250 0.628173496325994
300 0.815583498762626
350 0.940210202062288
400 0.991188871528858
420 0.996183756813851
};
\addlegendentry{ max without MA}
\addplot [semithick, blue, mark=diamond*, mark size=3, mark options={solid}]
table {%
200 0.00144392298895595
250 0.0445240948600489
300 0.0700605108585342
350 0.0908144552599831
400 0.106674183715621
420 0.105605880946619
};
\addlegendentry{ min without MA}
\addplot [semithick, green01270, mark=square, mark size=2, mark options={solid}]
table {%
210 0.0908788729785786
255 0.0990141142533244
300 0.101452524776527
360 0.116379634552024
420 0.130727360845191
435 0.133280958004132
450 0.138595653433926
465 0.145531901182544
480 0.149038230007658
495 0.151049396416884
510 0.156491064089658
555 0.165887062166934
600 0.176823338674526
660 0.189289458624989
705 0.186369750054444
765 0.216314219179776
810 0.212590410469704
855 0.213150116479205
900 0.222454187199964
945 0.251566350570634
990 0.258402496437618
};
\addlegendentry{ max with MA}
\addplot [semithick, goldenrod1911910, mark=square, mark size=2, mark options={solid}]
table {%
210 0.00110702366460076
255 0.00465980773062016
300 0.00457991934891421
360 0.0066687637417085
420 0.000744584189569745
435 0.000357651519881428
450 0.000349791048516737
465 0.001021314266649
480 0.00197912245016217
495 0.00162402650866085
510 0.00220609929135031
555 0.0023149707867347
600 0.00516902130996534
660 0.00712966164786685
705 0.0106885066757458
765 0.0111207416896434
810 0.0104996509939539
855 0.0192500094914435
900 0.0346762879698341
945 0.010332630937756
990 0.00969844285614872
};
\addlegendentry{ min with MA}
\end{axis}

\end{tikzpicture}
        \caption{Example 1: Comparison of minimum and maximum skewness over time steps in the corner region around $\bm{X}_c$ with radius $r=0.1$ for the problem with and without mesh adaptation (MA).}\label{fig:numerical_exmaple_1_skewness}
\end{figure}  
\begin{figure}[htbp]
    \centering
    \begin{subfigure}[b]{1\linewidth}
        \centering
        \scalebox{0.775}{\input{Example_1_woma}}
    \end{subfigure}\\
    \begin{subfigure}[b]{0.45\linewidth}
        \begin{tikzpicture}
            \node [] (a) at (0,0){};
        \end{tikzpicture}
        \caption{ without adaptation: step 425}\label{fig:example_1_woma_step425}
    \end{subfigure}
    \begin{subfigure}[b]{.45\linewidth}
        \begin{tikzpicture}
            \node [] (a) at (0,0){};
        \end{tikzpicture}
        \caption{ with adaptation: step 425}\label{fig:example_1_wma_step425}
    \end{subfigure}\\
    \centering
    \begin{subfigure}[b]{1\linewidth}\setcounter{subfigure}{2}
        \centering
        \scalebox{0.85}{\input{Example_1}}
    \end{subfigure}\\
    \begin{subfigure}[b]{0.45\linewidth}
        \begin{tikzpicture}
            \node [] (a) at (0,0){};
        \end{tikzpicture}
        \caption{ step 800}\label{fig:example_1_step800}
    \end{subfigure}
    \begin{subfigure}[b]{.45\linewidth}
        \begin{tikzpicture}
            \node [] (a) at (0,0){};
        \end{tikzpicture}
        \caption{ step 1000}\label{fig:example_1_step1000}
    \end{subfigure}
    \caption{ Example 1: deformed state at step (a) 425 without adaptation, (b) 425 with adaptation, (c) 800 with adaptation, and (b) 1000 with adaptation. }\label{fig:nuermical_exmaple_1_stepafter425}
\end{figure}
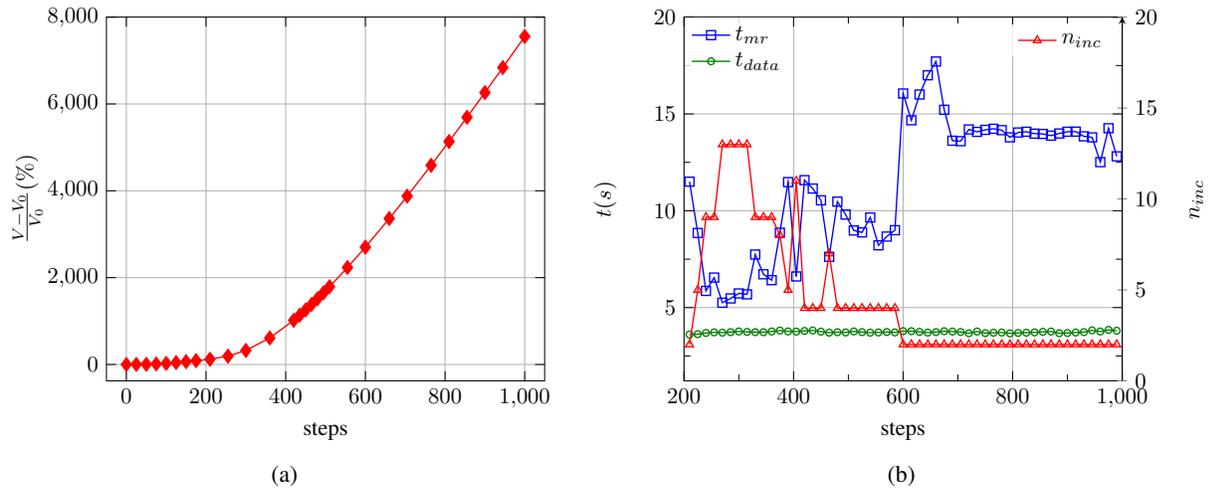
\begin{figure}[htbp] 
\hspace{-30pt}
    \begin{subfigure}[b]{0.55\textwidth}
        \centering 
\begin{tikzpicture}[scale=0.85]

\definecolor{darkgray176}{RGB}{176,176,176}

\begin{axis}[
xlabel={steps},
ylabel={$\frac{V-V_0}{V_0}(\%)$},
tick pos=both,
x grid style={darkgray176},
xmin=-50, xmax=1050,
xtick style={color=black},
y grid style={darkgray176},
ymin=-377.652164621685, ymax=8000.0,
ymajorgrids,
xmajorgrids,
ytick style={color=black}
]
\addplot [semithick, red, mark=diamond*, mark size=3, mark options={solid}]
table {%
0 0
25 1.19478034778181
50 4.95903109037181
75 12.2490896908826
100 24.9708550085234
125 43.5049071490958
150 65.0705312053657
175 87.8608718087286
210 121.233399834021
255 191.845006750095
300 322.761001910065
360 610.615087750952
420 1017.12888438897
435 1134.14182091833
450 1256.47746417478
465 1383.7460438606
480 1515.60403924056
495 1651.72280432822
510 1791.81575135771
555 2233.24318943215
600 2701.87364073671
660 3360.96560865646
705 3876.51475214796
765 4587.05969759748
810 5134.77655876619
855 5693.3813968146
900 6261.50361803684
945 6838.10689706966
1000 7553.04329243371
};
\end{axis}

\end{tikzpicture}
        \caption{}\label{fig:numerical_exmaple_1_volume_time}
    \end{subfigure}
\hspace{-30pt}
    \begin{subfigure}[b]{0.55\textwidth}
        \centering 
\begin{tikzpicture}[scale=0.85]

\definecolor{darkgray176}{RGB}{176,176,176}
\definecolor{green01270}{RGB}{0,127,0}

\begin{axis}[
legend cell align={left},
legend style={fill opacity=0.8, draw opacity=1, text opacity=1, draw=none,at={(0,1)},
  anchor=north west},
tick pos=both,
x grid style={darkgray176},
xmajorgrids,
ylabel={$t(s)$},
xlabel={steps},
xmin=200, xmax=1000,
xtick style={color=black},
y grid style={darkgray176},
minor tick num=1,
minor grid style={black},
ymajorgrids,
ymin=1.214582, ymax=20,
ytick style={color=black}
]
\addplot [semithick, blue,mark=square, mark size=2, mark options={solid}]
table {%
210 11.50311
225 8.85215
240 5.86099666666667
255 6.54952777777778
270 5.25157461538462
285 5.47279615384615
300 5.73020692307692
315 5.68361384615385
330 7.73806444444444
345 6.72051888888889
360 6.41601333333333
375 8.866545
390 11.47862
405 6.60763909090909
420 11.58104
435 11.14677
450 10.539085
465 7.62052142857143
480 10.4721
495 9.80292
510 8.98126
525 8.88248
540 9.65458
555 8.214535
570 8.66713
585 8.99723
600 16.05536
615 14.6718
630 16.00306
645 16.99636
660 17.70836
675 15.21529
690 13.61682
705 13.59317
720 14.19785
735 14.07514
750 14.17724
765 14.23402
780 14.16182
795 13.79028
810 14.04227
825 14.08086
840 13.98168
855 13.96776
870 13.87945
885 13.99051
900 14.08072
915 14.08706
930 13.84357
945 13.7894
960 12.51045
975 14.26101
990 12.81071
};
\addlegendentry{${t}_{mr}$}
\addplot [semithick, green01270, mark=o, mark size=1.5, mark options={solid}]
table {%
210 3.60599
225 3.62605
240 3.68767
255 3.71985
270 3.70481
285 3.73105
300 3.76487
315 3.74254
330 3.72938
345 3.72717
360 3.76292
375 3.81508
390 3.76632
405 3.75827
420 3.79276
435 3.81703
450 3.75609
465 3.70555
480 3.7268
495 3.71663
510 3.76719
525 3.73147
540 3.70357
555 3.71644
570 3.73737
585 3.71637
600 3.78219
615 3.779
630 3.74349
645 3.70759
660 3.73389
675 3.77916
690 3.75943
705 3.73858
720 3.6725
735 3.76341
750 3.67916
765 3.69953
780 3.71188
795 3.66617
810 3.69013
825 3.70199
840 3.71752
855 3.74714
870 3.761
885 3.67349
900 3.68688
915 3.71114
930 3.73198
945 3.8232
960 3.76125
975 3.84609
990 3.80514
};
\addlegendentry{${t}_{data}$}
\end{axis}

\begin{axis}[
    legend cell align={left},
    legend style={fill opacity=0.8, draw opacity=1, text opacity=1,draw=none},
    axis x line=none,
    axis y line=right,
    ylabel={$n_{inc}$},
    xmin=200, xmax=1000,
    ymin=0, ymax=20,
]
\addplot [semithick, red, mark=triangle, mark size=2, mark options={solid}]
table {%
210 2
225 5
240 9
255 9
270 13
285 13
300 13
315 13
330 9
345 9
360 9
375 8
390 5
405 11
420 4
435 4
450 4
465 7
480 4
495 4
510 4
525 4
540 4
555 4
570 4
585 4
600 2
615 2
630 2
645 2
660 2
675 2
690 2
705 2
720 2
735 2
750 2
765 2
780 2
795 2
810 2
825 2
840 2
855 2
870 2
885 2
900 2
915 2
930 2
945 2
960 2
975 2
990 2
};
\addlegendentry{$n_{inc}$}
\end{axis}

\end{tikzpicture}
        \caption{}\label{fig:numerical_exmaple_1_computation_time}
    \end{subfigure}
    \caption{Example 1: (a) percentage increase in total volume ($\frac{V-V_0}{V_0}$) vs. step. (b) computation time viz, time for mesh refitting step $t_{mr}$ per incrementation steps $n_{inc}$ and data mapping $t_{data}$ time for each mesh adaptation step.}
\end{figure}
\begin{figure}[htbp]
        \centering
        \input{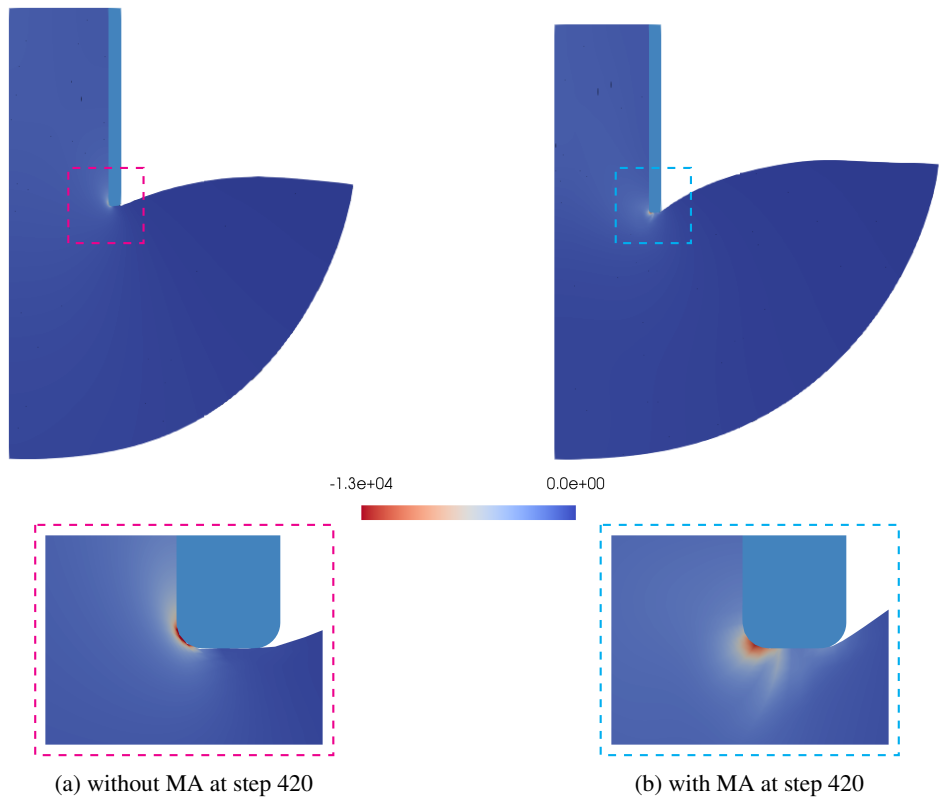}
        \caption{Example 1: Comparison of first principal Cauchy stress (maximum compressive) at step 420. Subfigures (a) without MA and (b) with MA.}\label{fig:numerical_exmaple_1_stress_comp}
\end{figure}  
\begin{figure}[htbp]
        \centering
        \input{Example_1_stress_plots}
        \caption{Example 1: von Mises equivalent stress and first principal Cauchy stress (maximum compressesive) at steps 600 and 990.}\label{fig:numerical_exmaple_1_stress}
\end{figure}

First, the results without adaption are investigated. As expected, once the expanded material passes the corner, the mesh quality reduces (see~\cref{fig:Example_1_mesh_at_corner_woma_300,fig:Example_1_mesh_at_corner_woma_350,fig:Example_1_mesh_at_corner_woma_400}). The element skewness around the corner $\bm{X}_c$ in the radius $0.1$ is plotted over time in~\cref{fig:numerical_exmaple_1_skewness}. The element skewness is computed as
\begin{align}\label{skewness_def}
    skewness = \text{max}\Big( \frac{\theta_{max} - 90}{180-90},  \frac{ 90-\theta_{min}}{90} \Big) \quad \in[0,1],
\end{align}
where $\theta_{max}$ and $\theta_{min}$ are the maximum and minimum included angle (in degree) between the edges. For a cuboid shape $\theta_{max}=\theta_{min}=90^{\circ}$ and the skewness is zero (optimal element quality). In contrast, when $\theta_{max,min}\to 0$, corresponding to a very skewed element, the skewness value is 1 (worst element quality). From~\cref{fig:numerical_exmaple_1_skewness} it can be seen that the skewness increases rapidly, starting from time step $200$ to $425$. After step $250$, the maximum skewness is greater than $0.6$, which may affect the accuracy of the solution. At time step $425$ (see~\cref{fig:example_1_woma_step425}), the elements are heavily distorted (skewness $\approx1$) such that computation can no longer be continued. However, a volume increase of about $863\%$ is achieved at this step.

Next, the results including the proposed mesh adaptation scheme are studied. The mesh around the corner is portrayed in~\cref{fig:Example_1_mesh_at_corner_wma_300,fig:Example_1_mesh_at_corner_wma_350,fig:Example_1_mesh_at_corner_wma_400,fig:example_1_wma_step425}. Compared to the results without mesh adaptation, a mesh of higher quality is maintained around the corner ($\bm{X}_c$) during the expansion. The skewness near the corner is greatly reduced throughout the simulation (see~\cref{fig:numerical_exmaple_1_skewness}). A slight increase in skewness can be attributed to the extreme volume expansion and shape change which can not be completely avoided. \cref{fig:example_1_wma_step425} shows the expanded state at step $425$, which has exhibits a significantly improved mesh quality as compared to the simulation without mesh adaptation (see~\cref{fig:example_1_woma_step425}). Moreover, further states at steps $800$ and $1000$, that could only be produced/reached when the mesh adaptation is activated, are portrayed in~\cref{fig:example_1_step800,fig:example_1_step1000}. At the end of step $1000$, there is a volume increase of approximately $7553\%$. The increase in volume over time steps is depicted in~\cref{fig:numerical_exmaple_1_volume_time}. The mesh quality at step 1000 with a volume expansion of $7553\%$ is still significantly better as compared to the simulation without MA at step 425 (at which the volume increase is only $863\%$). It has been tested that the expansion process could even be carried out further, which confirms the robustness of the overall numerical solution scheme.

To get insights into the physical behavior of the investigated material, the first principal Cauchy stress is plotted. The Cauchy stress is obtained from interpolated second Piola-Kirchhoff stress tensor (see~\cref{sec:problem_material}). The first principal Cauchy stress at step 420 resulting from simulations with and without MA is plotted in~\cref{fig:numerical_exmaple_1_stress_comp}. At step 420, mesh resulting from a simulation without MA is heavily distorted (c.f~\cref{fig:example_1_woma_step425}), leading to a rather unphysical stress distribution, namely very high peak stresses within distorted finite elements in boundary layer, which abruptly drop to significantly smaller stress values in next finite element layers (see~\cref{fig:Example_1_pp_420_woma}). In contrast, simulation with MA leads to physically more reasonable, i.e., smoother, stress distributions, but still with highest stress values occurring in boundary region near sharp corner at $\bm{X}_c$ (see~\cref{fig:Example_1_pp_420_wma}). Furthermore, first principal Cauchy stress at steps 600 and 990, plotted in~\cref{fig:Example_1_pp_600,fig:Example_1_pp_990}, shows a similar trend. As a result, von Mises equivalent stress exhibits its peak values in this boundary region, as depicted in~\cref{fig:Example_1_vm_600,fig:Example_1_vm_990}.

The computation time for the mesh adaptation is plotted in~\cref{fig:numerical_exmaple_1_computation_time}. The figure portrays the time for the mesh refitting $t_{mr}$ per incrementation step $n_{inc}$ and the data mapping time $t_{data}$ for every mesh adaptation step. Furthermore, the total time for the mesh adaptation step can be computed as $t_{mr}n_{inc}+t_{data}$. Time $t_{mr}$ is the total time spent for the MR including all the necessary setup. For $n_{inc}>1$, also the time spent on the unconverged Newton-Raphson iterations is included. On average $t_{mr}$ is $13s$. The data transfer time $t_{data}$ is $4s$ and approximately constant. In the initial phase, specifically between steps 200 to 400 (see~\cref{fig:numerical_exmaple_1_volume_time}) the number of incrementation steps range between 5 to 13. However, as we progress, the incrementation requirement decreases significantly. From steps 600 to 1000, only two steps are needed for each mesh adaption. During the simulation a total of 52 MA steps were performed with an average computational time per mesh adaptation step of $56s$. The total simulation time is $10,960s$, i.e.,  the mesh adaptation accounts for a share of approximately $27\%$.

\subsection{Expansion past a rigid obstruction and a deformable body}
\begin{figure}[htbp]
    \centering
    \begin{minipage}{0.4\textwidth}
        \centering
        {\begin{tikzpicture}[scale=0.75]
	\def\ax{1};
	\def\lx{3};
	\def\nx{3};

	\def\xshift{0}
	\def\yshift{3}
	\def\thic{1.25}
	\def\rad{0.35}
	\coordinate [](origin) at (\xshift,\yshift);
	\begin{scope}
		\fill [black!30] (origin) rectangle (\lx+\xshift,\lx+\yshift);

		\tikzset{d/.style={minimum width=4pt,inner sep=0pt,circle,fill=black}}

		\coordinate [] (1) at (origin) {};
		\coordinate [] (2) at (\lx+\xshift,\yshift) {};
		\coordinate [] (3) at (\lx+\xshift,\lx+\yshift) {};
		\coordinate [] (4) at (\xshift, \lx+\yshift) {};


		\foreach \i in {0,...,\nx}
		\draw [fill=black] ([xshift=-4pt]\xshift,0.75*\i+\yshift+0.35) circle [radius=3pt];
		\foreach \i in {0,...,\nx}
		\draw [fill=black] (0.75*\i+\xshift+0.35,0.125+\lx+\yshift) circle [radius=3pt];

		\draw[]([yshift=7pt,xshift=-7pt]4) -- ([yshift=7pt]3);
		\path[pattern=north west lines, shift={(3cm,3cm)}] ([yshift=7pt,xshift=-7pt]4) rectangle ([yshift=12pt]3);


		\draw[]([yshift=7pt,xshift=-7pt]4) -- ([xshift=-7pt,yshift=-150pt]1);
		\path[pattern=north west lines, shift={(3cm,3cm)}] ([yshift=7pt,xshift=-7pt]4) rectangle ([xshift=-12pt,yshift=-150pt]1);

		\draw[] (1) -- (2);
		\draw[] (2) -- (3);
		\draw[] (3) -- (4);
		\draw[] (4) -- (1);

		\coordinate [] (7) at (\lx+\xshift,0+\rad) {};
		\coordinate [] (8) at (\lx+\xshift+\thic,0+\rad) {};
		\coordinate [] (9) at (\lx+\xshift,\lx+\yshift) {};
		\coordinate [] (10) at (\lx+\xshift+\thic,\lx+\yshift) {};

		\coordinate [] (20) at (\lx+\xshift+\rad,0+\rad) {};
		\coordinate [] (21) at (\lx+\xshift+\thic-\rad,0+\rad) {};

		\coordinate [] (11) at (\lx+\xshift+\rad,0) {};
		\coordinate [] (12) at (\lx+\xshift+\thic-\rad,0) {};

		\coordinate [] (96) at (\lx+\xshift,0) {};
		\coordinate [] (97) at  (\lx+\xshift+\thic,0) {};
		\draw[rounded corners=8pt,fill=blue!30] (96) -- (97) -- (10) -- (9) -- cycle;
		\draw[fill=blue!30,blue!30] (9) rectangle ([xshift=0pt,yshift=-30pt]10);

		\draw[thick] (7) -- (9);
		\draw[thick] (9) -- (10);
		\draw[thick] (10) -- (8);
		\draw[thick] (7) arc  (0:90:-\rad);
		\draw[thick] (12) arc (90:180:-\rad);
		\draw[thick] (11) -- (12);


		\coordinate [] (22) at (\lx+\xshift-\rad+\thic,\lx+\yshift) {};
		\coordinate [] (23) at (\lx+\xshift-\rad+\thic,0) {};
		\dimline[color=black]{([shift={(10mm,0mm)}]22)}{([shift={(10mm,0mm)}]23)}{\rotatebox{90}{$h_o$}};

		\coordinate [] (24) at (\lx+\xshift,0) {};
		\coordinate [] (25) at (\lx+\xshift+\thic,0) {};

		\draw[fill=green!30] (0,0) rectangle (4*\lx,-\thic);
		\coordinate [] (26) at (4*\lx,0) {};
		\coordinate [] (27) at (4*\lx,-\thic) {};
		\path[pattern=north west lines, shift={(3cm,3cm)}] ([yshift=0pt,xshift=7pt]26) rectangle ([yshift=0pt]27);

		\foreach \i in {0,...,1}
		\draw [fill=black] ([xshift=-4pt]\xshift,-0.75*\i-0.25) circle [radius=3pt];

		\coordinate [] (28) at (0,-\thic) {};
		\dimline[color=black]{([shift={(0mm,-5mm)}]27)}{([shift={(0mm,-5mm)}]28)}{$2h_0$};

		\draw[-latex,color=blue] (0,0) -- (\ax,0) node[anchor=west]{$x$};
		\draw[-latex,color=blue] (0,0) -- (0,\ax) node[anchor=west]{$y$};
		

	    \draw[shorten <=2pt,very thin] (2,0) to[out=90,in=-90] ++ (-110:-0.8) node[above]{$\Gamma^{(1)}_c$};
	    
      \draw[shorten <=2pt,very thin] ([xshift=-40pt]2) to[out=-90,in=90] ++ (-110:0.8) node[below]{$\Gamma^{(2)}_c$};
	\end{scope}

\end{tikzpicture}}
        \caption{ Example 2: Problem setup with a grey body denoting the material in expansion. The blue and green bodies represent the rigid body, and the deformable body, respectively.}\label{fig:numerical_exmaple_1a_problem_dimension}
        \label{fig:numerical_example_1a_problem_dimension}
    \end{minipage}\hfill
    \begin{minipage}{0.4\textwidth}
        \centering
        \captionof{table}{Material properties of the deformable body}
        \label{material_parameters_deformable}
        \begin{tabular}[t]{cc}
            \hline
            Parameter & Value\\ 
            \hline
            Young's modulus ($E$) &$1\times10^4$\\
            Poisson's ratio ($\nu$) & $0.3$ \\
            Thermal conductivity ($k_0$) &$7.55\times 10^6$\\
            Heat capacity ($c_v$)  & $1.4\times10^{-2}$\\ 
            \hline
        \end{tabular}
    \end{minipage}
\end{figure}
\begin{figure}[htbp]
    \input{Example_1a_mesh.tex}
    \caption{Example 2: (a) and (c) show the mesh without mesh adaptation at steps 300 and 400, respectively; (b) and (d) show the corresponding meshes with mesh adaptation. (e) and (f) mesh at steps 450 and 500 with mesh adaptation, respectively.}\label{fig:example_2_mesh}
\end{figure}
\begin{figure}[htbp]
    \centering
\begin{tikzpicture}

\definecolor{darkgray176}{RGB}{176,176,176}
\definecolor{green01270}{RGB}{0,127,0}

\begin{axis}[
legend cell align={left},
legend style={fill opacity=0.8, draw opacity=1, text opacity=1, draw=none,at={(0,1)},
  anchor=north west},
tick pos=both,
x grid style={darkgray176},
xmajorgrids,
ylabel={$t\ (s)$},
xlabel={steps},
xmin=200, xmax=500,
xtick style={color=black},
y grid style={darkgray176},
ymajorgrids,
ymin=0, ymax=20,
ytick style={color=black},
]
\addplot [semithick, blue,mark=square, mark size=2, mark options={solid}]
table {%
200 7.21677
205 7.36028
210 9.7403
215 11.92769
220 14.14755
225 11.50223
230 10.417
235 10.51491
240 10.42378
245 10.49418
250 10.57209
255 10.74734
260 11.55187
265 11.59886
270 11.43722
275 11.4504
280 11.48196
285 12.64229
290 12.57934
295 12.64816
300 12.96995
305 13.11187
310 12.8501
315 12.88087
320 13.08912
325 13.08719
330 13.00632
335 13.86619
340 19.00029
345 19.06078
350 12.90479
355 16.68418
360 14.55306
365 12.20418
370 12.27107
375 12.15579
380 9.98029
385 10.01338
390 10.02654
395 12.26218
400 12.6461
405 12.55357
410 12.34648
415 12.38044
420 10.22837
425 10.27803
430 10.3043
435 10.52024
440 10.76494
445 10.76396
450 10.92011
455 10.68901
460 10.80605
465 11.0461
470 11.75079
475 11.15595
480 11.21183
485 11.01975
490 10.9833
495 10.98375
500 11.17406
};
\addlegendentry{${t}_{mr}$}
\addplot [semithick, green01270, mark=o, mark size=1.5, mark options={solid}]
table {%
200 2.79113
205 2.79072
210 2.8486
215 2.86591
220 2.86735
225 2.87952
230 2.85325
235 2.99634
240 2.84957
245 2.85712
250 2.87041
255 2.90646
260 2.86113
265 2.87294
270 2.96763
275 2.9136
280 2.88784
285 2.86731
290 2.95451
295 2.92974
300 2.90675
305 2.86818
310 2.90785
315 2.98268
320 2.86768
325 2.92471
330 2.85518
335 2.88726
340 2.85431
345 2.89162
350 2.84376
355 2.92182
360 2.82444
365 2.84472
370 2.88283
375 2.88231
380 2.95431
385 2.87092
390 2.89096
395 2.84692
400 2.8258
405 2.83113
410 2.85132
415 2.87676
420 2.86273
425 2.87187
430 2.8737
435 2.88656
440 2.87966
445 2.87984
450 2.83399
455 2.83009
460 2.85245
465 2.8757
470 2.85841
475 2.88295
480 2.85867
485 2.90725
490 2.8658
495 2.84055
500 2.87564
};
\addlegendentry{${t}_{data}$}
\end{axis}

\begin{axis}[
    legend cell align={left},
    legend style={fill opacity=0.8, draw opacity=1, text opacity=1,draw=none},
    axis x line=none,
    axis y line=right,
    ylabel={$n_{inc}$},
    xmin=200, xmax=500,
    ymin=0, ymax=20,
]
\addplot [semithick, red, mark=triangle, mark size=2, mark options={solid}]
table {%
200 1
205 1
210 1
215 1
220 1
225 2
230 2
235 2
240 2
245 2
250 2
255 2
260 2
265 2
270 2
275 2
280 2
285 2
290 2
295 2
300 2
305 2
310 2
315 2
320 2
325 2
330 2
335 2
340 1
345 1
350 2
355 1
360 1
365 1
370 1
375 1
380 1
385 1
390 1
395 1
400 1
405 1
410 1
415 1
420 1
425 1
430 1
435 1
440 1
445 1
450 1
455 1
460 1
465 1
470 1
475 1
480 1
485 1
490 1
495 1
500 1
};
\addlegendentry{$n_{inc}$}
\end{axis}
\end{tikzpicture}
     \caption{Example 2: 
        Computation time for mesh refitting step $t_{mr}$ per incrementation steps $n_{inc}$ and data mapping $t_{data}$ time for each mesh adaptation step.}\label{fig:example_2_computation_time}
\end{figure}

This example is an extension of the previous example, and the problem setup is illustrated in~\cref{fig:numerical_exmaple_1a_problem_dimension}. The dimensions, material properties, boundary conditions, and discretization of the expanding and rigid body are the same as in the previous example (see~\cref{fig:numerical_exmaple_1_problem_dimension},~\cref{material_parameters}). The length of the deformable body (green) is twice the length of the rigid body, has the same thickness as the rigid body, and boundary conditions as illustrated in~\cref{fig:numerical_exmaple_1a_problem_dimension}. The material of the deformable body is of Neo-Hookean type with parameters as given in~\cref{material_parameters_deformable}. For the thermomechanical mortar contact between the expanding and deformable body, the boundary of the expanding body is chosen as the master surface $\Gamma^{(2)}_c$ and the boundary of the deformable body as the slave surface $\Gamma^{(1)}_c$ (see~\cref{fig:numerical_exmaple_1a_problem_dimension}). Furthermore, the penalty parameter for this contact pair is set to $1\times10^7$, and the interface conductivity is $\beta_c=0$, i.e., adiabatic contact. The deformable body is discretized with 420 linear hexahedral elements. The thermo-mechanical problem is analyzed for $500$ time steps with a step size of $\Delta t=0.05$. The nonlinear system of equations resulting in each time step is solved using a Newton-Raphson scheme with a tolerance on the combined residual and increment of $10^{-8}$. Again, the linearized monolithic thermomechanical system is solved using the iterative GMRES method with AMG(BGS) preconditioner with a tolerance of $10^{-10}$.

The mesh adaptation problem is formulated as follows: Like in the previous example, to achieve a high mesh quality around the corner $\bm{X}_c$, we employed a mesh localization according to~\cref{mesh_localization,spatial_exp_average_length} with $c=100$. The target element edge length ($l^i_{\text{r0}}$) in a time step are estimated as in~\cref{average_lengths}. The mesh refitting parameters are the same as for the previous example and are listed in~\cref{meshapatation_parameters}. The convergence tolerance for the residual and displacement increment is chosen to $10^{-5}$ and the linearized system is solved with the "SuperLU" solver. Again, the data transfer parameters remain the same as in the previous example. The mesh adaptation is carried out every $5^{\text{th}}$ step starting from step 200. Finally, the computation is carried out on 3 nodes (72 CPUs) of a computing cluster with Intel Xeon E5-2680v3 2.5GHz processors.

The mesh resulting from a simulation of the expansion process without and with mesh adaptation is depicted for the time steps 300 and 400 in~\cref{fig:example_2_mesh_a,fig:example_2_mesh_b,fig:example_2_mesh_c,fig:example_2_mesh_d}, respectively. The mesh near the corner point $\bm{X}_c$ (not plotted) shows the same trend as in the previous example (see~\cref{fig:Example_1_mesh_at_corner}). The simulation without mesh adaptation in this example fails at step 400 due to a heavily distorted mesh near $\bm{X}_c$, leading to non-convergence of the Newton-Raphson scheme. The deformed states resulting from the simulation with mesh adaptation at time steps 450 and 500 are depicted in~\cref{fig:example_2_mesh_e,fig:example_2_mesh_f}, respectively. Also in this example, the proposed MA approach results in a significantly improved mesh quality compared to simulations without MA. The additional presence of the deformable body in this second example is motivated by a practical application scenario within our broader research interests, considering seals for such expandable foams. It is demonstrated that even for this highly challenging scenario, i.e., extreme volume expansion of a foam squeezed through the gap between a rigid and a deformable body including mutual thermo-mechanical contact interaction, the proposed MA approach allows for a high mesh quality and robust simulations.

Finally, the computational costs for the mesh adaptation procedure are presented in~\cref{fig:example_2_computation_time}. In contrast to the previous example, less incrementation steps ($\leq 2$) are necessary as the mesh adaption frequency is higher. The data transfer time is $\leq 3s$ and $t_{mr}$ is $15s$ on average. Finally, a total of 60 MA steps are performed during the simulation with an average computation time of $20s$ spent for MA. The total simulation time is $5,621s$ for this example, i.e., the mesh adaptation accounts for a share of $\approx20\%$.

\subsection{Inward expansion of a hollow cylinder past a rigid obstruction}
\begin{figure}[htbp]
     \input{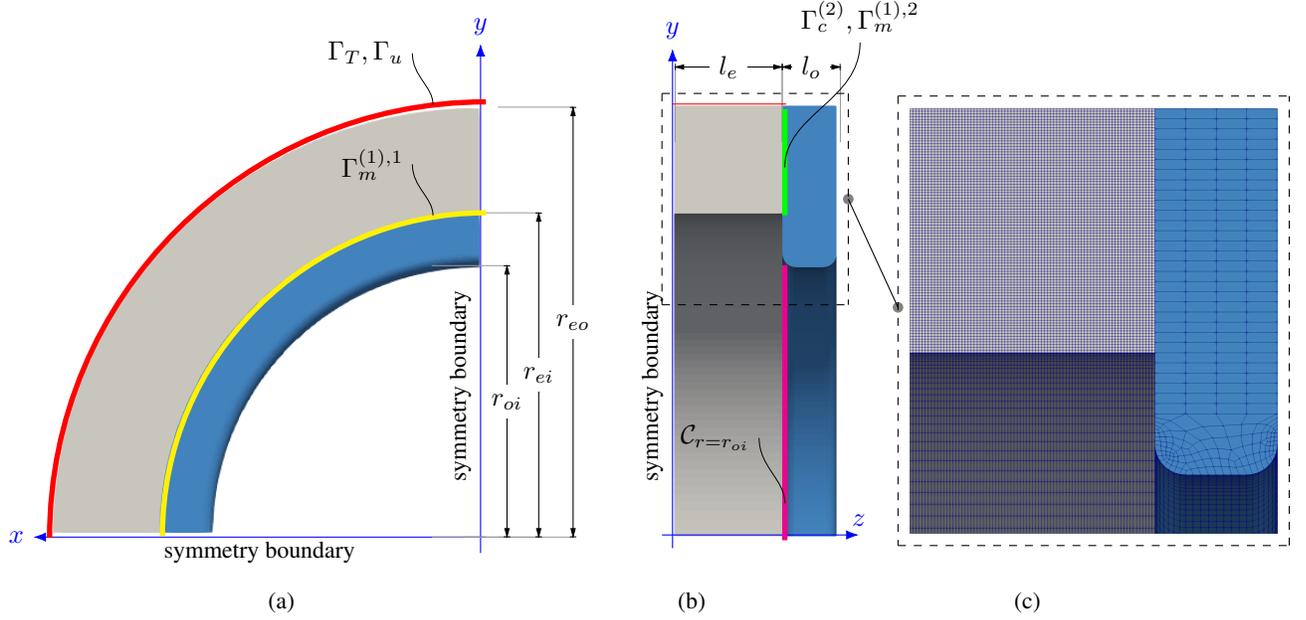}
    \caption{ Example 3: expanding material (gray), obstruction (blue), and auxiliary boundary (green) (a) +z-plane view (b) +x-plane view (same as +y-plane due to symmetry). (c) mesh at reference configuration. The temperature and displacement surface Dirichlet boundaries ($\Gamma_T,\Gamma_u$) are represented by a red line. Furthermore, all other boundaries of expanding material are modeled adiabatic. The master side ($\Gamma^{(2)}_c$) of the thermo-mechanical contact is denoted by green line. The mesh sliding surfaces ($\Gamma^{(1)}_m$) for the mesh adaptation are represented by green and yellow lines. Finally, magenta line represent a curve $\mathcal{C}_{r=r_{oi}}$.
    }\label{fig:exmaple_3_problem_setup}
\end{figure}

\newcounter{mycounter2}
\setcounter{mycounter2}{1}
\renewcommand{\thefigure}{\arabic{figure}.\Alph{mycounter2}}
\begin{figure}[htbp]
    \input{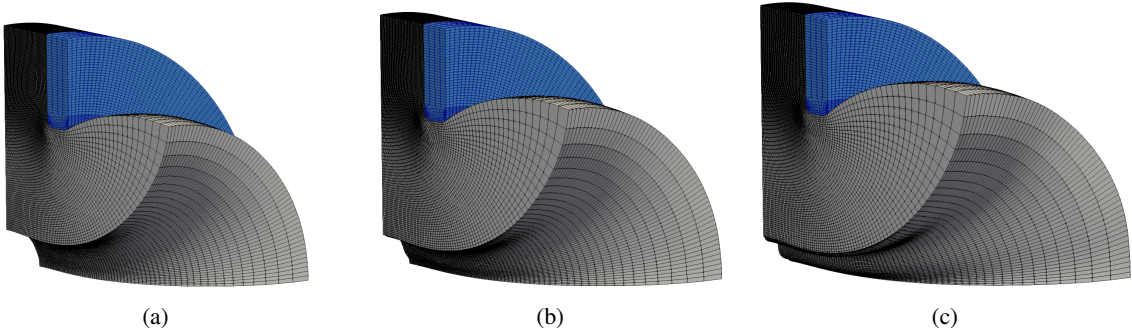}
    \caption{Example 3: deformed state at step (a) 360 (b) 400 (c) 428.}
    \label{fig:numerical_exmaple_2_result_p1}
\end{figure}
\addtocounter{mycounter2}{+1}
\addtocounter{figure}{-1}
\begin{figure}[htbp]
    \centering
    \input{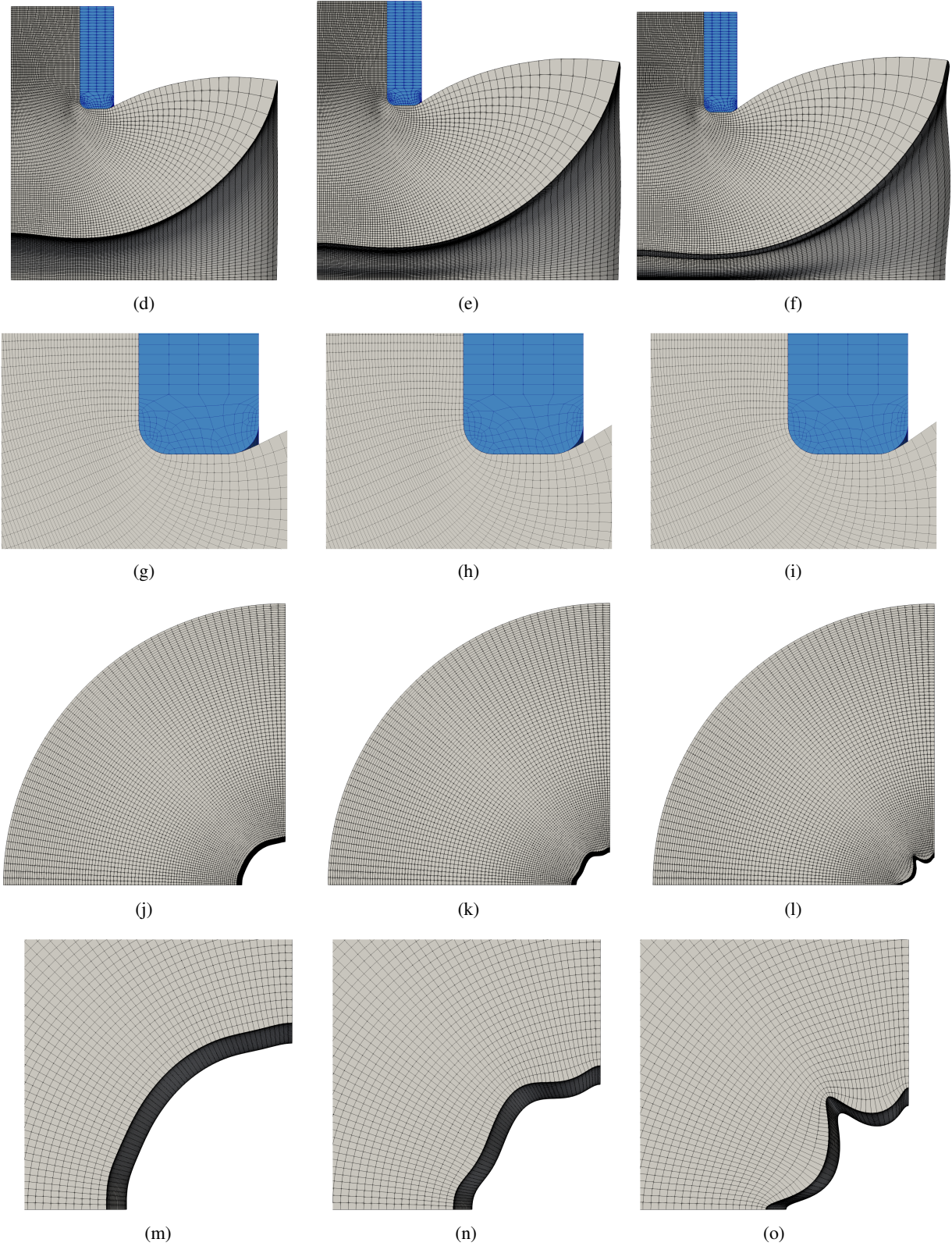}
    \caption{Example 3: deformed states at steps 360, 400, and 428 are illustrated in columns 1, 2, and 3, respectively. Row 1: +x-plane view, Row 2: +x-plane detailed view around the fillet, Row 3: +z-plane view, and Row 4: +z-plane detailed view at the center (c.f.~\cref{fig:numerical_exmaple_2_problem_geo,fig:numerical_exmaple_2_problem_geo_side}).}
    \label{fig:numerical_exmaple_2_result_p3}
\end{figure}
\renewcommand{\thefigure}{\arabic{figure}}
\begin{figure}[htbp]
    \centering
    \input{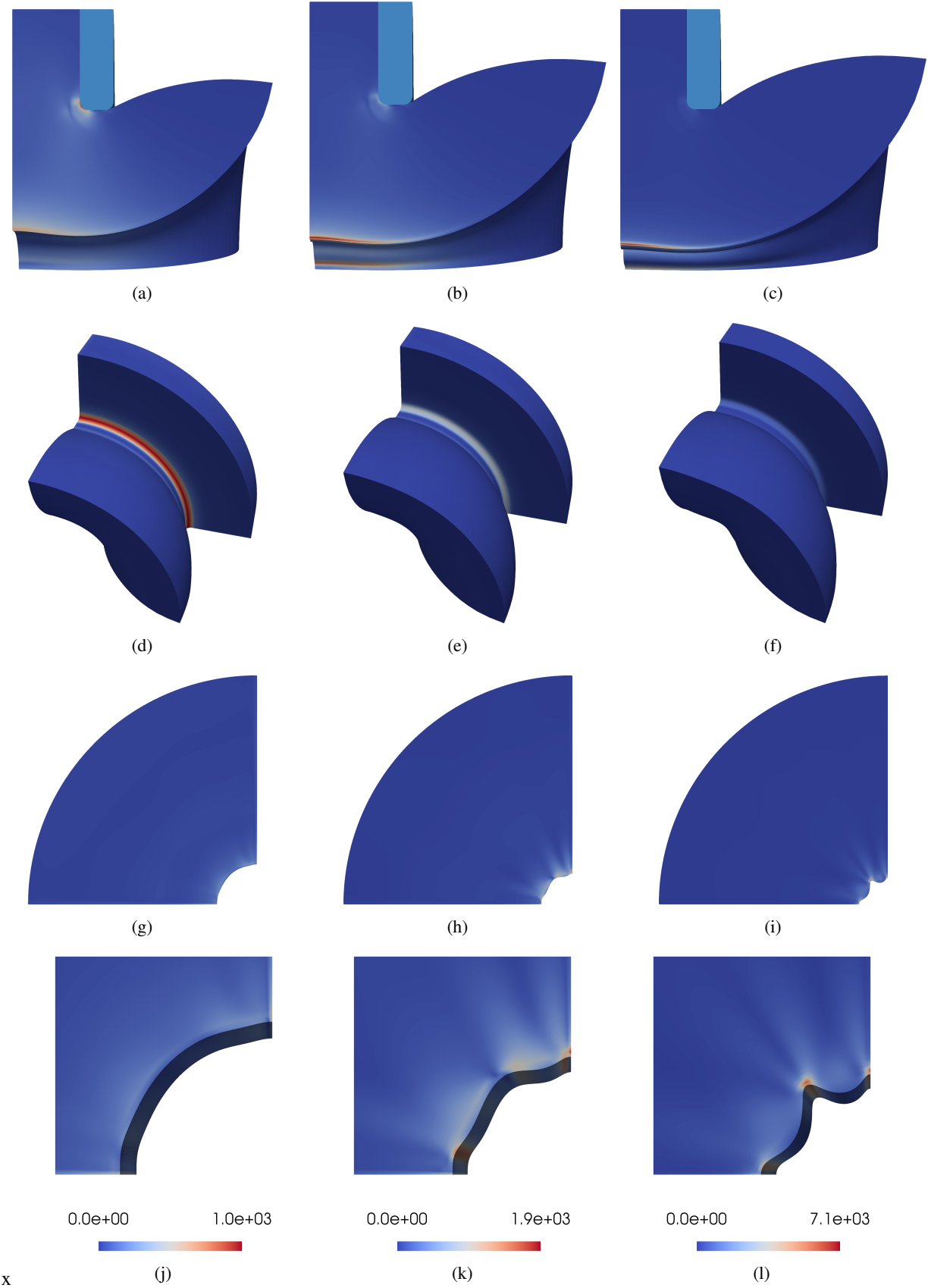}
    \caption{Example 3: von Mises equivalent stress at steps 360, 400, and 425 are illustrated in column 1, 2, and 3, respectively. Row 1: +x-plane view, Row 2: 3D view, Row 3: +z-plane view, and , Row 4: +z-plane detailed view at the center (c.f.~\cref{fig:numerical_exmaple_2_problem_geo,fig:numerical_exmaple_2_problem_geo_side}).}
    \label{fig:numerical_exmaple_3_vonmises}
\end{figure}

Next, the proposed MA approach shall be applied to a complex 3D problem. Thereto, the inward expansion of a hollow cylinder past a rigid obstruction is investigated. Consider the problem setup depicted in~\cref{fig:exmaple_3_problem_setup}. Due to symmetry, only a quarter portion of the system is simulated. The hollow quarter cylinder has an outer radius of $r_{eo} = 4$, an inner radius of $r_{ei} = 3$, and a length of $l_e=1$. The rigid body has the same outer radius as the hollow cylinder's, an inner radius of $r_{oi}=2.5$, and a length of $l_o=0.5$. The corner radius of the rigid obstruction has the same dimensions as in Example 1 (see~\cref{{fig:numerical_exmaple_1_problem_dimension}}). Also the material model for the expanding body is the same as in Example 1 (\cref{sec:example_1}). The initial temperature $T_0$ is set to $198$ and a temperature surface Dirichlet boundary condition according to $T= 148+ 345 \log_{10}(1+[(8\times(t+3))/60])$ is applied at the outer surface of the expanding cylinder denoted by $\Gamma_T$. Moreover, all other boundaries of expanding material are modeled adiabatic. Apart from the symmetry boundary conditions (see~\cref{fig:exmaple_3_problem_setup}), for the structural problem also the outer surface of the expanding cylinder (denoted by $\Gamma_u$) is fixed. Both, the expanding and the rigid body are discretized with 8-noded hexahedral elements with $432000$ elements (449631 nodes) and $14700$ elements (18178 nodes), respectively. For the mortar thermo-mechanical contact, the outer surface of the rigid body is chosen as the slave side and the surface of the expanding material as the master side ($\Gamma^{(2)}_c$). The contact interface is discretized with 4-noded quadrilateral elements. Moreover, the penalty parameter is set to $\epsilon_c=10^8$ and the interface conductivity to $\beta_c=0$, i.e., adiabatic. The thermo-mechanical problem is analyzed for $500$ steps with time step size $\Delta t=0.05$ using Newton-Raphson scheme with a convergence tolerance of $10^{-8}$ on the combined residual and increment. Again, the linearized monolithic thermomechanical system is solved using the iterative GMRES method with AMG(BGS) preconditioner with tolerance $10^{-10}$.

The mesh adaptation problem is formulated as follows: To achieve a high mesh quality around the corner edge $\mathcal{C}_{r=r_{oi}}$ (see~\cref{fig:exmaple_3_problem_setup}), we employ a mesh localization according to~\cref{mesh_localization} with ${l}^i_{\text{r0}}$ as in~\cref{average_lengths}. The function $f(\cordX)$ as shown in~\cref{{spatial_exp_average_length}} is reformulated according to 
\begin{align}\label{dist_cylinderical_coords}
    f(\cordX) =1+\exp{ \big( -c\ (r_n^2 + r_{ei}^2 - 2 r_n r_{ei}+(z-l_e)^2) \big)},
\end{align}
with $r_n=x^2+y^2$ and $c=150$. The mesh refitting parameters are the same as in the previous example (see~\cref{meshapatation_parameters}). The mesh sliding surfaces are denoted by $\Gamma^{(1),1}_m$ and $\Gamma^{(1),2}_m$ in~\cref{fig:exmaple_3_problem_setup}. The convergence tolerance of the Newton-Raphson scheme is set to $10^{-5}$. In contrast to previous examples, the linearized system is solved iteratively using the GMRES method with an AMG preconditioner. Furthermore, the convergence tolerance of the linear solver is set to $10^{-10}$. The data transfer for scalars is done using a moving least squares scheme with trilinear shape functions and for tensors using the 'R-MLS' variant with trilinear basis as described in Section~\ref{sec:tensor_interpolation}. The mesh adaptation is carried out every $5^{\text{th}}$ step starting from step 100. The computation is carried out on 9 nodes (216 CPUs) of a computing cluster with Intel Xeon E5-2680v3 2.5GHz processors.

The deformed states at time steps 360, 400, and 428 are depicted in~\cref{fig:numerical_exmaple_2_result_p1,fig:numerical_exmaple_2_result_p3}. In step 428, mechanical instabilities, i.e. local buckling phenomena, are observed in the system, an effect that is particularly challenging with respect to mesh quality. At this step, the expanded material almost closes the annular opening leading to a minimum inner radius of approximately $0.3545$ ($=0.118\ r_{ei}$). The von Mises equivalent stress at steps 360, 400, and 425 is depicted in~\cref{fig:numerical_exmaple_3_vonmises}. Like in the previous example, the equivalent stress is initially higherin the region near the fillet~$\mathcal{C}_{r=r_{oi}}$ (see~\cref{fig:example_2_vm_step_x_1,fig:example_2_vm_step_xyz_1,fig:example_2_vm_step_z_1,fig:example_2_vm_step_z_zo_1}). But once mechanical instabilities start to form, the position of the peak value of the equivalent stress shifts towards the kinks resulting from the buckling as portrayed in
~\cref{fig:example_2_vm_step_x_2,fig:example_2_vm_step_xyz_2,fig:example_2_vm_step_z_2,fig:example_2_vm_step_z_zo_2,fig:example_2_vm_step_x_3,fig:example_2_vm_step_xyz_3,fig:example_2_vm_step_z_3,fig:example_2_vm_step_z_zo_3}.

\begin{figure}[htbp]
\begin{subfigure}[b]{0.45\textwidth}
            \centering
\begin{tikzpicture}[scale=0.85]

\definecolor{darkgray176}{RGB}{176,176,176}

\begin{axis}[
xlabel={steps},
ylabel={$\frac{V-V_0}{V_0}(\%)$},
tick pos=both,
x grid style={darkgray176},
xmin=0, xmax=450,
xtick style={color=black},
y grid style={darkgray176},
ymin=0, ymax=550,
ymajorgrids,
xmajorgrids,
ytick style={color=black}
]
\addplot [semithick, red, mark=diamond*, mark size=3, mark options={solid}]
table {%
0 0
25 1.58282151045263
50 6.57087763422902
75 16.260835213811
100 29.5278664240146
125 43.9819199636276
150 61.4732158325772
175 83.6953056778997
200 110.88255397634
225 141.809689481327
250 176.417711301181
275 214.623596360336
300 256.267887989086
325 301.245183566872
340 329.821647643321
360 369.786231737945
380 411.906983706999
400 456.241051801627
420 502.974787630554
428 525.002582227488
};
\end{axis}

\end{tikzpicture}
         \caption{}\label{fig:exmaple_3_volume_time}
\end{subfigure}
\begin{subfigure}[b]{0.45\textwidth}
            \centering
\begin{tikzpicture}[scale=0.85]

\definecolor{darkgray176}{RGB}{176,176,176}
\definecolor{goldenrod1911910}{RGB}{191,191,0}
\definecolor{green01270}{RGB}{0,127,0}
\begin{axis}[
legend cell align={right},
legend style={
  fill opacity=0.2,
  draw opacity=1,
  text opacity=1,
  draw=white
},
xmajorgrids,
ymajorgrids,
tick pos=both,
x grid style={darkgray176},
xmin=100, xmax=425,
xtick style={color=black},
ylabel={skewness},
xlabel={steps},
y grid style={darkgray176},
ymin=-0.0, ymax=0.15,
yticklabel={\pgfmathprintnumber[fixed,precision=3]{\tick}},
ytick style={color=black}
] 
\addplot[semithick, green01270, mark=square, mark size=1.5, mark options={solid}]
table {%
105 0.0134357180734158
110 0.0263598264034245
115 0.0431534704952111
120 0.0760984572578558
125 0.109707071495478
130 0.126443727434216
135 0.115399912629802
140 0.0987970741228305
145 0.0861649978269172
150 0.0848587311216955
155 0.0931351510517078
160 0.0992879540264902
165 0.102404208509452
170 0.106782929705314
175 0.108425077596739
180 0.109714911829983
185 0.110938263130568
190 0.111511542209527
195 0.110519555204176
200 0.110960604938815
205 0.111148099393457
210 0.110900165678347
215 0.111790818627519
220 0.11138858437392
225 0.112677344975423
230 0.113409091112831
235 0.115177633098389
240 0.115433737840179
245 0.114638600013214
250 0.113923453271869
255 0.115425053944672
260 0.116756157087498
265 0.114452232976636
270 0.11422437708693
275 0.1165851991623
280 0.115866470156757
285 0.116478471396513
290 0.116872292730997
295 0.114954106841571
300 0.116886706711828
305 0.114846890265704
310 0.117419589519743
315 0.115299004911545
320 0.11743248786642
325 0.114745770415016
330 0.118020437159999
335 0.116796085578115
340 0.117358796638324
345 0.118062371867599
350 0.114878437107583
355 0.117126947504013
360 0.116756074250421
365 0.114646432465058
370 0.117262787265951
375 0.117158790198072
380 0.115183326342677
385 0.115940345695478
390 0.117561387238757
395 0.116763914597023
400 0.114563549429549
405 0.115119692295734
410 0.11554643828565
415 0.116010779206509
420 0.114466736835986
};
\addlegendentry{max}
\addplot[semithick, goldenrod1911910, mark=square, mark size=1.5, mark options={solid}]
table {%
105 0.0044226995585785
110 0.00563282832047523
115 0.00608095962746028
120 0.00617925357327746
125 0.00550487340834643
130 0.00585029138483545
135 0.0061761035608525
140 0.00569021854199924
145 0.00573712054995023
150 0.00584721578248693
155 0.00596933471546035
160 0.00592175038578723
165 0.00671320589799728
170 0.00666991755369126
175 0.00678257624547456
180 0.00628860241595311
185 0.00631820422778952
190 0.00666143235534741
195 0.00643698198535926
200 0.00623707536427689
205 0.00649724540935921
210 0.00675418626347171
215 0.00673936283219234
220 0.00676368038239635
225 0.00609857688110919
230 0.00674946123994241
235 0.00682804903974746
240 0.00686661458633056
245 0.00654180997630662
250 0.00687605543293775
255 0.00684245063987182
260 0.00717407889150239
265 0.00695056689367577
270 0.00689721352271704
275 0.00687043866656201
280 0.00711732943932601
285 0.00681807018485582
290 0.00700006634722822
295 0.00729535460200436
300 0.00686174106770241
305 0.00702659444207193
310 0.00656177138152761
315 0.00716779341300493
320 0.00717870308827394
325 0.00697623356360187
330 0.00692978301435435
335 0.00698428113151029
340 0.00711575492571891
345 0.00705558149835592
350 0.0071964572245365
355 0.00713454716008341
360 0.00707132026526141
365 0.00709791124629441
370 0.00689256143579308
375 0.00745836363776202
380 0.0071827030333291
385 0.0072810702546145
390 0.00735136812057959
395 0.00706734634296787
400 0.00715466276703233
405 0.00686044369085421
410 0.00727196349687315
415 0.00733268318651691
420 0.00735207395356383
};
\addlegendentry{min}
\end{axis}

\end{tikzpicture}
        \caption{}\label{fig:exmaple_3_skewness}
\end{subfigure}
        \caption{Example 3: (a) percentage change in volume over time. (b) minimum and maximum skweness near the corner.}\label{fig:exmaple_3_skewness_and_volume}
\end{figure}
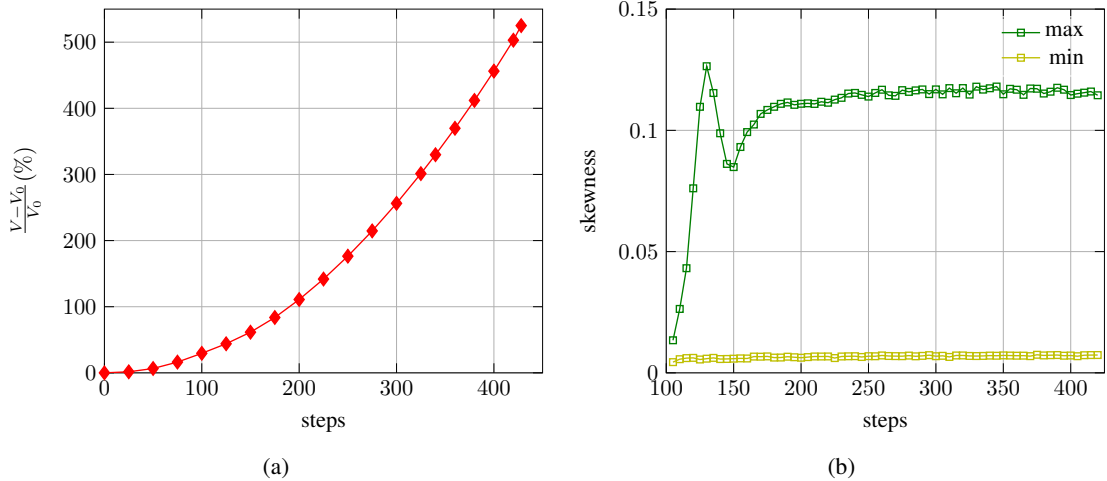
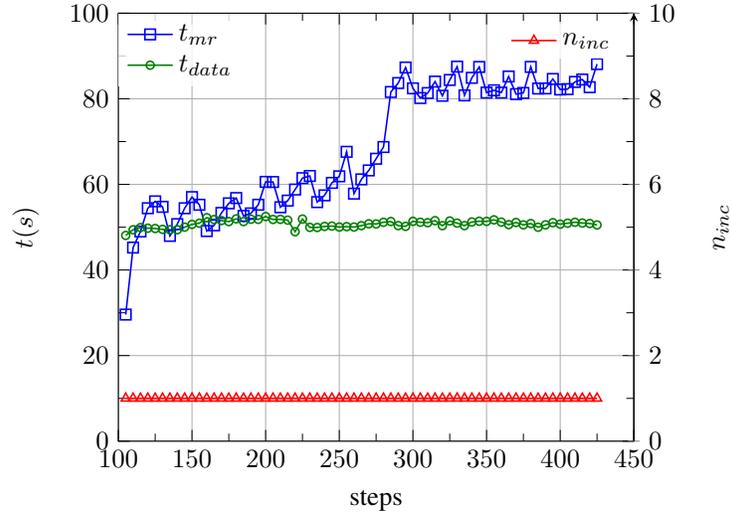
\begin{figure}[htbp]
    \centering
\begin{tikzpicture}

\definecolor{darkgray176}{RGB}{176,176,176}
\definecolor{green01270}{RGB}{0,127,0}

\begin{axis}[
legend cell align={left},
legend style={fill opacity=0.8, draw opacity=1, text opacity=1, draw=none,at={(0,1)},
  anchor=north west},
tick pos=both,
x grid style={darkgray176},
xmajorgrids,
ylabel={$t(s)$},
xlabel={steps},
xmin=100, xmax=450,
xtick style={color=black},
y grid style={darkgray176},
minor tick num=1,
minor grid style={black},
ymajorgrids,
ymin=0.0, ymax=100,
ytick style={color=black}
]
\addplot [semithick, blue,mark=square, mark size=2, mark options={solid}]
table {%
105 29.5796
110 45.2374
115 48.995
120 54.4299
125 56.0358
130 54.753
135 47.968
140 50.7163
145 54.4243
150 57.0336
155 55.2407
160 49.1015
165 50.4322
170 53.3298
175 55.5737
180 56.8137
185 52.6331
190 53.2259
195 55.2429
200 60.587
205 60.5611
210 54.6577
215 56.208
220 58.799
225 61.4608
230 61.9683
235 55.8611
240 57.4375
245 60.3525
250 61.8898
255 67.6138
260 57.8393
265 61.1765
270 63.2606
275 65.9931
280 68.7332
285 81.5942
290 83.6783
295 87.3146
300 82.4427
305 80.1803
310 81.3456
315 84.0234
320 80.7119
325 84.4023
330 87.5136
335 80.7967
340 84.9047
345 87.4447
350 81.4503
355 81.9912
360 81.4476
365 85.2082
370 81.0986
375 81.3707
380 87.4536
385 82.401
390 82.4302
395 84.6324
400 82.1746
405 82.2661
410 83.9434
415 84.5054
420 82.7178
425 88.1131
};
\addlegendentry{${t}_{mr}$}
\addplot [semithick, green01270, mark=o, mark size=1.5, mark options={solid}]
table {%
105 48.074
110 49.4006
115 49.911
120 49.7591
125 49.6932
130 49.491
135 49.4151
140 49.4077
145 50.0047
150 50.6484
155 50.9283
160 52.2225
165 51.7788
170 51.5432
175 51.3083
180 51.9453
185 51.3019
190 51.8511
195 51.8291
200 52.49
205 51.8009
210 51.8273
215 51.645
220 48.942
225 51.9002
230 49.9587
235 49.9319
240 50.1825
245 50.2585
250 50.0302
255 50.1262
260 50.0327
265 50.3575
270 50.7904
275 50.7549
280 51.1438
285 51.3238
290 50.3647
295 50.1934
300 51.3703
305 51.1527
310 51.0514
315 51.5246
320 50.3541
325 51.4537
330 50.9514
335 50.3653
340 51.2143
345 51.3913
350 51.3567
355 51.7118
360 51.1874
365 50.5928
370 51.1034
375 50.5393
380 50.8474
385 49.986
390 50.5428
395 51.0536
400 50.7064
405 50.9289
410 51.1526
415 50.9726
420 50.8592
425 50.5249
};
\addlegendentry{${t}_{data}$}
\end{axis}

\begin{axis}[
    legend cell align={left},
    legend style={fill opacity=0.8, draw opacity=1, text opacity=1,draw=none},
    axis x line=none,
    axis y line=right,
    ylabel={$n_{inc}$},
    xmin=100, xmax=450,
    ymin=0, ymax=10,
]
\addplot [semithick, red, mark=triangle, mark size=2, mark options={solid}]
table {%
105 1
110 1
115 1
120 1
125 1
130 1
135 1
140 1
145 1
150 1
155 1
160 1
165 1
170 1
175 1
180 1
185 1
190 1
195 1
200 1
205 1
210 1
215 1
220 1
225 1
230 1
235 1
240 1
245 1
250 1
255 1
260 1
265 1
270 1
275 1
280 1
285 1
290 1
295 1
300 1
305 1
310 1
315 1
320 1
325 1
330 1
335 1
340 1
345 1
350 1
355 1
360 1
365 1
370 1
375 1
380 1
385 1
390 1
395 1
400 1
405 1
410 1
415 1
420 1
425 1
};
\addlegendentry{$n_{inc}$}
\end{axis}
\end{tikzpicture}
     \caption{Example 3: 
        Computation time viz, time for mesh refitting step $t_{mr}$ per incrementation steps $n_{inc}$ and data mapping $t_{data}$ time for each mesh adaptation step.}\label{fig:example_3_computation_time}
\end{figure}

The volume increase during expansion is showcased in~\cref{fig:exmaple_3_volume_time} with a final volume increase of $525\%$. Again, similar to the previous examples, the quality of the mesh in the vicinity of the rigid body (\cref{fig:example_2_step_x_zo_1,fig:example_2_step_x_zo_2,fig:example_2_step_x_zo_3}) is preserved. The minimum and maximum skewness (see~\cref{skewness_def}) change in the region around the curved edge $\mathcal{C}_{r=r_{oi}}$ over time is depicted in~\cref{fig:exmaple_3_skewness}. To isolate this most critical region around the curved edge for the post-processing, only contributions from elements satisfying the condition $0.8<f(\cordX_e)<1.0$ were considered in~\cref{fig:exmaple_3_skewness}, where $f(\cordX_e)$ is given by~\cref{dist_cylinderical_coords} with $c=1$ and $\cordX_e$ is the element centroid position. It can be seen that the skewness is $<0.15$ in this critical region during the entire expansion process, which indicates a very high mesh quality.

Finally, the computational costs are shown in~\cref{{fig:example_3_computation_time}}. The mesh adaptation problem is solved in a single step, i.e., $n_{inc}=1$. The timing for one data transfer step is approximately $50s$ on average, and the average timing for a mesh refitting step is around $t_{mr}=70s$. Finally, the total simulation time is approximately $27h$ for this example, wherein the total computational time required for mesh adaptation accounts for a share of only $\approx 5\%$. This underlines again the efficiency of the proposed mesh adaptation approach, in particular when comparing it to remeshing schemes.

\section{Conclusion}\label{sec:conclusion}
In the present work, a novel mesh adaptation scheme has been proposed for finite element-based models of mechanical, or more general multi-physics, problems involving a strong mesh distortion. The central building block of this mesh adaptation scheme is a novel mesh refitting approach, also denoted as mesh regularization, based on the definition of an element distortion potential considering contributions from different distortion modes such as skewness and aspect ratio of the elements. The regularized mesh is obtained by minimizing this potential. Moreover, based on the concept of spatial localization functions, the method allows to specify tailored requirements on mesh resolution and quality for regions with strongly localized mechanical deformation and mesh distortion. To address also problems involving significant surface deformation, the usage of a mortar mesh-sliding scheme has been proposed to allow for a tangential motion of boundary nodes without changing the boundary topology. To transfer tensor-valued history data from the old to the new mesh, the novel mesh refitting approach is combined with structure-preserving tensor interpolation schemes as proposed in our previous work~\cite{satheesh}.

Based on two elementary test cases, i.e. large deformation mechanical problems involving frictional contact interaction, it has been demonstrated that the mesh refitting approach together with the mesh-sliding scheme enables a significantly improved mesh relaxation as compared to approaches with fixed boundary nodes. Moreover, as a practically relevant application scenario, the thermo-mechanical expansion of materials such as foams involving extreme volume changes by up to two orders of magnitude along with large and strongly localized strains as well as thermo-mechanical contact interaction has been considered. For this scenario it has been demonstrated that the proposed regularization approach preserves a high mesh quality with a maximal element skewness below 30\%. Moreover, in the investigated numerical examples, the computation time for mesh adaptation was typically in the order of only a few percent of the total simulation time. In contrast , simulations without mesh adaption have been shown to lead to significant mesh distortion with larger element aspects ratios and a maximal element skewness close to 100\%, i.e., neighboring element edges that are almost parallel, and eventually, to non-convergence of the numerical solution scheme.

In cases were the global shape change of the discretized mechanical body is very anisotropic, an increase of the element aspect ratios can be reduced, but not completely avoided, with regularization schemes that preserve the mesh connectivity. For such scenarios, a future combination of the proposed mesh regularization scheme with element subdivision procedures is considered promising.

\section*{Acknowledgments}
The authors acknowledge  the financial support from the European Union's Horizon 2020 research and
innovation program under the Marie Skłodowska-Curie grant agreement No 764636.

\setcounter{equation}{0}
\renewcommand{\theequation}{\Alph{section}.\arabic{equation}}
\setcounter{figure}{0}
\renewcommand{\thefigure}{\Alph{section}.\arabic{figure}}

\setcounter{equation}{0}
\renewcommand{\theequation}{\Alph{section}.\arabic{equation}}
\setcounter{figure}{0}
\renewcommand{\thefigure}{\Alph{section}.\arabic{figure}}

\appendix
\section{Residual and system matrix: Distortion potential}\label{appendix_rhs_sysmat}
For sake of simplicity, the distortion potential in~\cref{distortion_potential} is denoted in a abstract from as 
\begin{align}
    \pi_m=\sum_{k=1}^{nco} \frac{1}{2}\ \varepsilon_k\  G_k^2 (\mathbf{d}),
\end{align}
where $\varepsilon_k$ can be $\{\bar{\varepsilon}_{E},  \widehat{\varepsilon}_E, \varepsilon_{A} \}$, $G_k\in \{\bar{G}^i_{E},
    (\widehat{G}_{E})_j,  G^{mn}_{A}\}$, and $nco$ is the number of total constraints. The residual of the distortion potential is given as
\begin{align}
    \mathbf{f}^d_u=\frac{\partial \pi_m}{\partial \mathbf{d}}=\sum_{k=1}^{nco} \varepsilon_k\ G_k \frac{\partial G_k}{\partial \mathbf{d}}.
\end{align}
Exemplary the first derivative of $\bar{G}^i_{E}$~\cref{constraint_avg}, $(\widehat{G}_{E})_j$~\cref{constraint_equal_edges}, and $G^{12}_{A}$~\cref{constratint_angle} reads
\begin{align}
    \frac{\partial \bar{G}^i_{E} }{\partial \mathbf{d}} &= \frac{\bar{\bm{v}}^i}{(\bar{\bm{v}}^i\cdot \bar{\bm{v}}^i)^{1/2}}  \frac{\partial \bar{\bm{v}}^i }{\partial \mathbf{d}},\\
    \frac{\partial (\widehat{G}_{E})_j }{\partial \mathbf{d}} &= \frac{1}{(\bar{\bm{v}}^i\cdot \bar{\bm{v}}^i)} \frac{\partial (\bm{v}_j^i\cdot \bm{v}_j^i)}{\partial \mathbf{d}} - \frac{((\widehat{G}_{E})_j +1)}{(\bar{\bm{v}}^i\cdot \bar{\bm{v}}^i)} \frac{\partial (\bar{\bm{v}}^i\cdot \bar{\bm{v}}^i)}{\partial \mathbf{d}}, \\
     \frac{\partial G^{12}_{A} }{\partial \mathbf{d}} &=
     \frac{1}{\parallel\bm{v}^1_1\parallel\parallel\bm{v}^2_1\parallel}\frac{\partial (\bm{v}^1_1 \cdot \bm{v}^2_1) }{\partial \mathbf{d}}-\frac{1}{2}\frac{(\bm{v}^1_1 \cdot \bm{v}^2_1)}{(\bm{v}^1_1 \cdot \bm{v}^1_1)^{3/2}\parallel\bm{v}^2_1\parallel}\frac{\partial (\bm{v}^1_1 \cdot \bm{v}^1_1) }{\partial \mathbf{d}}-\frac{1}{2}\frac{(\bm{v}^1_1 \cdot \bm{v}^2_1)}{(\bm{v}^2_1 \cdot \bm{v}^2_1)^{3/2}\parallel\bm{v}^1_1\parallel}\frac{\partial (\bm{v}^2_1 \cdot \bm{v}^2_1) }{\partial \mathbf{d}},
\end{align}
respectively. Next, linearization yields the system matrix $\mathbf{K}^d$ as
\begin{align}
   \mathbf{K}^d = \frac{\partial \mathbf{f}^d_u }{\partial \mathbf{d}}=\sum_{k=1}^{nco} \varepsilon_k\ \frac{\partial G_k}{\partial \mathbf{d}} \frac{\partial G_k}{\partial \mathbf{d}} +\sum_{k=1}^{nco} \varepsilon_k\ G_k \frac{\partial^2 G_k}{\partial \mathbf{d}^2}.
\end{align}
Finally, discrete matrix vector system:
\begin{align}
     \mathbf{K}^d \Delta \mathbf{d}= -\mathbf{f}^d_u.
\end{align}

\bibliographystyle{unsrt}
\bibliography{references}

\end{document}